\documentclass[final,5p,times,twocolumn,nofootinbib,nopreprintline]{elsarticle}
\usepackage{amsmath,amssymb,amsfonts}
\usepackage{graphicx}
\usepackage{hyperref}
\usepackage{slashed}
\usepackage{booktabs,tabulary,longtable, multirow,multicol}
\usepackage{mciteplus}
\mciteErrorOnUnknownfalse
\usepackage{xcolor}
\usepackage{soul}
\usepackage{lipsum}

\usepackage{bibentry}
\newcommand{\ignore}[1]{}
\newcommand{\nobibentry}[1]{{\let\nocite\ignore\bibentry{#1}}}

\makeatletter
\def\bibinfo@X@title#1,{\ignorespaces}
\makeatother

\def\azero{\ensuremath{a_{0}(980)} }
\def\atwo{\ensuremath{a_{2}(1320)} }

\def\ftwo{\ensuremath{f_{2}(1270)} }

\def\ggpieta{\ensuremath{\gamma\gamma\to \pi^0\eta} }
\def\ggksks{\ensuremath{\gamma\gamma\to K_S K_S} }
\def\ggkoko{\ensuremath{\gamma\gamma\to K^0 \bar{K}^0} }
\def\ggkk{\ensuremath{\gamma\gamma\to K \bar{K}} }

\def\ggkpkm{\ensuremath{\gamma\gamma\to K^+ K^-} }
\def\ggpietakk{\ensuremath{\gamma\gamma\to \{\pi\eta,KK\}} }

\begin{document}

\title{A dispersive estimate of the $a_0(980)$ contribution to $(g-2)_\mu$}
	
\author{Oleksandra Deineka}
\author{Igor Danilkin}
\author{Marc Vanderhaeghen}
\address{Institut f\"ur Kernphysik \& PRISMA$^+$  Cluster of Excellence, Johannes Gutenberg Universit\"at,  D-55099 Mainz, Germany}
	
\date{\today}

\begin{abstract}

A dispersive implementation of the $a_0(980)$ resonance to $(g-2)_\mu$ requires the knowledge of the double-virtual $S$-wave $\gamma^*\gamma^*\to\pi\eta/ K\bar{K}_{I=1}$ amplitudes. To obtain these amplitudes, we used a modified coupled-channel Muskhelishvili–Omn\`es formalism, with input from the left-hand cuts and the hadronic Omn\`es matrix. The latter was derived using a data-driven $N/D$ method, where the hadronic left-hand cuts were approximated via a conformal expansion. Due to the absence of direct hadronic data in the $\pi\eta$ channel, the expansion coefficients were fitted to various experimental data sets on two-photon fusion processes with $\pi\eta$ and $K\bar{K}$ final states. The resulting dispersive estimate for the $a_0(980)$ contribution to $(g-2)_\mu$ is $a_\mu^{\text{HLbL}}[a_0(980)]_{\text{resc.}}=-0.44(5)\times 10^{-11}$, which presents an order of magnitude improvement in precision over the narrow resonance approximation.
\end{abstract}
	
\maketitle

\section{Introduction}\label{sec:intro}
For decades, the anomalous magnetic moment of the muon $(g-2)_\mu$ has been in the spotlight as a precision tool for testing the Standard Model, due to the progress made in its experimental determination. The latest results from Fermilab achieved a remarkable precision of 0.20 ppm \cite{Muong-2:2023cdq, Muong-2:2024hpx}, leading to the updated world average
\begin{equation}
a_\mu^\text{Exp}= 116\,592\,059(22)\times 10^{-11}\,.
\end{equation}
From the theoretical perspective, the ongoing efforts \cite{Colangelo:2022jxc} are focused on improving upon the Standard Model (SM) result estimated in the 2020 White Paper (WP) \cite{Aoyama:2020ynm,*Aoyama:2012wk,*Aoyama:2019ryr,*Czarnecki:2002nt,*Gnendiger:2013pva,*Davier:2017zfy,*Keshavarzi:2018mgv,*Colangelo:2018mtw,*Hoferichter:2019mqg,*Davier:2019can,*Keshavarzi:2019abf,*Kurz:2014wya,*Melnikov:2003xd,*Masjuan:2017tvw,*Hoferichter:2018kwz,*Gerardin:2019vio,*Bijnens:2019ghy,*Colangelo:2019uex,*Blum:2019ugy,*Colangelo:2014qya,Colangelo:2017qdm,*Colangelo:2017fiz} with an ultimate goal of achieving a comparable uncertainty level.

The present theoretical uncertainty mainly arises from the hadronic contributions, namely hadronic vacuum polarization (HVP) and hadronic light-by-light scattering (HLbL). Compared to HVP, the HLbL contribution is suppressed by an additional power of the electromagnetic coupling constant. Hence, a lot of theoretical efforts in recent years have been dedicated to resolving the tensions in estimating the dominant HVP contribution,  arising both from the data-driven approaches, particularly in the light of the lattest $e^+e^-\to \pi^+\pi^-$ CMD-3 measurement \cite{CMD-3:2023rfe, CMD-3:2023alj}, as well as from lattice QCD results \cite{Boccaletti:2024guq, Borsanyi:2020mff,RBC:2024fic,Djukanovic:2024cmq,Bazavov:2024eou}. 

Nonetheless, to match the precision of the final experimental result from Fermilab, which may surpass the original design sensitivity of 0.14 ppm \cite{Muong-2:2015xgu}, it remains crucial to validate and improve upon the WP estimate of the HLbL contribution \cite{Aoyama:2020ynm,*Aoyama:2012wk,*Aoyama:2019ryr,*Czarnecki:2002nt,*Gnendiger:2013pva,*Davier:2017zfy,*Keshavarzi:2018mgv,*Colangelo:2018mtw,*Hoferichter:2019mqg,*Davier:2019can,*Keshavarzi:2019abf,*Kurz:2014wya,*Melnikov:2003xd,*Masjuan:2017tvw,*Hoferichter:2018kwz,*Gerardin:2019vio,*Bijnens:2019ghy,*Colangelo:2019uex,*Blum:2019ugy,*Colangelo:2014qya,Colangelo:2017qdm,*Colangelo:2017fiz}  
\begin{equation}
a_\mu^\text{HLbL}=92(19)\times 10^{-11}\,.
\end{equation}

Apart from the dominant pseudoscalar $\pi^0, \eta, \eta'$ pole contributions, additional nontrivial contributions to HLbL arise from two-particle intermediate states such as $\pi\pi$, $\pi\eta$, and $K\bar{K}$. Currently, only the contributions from the $\pi\pi_{I=0,2}$ and $K\bar{K}_{I=0}$ channels have been considered in a dispersive framework \cite{Colangelo:2017qdm,*Colangelo:2017fiz, Danilkin:2021icn}. The isospin-0 part of this result can be understood as a model-independent implementation of the contributions from the $f_0(500)$ and $f_0(980)$ resonances. In this paper, we aim to evaluate the contribution from the $a_0(980)$ resonance, which arises from the rescattering of the $\pi\eta/K\bar{K}_{I=1}$ states. In contrast to the $\pi\pi/ K\bar{K}_{I=0}$ systems, a precise analysis of hadronic $\pi\eta/K\bar{K}_{I=1}$ rescattering is hindered by the lack of direct hadronic experimental data. Furthermore, the calculation of the $a_0(980)$ contribution requires knowledge of the double-virtual processes $\gamma^*\gamma^*\to\pi\eta/ K\bar{K}_{I=1}$. Thus, it is natural to utilize the available experimental data on two-photon fusion reactions, which is currently provided by the Belle Collaboration for the real photon case with $\pi\eta$ and $K_S K_S$ final states \cite{Belle:2009xpa, Belle:2013eck}. Additionally, measurements of photon-fusion processes with a single-tagged photon are part of the ongoing two-photon physics program of the BESIII Collaboration \cite{Redmer:2018gah}.

In order to provide theoretical predictions for the single- and double-virtual light-by-light scattering processes, and estimate the contribution of the $a_0(980)$ resonance to HLbL, it is necessary to construct an approach that adheres to the fundamental properties of the $S$-matrix, namely analyticity and coupled-channel unitarity. Such an approach, based on the modified Muskhelishvili-Omnès formalism, has already proven to be effective in describing the $\gamma\gamma\to\pi\pi/K\bar{K}_{I=0}$ \cite{Garcia-Martin:2010kyn, Moussallam:2011zg, Dai:2014zta,*Dai:2014lza, Danilkin:2018qfn,*Danilkin:2019opj, Stamen:2024ocm} and $\gamma\gamma\to\pi\eta/K\bar{K}_{I=1}$ \cite{Danilkin:2017lyn, Lu:2020qeo,Schafer:2023qtl} reactions. Here, we extend the work of \cite{Danilkin:2017lyn} to $\gamma^*\gamma^*\to\pi\eta/K\bar{K}_{I=1}$ by including the relevant kinematic constraints and, at the same time, relaxing some of the chiral constraints to allow for a more data-driven approach.

This paper is organized as follows. First, we present in Sec.~\ref{subsec:hlbl} the formalism for calculating the $a_0(980)$ HLbL contribution to $(g-2)_\mu$, along with the kinematic constraints on the partial-wave amplitudes. In Secs. \ref{subsec:dr} and \ref{subsec:omnes} we focus on the dispersion relations for photon fusion and purely hadronic processes. For the former, we consider possible left-hand cut contributions in Sec. \ref{subsec:lhc} and provide the approximations used for the D-wave amplitudes in Sec. \ref{subsec:dwave}. In Sec. \ref{sec:results}, we discuss the experimental input and present our numerical fits to the cross-section data. Finally, we discuss the pole position corresponding to the $a_0(980)$ and calculate its contribution to HLbL.

\section{Formalism}\label{sec:form}

\subsection{$a_0(980)$ contribution to HLbL scattering}\label{subsec:hlbl}
To compute the HLbL contribution of $a_0(980)$ to $(g-2)_\mu$, we adopt the formalism outlined in \cite{Colangelo:2017qdm,*Colangelo:2017fiz}. This approach yields the following master formula:
\begin{align}\label{eq:master_formula}
a_\mu^{\text{HLbL}} = &\frac{2\alpha^3}{3\pi^2}\int\limits_0^\infty dQ_1\int\limits_0^\infty dQ_2\int\limits_{-1}^1 d\tau\sqrt{1-\tau^2}\,Q_1^3\,Q_2^3\nonumber\\
&\times\sum_{i=1}^{12}T_i(Q_1,Q_2,Q_3)\,\bar{\Pi}_i(Q_1,Q_2,Q_3)\,,
\end{align}
where $\alpha = e^2/(4\pi)$, $\bar{\Pi}_i$ are scalar functions containing the dynamics of the HLbL amplitude, and $T_i$ denote known kernel functions (its explicit form is given in \cite{Colangelo:2017qdm,*Colangelo:2017fiz}). In Eq.(\ref{eq:master_formula}) $\tau$ is defined as $Q_3^2=Q_1^2+2Q_1Q_2\tau+Q_2^2$, where $Q_i^2=-q_i^2$ are the space-like virtualities of photons in the light-by-light scattering. For the $S$-wave, the only contributing scalar functions can be written as (see Eqs. (2.22), and (2.92) of Ref.\cite{Colangelo:2017qdm,*Colangelo:2017fiz})
\begin{align}\label{eq:Pi3Pi9}
&\bar{\Pi}^{J=0}_3(Q_1,Q_2,Q_3)=\frac{1}{\pi}\int\limits_{s_{th}}^\infty ds' \frac{-2}{\lambda_{12}(s')(s'+Q_3^2)^2}\nonumber\\ 
&\times\left(4s'\text{Im}h_{++,++}^{(0)}(s')-(s'-Q_1^2+Q_2^2)(s'+Q_1^2-Q_2^2)\text{Im}h^{(0)}_{00,++}(s')\right)\,,\nonumber\\
&\bar{\Pi}^{J=0}_9(Q_1,Q_2,Q_3)=\frac{1}{\pi}\int\limits_{s_{th}}^\infty ds' \frac{4}{\lambda_{12}(s')(s'+Q_3^2)^2}\nonumber\\ 
&\times\left(2\,\text{Im}h_{++,++}^{(0)}(s')-(s'+Q_1^2+Q_2^2)\,\text{Im}h^{(0)}_{00,++}(s')\right),\nonumber\\
&\bar{\Pi}^{J=0}_4(Q_1,Q_2,Q_3)=\bar{\Pi}^{J=0}_3(Q_1,Q_3,Q_2)\,,\nonumber\\
&\bar{\Pi}^{J=0}_8(Q_1,Q_2,Q_3)=\bar{\Pi}^{J=0}_3(Q_3,Q_2,Q_1)\,,
\end{align}
where $\lambda_{12}(s) \equiv \lambda(s,Q_1^2,Q_2^2)$ is a K\"all\'en triangle function. In Eq.(\ref{eq:Pi3Pi9}), the hadronic light-by-light partial-wave helicity amplitudes are defined as $h^{(J)}_{\lambda_1\lambda_2,\lambda_3\lambda_4}(s')\equiv h^{(J)}_{\lambda_1\lambda_2,\lambda_3\lambda_4}(s',Q_1^2,Q_2^2,Q_3^2)$.

Since $a_0(980)$ is known to have a dynamical coupled-channel $\pi\eta/K\bar{K}_{I=1}$ origin, the inclusion of $K\bar{K}$ intermediate states is necessary. In this case, the unitarity relation implies 
\begin{align}\label{Eq:HLbL_Unitarity}
\text{Im}\,h^{(0),I=1}_{\lambda_1\lambda_2,\lambda_3\lambda_4}&=\bar{h}^{(0)}_{1,\lambda_1\lambda_2}(s,Q_1^2,Q_2^2)\,\rho_{\pi\eta}(s)\,\bar{h}^{(0)*}_{1,\lambda_3\lambda_4}(s,Q_3^2,0)\nonumber\\
&+\bar{k}^{(0)}_{1,\lambda_1\lambda_2}(s,Q_1^2,Q_2^2)\,\rho_{KK}(s)\,\bar{k}^{(0)*}_{1,\lambda_3\lambda_4}(s,Q_3^2,0)\, ,
\end{align}
where $\rho_{\pi\eta}(\rho_{KK})$ denotes the phase space factor of the $\pi\eta\,(K\bar{K})$ system
\begin{equation}
\rho_{a}(s)=\frac{p_{a}(s)}{8\,\pi\sqrt{s}}\,\theta(s-s^{a}_{th})\, , \quad a = \pi\eta, K\bar{K}
\end{equation}
with $p_{a}$ and $s^a_{th}$ being the center-of-mass (c.m.) three momentum and threshold of the corresponding two-meson system. In Eq.(\ref{Eq:HLbL_Unitarity})  $\bar{h}^{(0)}_{1,\lambda\lambda'}(\bar{k}^{(0)}_{1,\lambda\lambda'})$ denotes the $I=1$, $J=0$ Born subtracted\footnote{i.e., $\bar{h}\equiv h$ and $\bar{k}\equiv k-k^{\text{ Born}}$} partial-wave (p.w.) helicity amplitudes of the $\gamma^*(Q_1^2)\gamma^*(Q_2^2)\to \pi\eta (K\bar{K})$ processes\footnote{To maintain consistency with Eq.(\ref{eq:Pi3Pi9}) we follow the conventions from \cite{Colangelo:2017qdm,*Colangelo:2017fiz}, which slightly differ from those in \cite{Danilkin:2018qfn,*Danilkin:2019opj}.} \cite{Jacob:1959at,Garcia-Martin:2010kyn}:
 \begin{align}\label{eq:helicity}
&H_{I,\lambda_1\lambda_2}(s,t)=\kappa_{\lambda_1}^{1}\kappa_{\lambda_2}^{2}\sum_{J}(2J+1)\,h^{(J)}_{I,\lambda_1,\lambda_2}(s)\,d_{\lambda_1-\lambda_2,0}^{J}(\theta)\,,\nonumber \\
&K_{I,\lambda_1\lambda_2}(s,t)=\kappa_{\lambda_1}^{1}\kappa_{\lambda_2}^{2}\sum_{J}(2J+1)\,k^{(J)}_{I,\lambda_1,\lambda_2}(s)\,d_{\lambda_1-\lambda_2,0}^{J}(\theta)\,,\\
&\kappa_{\pm}^i=1,\quad \kappa_{0}^i=\frac{q_i^2}{\xi_i},\quad \xi_i=\sqrt{q_i^2}
\end{align}
where $d_{\lambda,\bar{\lambda}}^{J}(\theta)$ are Wigner rotation functions and $\theta$ denotes the c.m. scattering angle.

The p.w. amplitudes may have kinematic singularities or obey kinematic constraints \cite{Colangelo:2017qdm,*Colangelo:2017fiz, Danilkin:2018qfn,*Danilkin:2019opj}. Therefore, it is crucial to find a transformation to a new basis of amplitudes in order to use the p.w. dispersion relations. For the $S$-wave, the amplitudes free from kinematic constraints can be written as \cite{Colangelo:2017qdm,*Colangelo:2017fiz}
\begin{equation}\label{Eq:constraints}
\bar{h}_{i=1,2}^{(0)} = \frac{\bar{h}_{++}^{(0)}\mp Q_1 Q_2\bar{h}_{00}^{(0)}}{s-s_\text{kin}^{(\mp)}}\, ,\quad s_\text{kin}^{(\pm)}\equiv - (Q_1\pm Q_2)^2\,,
\end{equation}
with $Q_i\equiv \sqrt{Q_i^2}$, suppressing the isospin index for simplicity. Similar relations holds for $\bar{k}^{(0)}_i$. In the case of single virtual or real photons, this constraint arises from the requirement of the soft-photon theorem \cite{Low:1958sn}. For the $D$-wave, the kinematic constraints are more complicated and can be found in \cite{Danilkin:2018qfn,*Danilkin:2019opj, Hoferichter:2019nlq}.

\subsection{Dispersion relations}\label{subsec:dr}
The new set of amplitudes, $\bar{h}^{(0)}_i$ and $\bar{k}^{(0)}_i$, contain only dynamical singularities, namely the right- and left-hand cuts. Therefore, it is possible to write a dispersion relation (modulo subtractions, which will be discussed later) as 
\begin{align}\label{eq:DRGeneral}
&\left(\begin{array}{c}\bar{h}^{(0)}_{i}(s,Q_1^2,Q_2^2)\\
\bar{k}^{(0)}_{i}(s,Q_1^2,Q_2^2)\end{array}\right)=
\int_{L}\frac{d s'}{\pi}\frac{1}{s'-s}\,
\text{Im}\left(\begin{array}{c}\bar{h}^{(0)}_{i}(s',Q_1^2,Q_2^2)\nonumber\\
\bar{k}^{(0)}_{i}(s',Q_1^2,Q_2^2)\end{array}\right)\\
&\hspace{1cm}+\int_{R} \frac{ds'}{\pi}\frac{1}{s'-s}\,\text{Im}\left(\begin{array}{c}\bar{h}^{(0)}_{i}(s',Q_1^2,Q_2^2)\\
\bar{k}^{(0)}_{i}(s',Q_1^2,Q_2^2)\end{array}\right)\,,
\end{align}
where the unitarity condition over the right-hand cut has the following form
\begin{equation}\label{eq:hUnitarity}
\text{Im}\left(\begin{array}{c}\bar{h}^{(0)}_{i}\\
\bar{k}^{(0)}_{i}\end{array}\right)= t^{(0)*}\,\rho\,\left(\begin{array}{c}\bar{h}^{(0)}_{i}\\
\bar{k}^{(0)}_{i}\end{array}\right)\,,
\end{equation}
with $\rho = \text{diag}\{\rho_{\pi\eta}, \rho_{K\bar{K}}\}$.
In Eq.(\ref{eq:hUnitarity}), $t^{(0)}$ is the  hadronic coupled-channel $\pi\eta/ K\bar{K}_{I=1}$ scattering amplitude, normalized such that $\text{Im} (t^{(0)})^{-1}=-\rho$. The solution to Eq.~(\ref{eq:DRGeneral}) is provided by the Muskhelishvili-Omn\`es (MO) method \cite{Omnes:1958hv}. We apply the modified MO representation, which is based on writing a dispersion relation for $(\Omega^{(0)})^{-1}\left(\begin{array}{c}\bar{h}^{(0)}_{i}\\
\bar{k}^{(0)}_{i}\end{array}\right)$ \cite{Pennington:2006dg, Garcia-Martin:2010kyn}, where $\Omega^{(0)}$ is the Omn\`es matrix. This matrix satisfies the following unitarity constraint
\begin{equation}\label{eq:OmnesUnitarity}
\text{Im}\,\Omega_{ab}^{(0)}(s)=\sum_c t_{ac}^{(0)*}(s)\,\rho_c(s)\,\Omega_{cb}^{(0)}(s)\,.
\end{equation} 
The resulting dispersion relation for the $S$-wave amplitudes reads as
\begin{align}\label{eq:DRCC}
&\left(\begin{array}{c}h^{(0)}_{i}(s,Q_1^2,Q_2^2)\\
k^{(0)}_{i}(s,Q_1^2,Q_2^2)\end{array}\right)=\left(\begin{array}{c}0\\
k^{(0), \text{ Born}}_{i}(s,Q_1^2,Q_2^2)\end{array}\right)+\Omega^{(0)}(s) \\ &\hspace{1cm}\times\Bigg[\int\limits_{L}\frac{ds'}{\pi}\,\frac{(\Omega^{(0)}(s'))^{-1}}{s'-s}\text{Im}\left(\begin{array}{c}\bar{h}^{(0)}_{i}(s',Q_1^2,Q_2^2)\\ \bar{k}^{(0)}_{i}(s',Q_1^2,Q_2^2)\end{array}\right)\nonumber\\
&\hspace{1cm} -\int\limits_{R}\frac{ds'}{\pi}\,\frac{\text{Im}(\Omega^{(0)}(s'))^{-1}}{s'-s}\left(\begin{array}{c}0\\ k^{(0),\text{ Born}}_{i}(s',Q_1^2,Q_2^2)\end{array}\right)
\Bigg]\,,\nonumber
\end{align}
where the right-hand cut begins at $s_{th}\equiv s_{th}^{\pi\eta}$.
In principle, the presented form of the MO representation is not unique. As discussed in \cite{Garcia-Martin:2010kyn, Moussallam:2021dpk}, the solution to Eq.~(\ref{eq:DRGeneral}) can be obtained using various forms of MO representations, including those with integrations over only right-hand cuts or both right-hand and left-hand cuts, with equivalence demonstrated in \cite{Hoferichter:2019nlq}. The modified version used in this work, however, has an advantage as it effectively separates the well-known Born left-hand cut from the heavier contributions, which are usually known to a much lesser extent. As a result, it only involves the imaginary part of the Born-subtracted left-hand cut and ensures better high-energy behavior in its partial waves. We will discuss the role of the left-hand cut contributions in more detail in Sec.~\ref{subsec:lhc}.

\subsection{Hadronic Omn\`es matrix}\label{subsec:omnes}

To derive the Omn\`es matrix $\Omega^{(0)}(s)$, which accounts for the $\pi\eta/ K\bar{K}_{I=1}$ rescattering effects, we employ a once-subtracted dispersion relation for the hadronic $S$-wave amplitude
\begin{align}\label{eq:drhadronic}
t^{(0)}_{ab}(s) = U^{(0)}_{ab}(s)+\frac{s}{\pi}\int_{s_{thr}}^\infty\frac{ds'}{s'}\frac{\text{Im }t^{(0)}_{ab}(s')}{s'-s}\, , 
\end{align}
The asymptotically bounded function $U^{(0)}_{ab}(s)$ incorporates the contribution from the subtraction constant and left-hand cuts. The solution to Eq.~(\ref{eq:drhadronic}) can be found using the $N/D$ ansatz \cite{Chew:1960iv}, such that
\begin{equation}\label{eq:nd}
t^{(0)}_{ab}(s)=\sum_c (D^{(0)}_{ac}(s))^{-1}\,N^{(0)}_{cb}(s)\,,
\end{equation}
where the left- and right-hand cuts contributions are separated into the $N$ and $D$ functions, respectively. In the absence of Castillejo-Dalitz-Dyson (CDD) ambiguities \cite{Castillejo:1955ed} or bound states \cite{Danilkin:2020pak}, this ansatz leads to a system of linear integral equations \cite{Luming:1964,* Johnson:1979jy}
\begin{align}\label{eq:ndequations}
	N^{(0)}_{ab}(s)&=U^{(0)}_{ab}(s)+ \nonumber\\
	&\frac{s}{\pi} \sum_{c} \int_{s_{th}}^{\infty}\frac{d s'}{s'}\frac{N^{(0)}_{ac}(s')\,\rho_{c}(s')\,(U^{(0)}_{cb}(s')-U^{(0)}_{cb}(s))}{s'-s}\,, \nonumber\\
	D^{(0)}_{ab}(s)&=\delta_{ab}- \frac{s}{\pi} \int_{s_{th}}^{\infty}\frac{d s'}{s'}\frac{N^{(0)}_{ab}(s')\,\rho_{b}(s')}{s'-s}\,.
\end{align}
The $D^{(0)}_{ab}(s)$ function obtained from Eq.~(\ref{eq:ndequations}) is related to the Omn\`es function as
\begin{equation}\label{Omnes_Dfun}
\Omega^{(0)}_{ab}(s)=\left(D^{(0)}_{ab}(s)\right)^{-1}\,,
\end{equation}
which satisfies the following dispersion relation
\begin{equation}\label{eq:Omnes_direct_sub}
    \Omega_{ab}^{(0)}(s)=\delta_{ab}+\frac{s}{\pi}\int_{s_{th}}^{\infty}\frac{ds'}{s'}\frac{1}{s'-s}\,\text{Im}\,\Omega_{ab}^{(0)}(s')\,.
\end{equation}
The $N/D$ system in Eq.~(\ref{eq:ndequations}) can be solved numerically, with the input of $U^{(0)}_{ab}(s)$ on the right-hand cut.

In a general scattering problem, little is known about the left-hand cuts, aside from their analytic structure in the complex plane. Progress has been made in \cite{Gasparyan:2010xz,*Danilkin:2010xd,*Gasparyan:2011yw,*Gasparyan:2012km} by analytically continuing $U^{(0)}_{ab}(s)$ to the physical region using an expansion in a suitably contracted conformal mapping variable $\xi(s)$ 
\begin{align}
\label{eq:confexpansion}
U^{(0)}_{ab}(s)&\equiv t^{(0)}_{ab}(0)+\frac{s}{\pi}\int_{L}\frac{ds'}{s'}\frac{\text{Im }t^{(0)}_{ab}(s')}{s'-s}\nonumber\\
&\approx \sum_{n=0}^\infty C_{n, ab}\,(\xi_{ab}(s))^n\,,
\end{align}
which is chosen to map the left-hand cut plane onto the unit circle \cite{Frazer:1961zz}. 
The form of $\xi(s)$ for each channel is dictated by the position of the closest left-hand cut branching point, $s_L$, and a point, $s_E$, around which the series is expanded. To ensure the convergence of the conformal expansion in the region where the $S$-wave amplitude dominates, we adopt the following
\begin{equation}\label{Eq:s_E}
\sqrt{s_E}=\frac{1}{2}\,\left(\sqrt{s_{th}}+\sqrt{s_{max}}\right)\,,
\end{equation}
with $\sqrt{s_{max}} = 1.1$ GeV. Note that in the coupled-channel case, $s_{th}$ in Eq. (\ref{Eq:s_E}) refers to the physical threshold for the diagonal terms of $U_{ab}(s)$, while for the off-diagonal terms, it corresponds to the lowest threshold.

In the $\pi\eta,KK$ system, there are three types of left-hand cuts \cite{Albaladejo:2015aca,PhysRev.126.1596}. Each channel features a real-axis cut which runs from $-\infty < s < s_L$, where $s_L(\pi\eta\to\pi\eta) =(m_\eta-m_\pi)^2$,  $s_L(K\bar{K}\to K\bar{K}) = 4(m_K^2-m_\pi^2)$ and $s_L(\pi\eta\to K\bar{K}) = 0$. Additionally, the $\pi\eta\to\pi\eta$ channel has a complex circular cut located at $|s| = m_\eta^2 - m_\pi^2$. In the off-diagonal $\pi\eta \to K\bar{K}$ channel, there is a quasi-circular cut intersecting the real axis at $-(m_\eta^2-m_\pi^2)m_K/(m_K+m_\eta)$ and $m_K(m_\eta^2-m_\pi^2)/(m_\pi+m_K)$. Therefore, for the $\xi_{22}$ function, we use the following conformal map
\begin{align}\label{xi-1}
&\xi_{22}(s)=\frac{\sqrt{s-s_L}-\sqrt{s_E-s_L}}{\sqrt{s-s_L}+\sqrt{s_E-s_L}}\,,\\
&s_L=4(m_K^2-m_\pi^2)\,.\nonumber
\end{align}
For $\xi_{11}$, the conformal map satisfying the above constraints is given by:
\begin{align}\label{xi-2}
&\xi_{11}(s)=-\frac{(\sqrt{s}-\sqrt{s_E})(\sqrt{s} \sqrt{s_E}+s_L)}{(\sqrt{s}+\sqrt{s_E}) (\sqrt{s} \sqrt{s_E}-s_L)}\,,\\
&s_L=m_\eta^2-m_\pi^2\,.\nonumber
\end{align}
Finally, for $\xi_{12}$, we use the conformal map of Eq.~(\ref{xi-2}), but with $s_L = m_K(m_\eta^2-m_\pi^2)/(m_\pi+m_K)$.

In \cite{Danilkin:2020pak}, we performed the extraction of the coupled-channel Omn\`es matrix for the $\pi\pi/K\bar{K}_{I=0}$ system. The fit was based on the latest Roy and Roy--Steiner results for $\pi\pi \to \pi\pi$ \cite{Garcia-Martin:2011iqs} and $\pi\pi \to \bar{K}K$ \cite{Pelaez:2020gnd}, respectively. By analytic continuation to the complex plane, this solution reproduces the $f_0(500)$ and $f_0(980)$ poles in good agreement with Refs.~\cite{Moussallam:2011zg,Caprini:2005zr,Garcia-Martin:2011nna}.

For the $\pi\eta/K\bar{K}_{I=1}$ system, however, experimental data is lacking. In this case, the coefficients in the conformal expansion of Eq.~(\ref{eq:confexpansion}) can be theoretically estimated using chiral perturbation theory $\chi$PT, as demonstrated in \cite{Danilkin:2011fz, Danilkin:2012ua, Danilkin:2017lyn}. However, for the $\pi\eta/K\bar{K}_{I=1}$ system, SU(3) $\chi$PT converges slowly. Instead of relying solely on these estimates, our strategy in this work is to determine the unknown coefficients by fitting them to experimental data from the processes $\gamma\gamma \to \pi\eta/K_S K_S$ \cite{Belle:2009xpa, Belle:2013eck} (see Sec. \ref{subsec:experimental}) and to use the $\chi$PT predictions only as additional constraints. 

In particular, for the $\pi\eta \to K\bar{K}$ channel, we impose an Adler zero \cite{GomezNicola:2001as, Albaladejo:2015aca}. At leading order (LO), it is defined as:
\begin{equation}\label{eq:Adler}
    s_A^\text{LO} = \frac{1}{9}(m_\pi^2+8m_K^2+3 m_\eta^2) \, .
\end{equation}
We also ensure that the $\pi\eta\to\pi\eta$ and $\pi\eta\to K\bar{K}$ amplitudes remain consistent with $\chi$PT at $s_{th}=(m_\pi+m_\eta)^2$. Since the next-to-leading order (NLO) $\chi$PT amplitudes depend on low-energy constants (LECs), we take the average between LO and NLO results, ensuring that the uncertainty range covers all possible LEC values \cite{Bijnens:2014lea, Gasser:1984gg}. Specifically, we find:
\begin{align}\label{eq:chiral_const}
t^{(0)}_{\pi\eta\to \pi\eta}(s_{th}) = +0.03\pm 0.03\,,\nonumber\\
t^{(0)}_{\pi\eta\to KK}(s_{th}) = -0.36\pm 0.14\,.
\end{align}
Similarly, for the Adler zero constraint, we take the range between the LO value from Eq.~(\ref{eq:Adler}) and the NLO value, yielding 
\begin{align}\label{eq:AdlerNLO}
&s_A = 0.251 \pm 0.069\,\text{GeV}^2\,,\nonumber\\
&t^{(0)}_{\pi\eta\to KK}(s_{A})=0\,.
\end{align}

\subsection{Left-hand cuts for \ggpietakk}\label{subsec:lhc}
\begin{figure*}[t]
	\centering
	 \includegraphics[width =0.35\textwidth ]{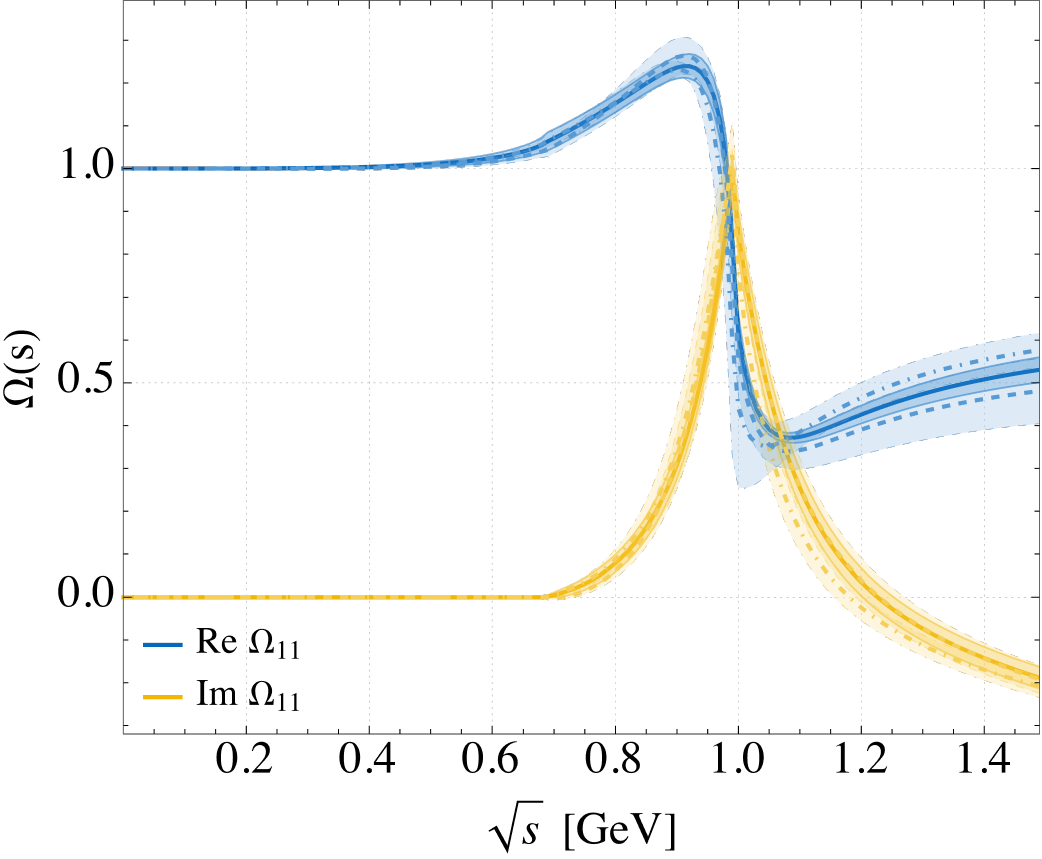}
	 \includegraphics[width =0.36\textwidth ]{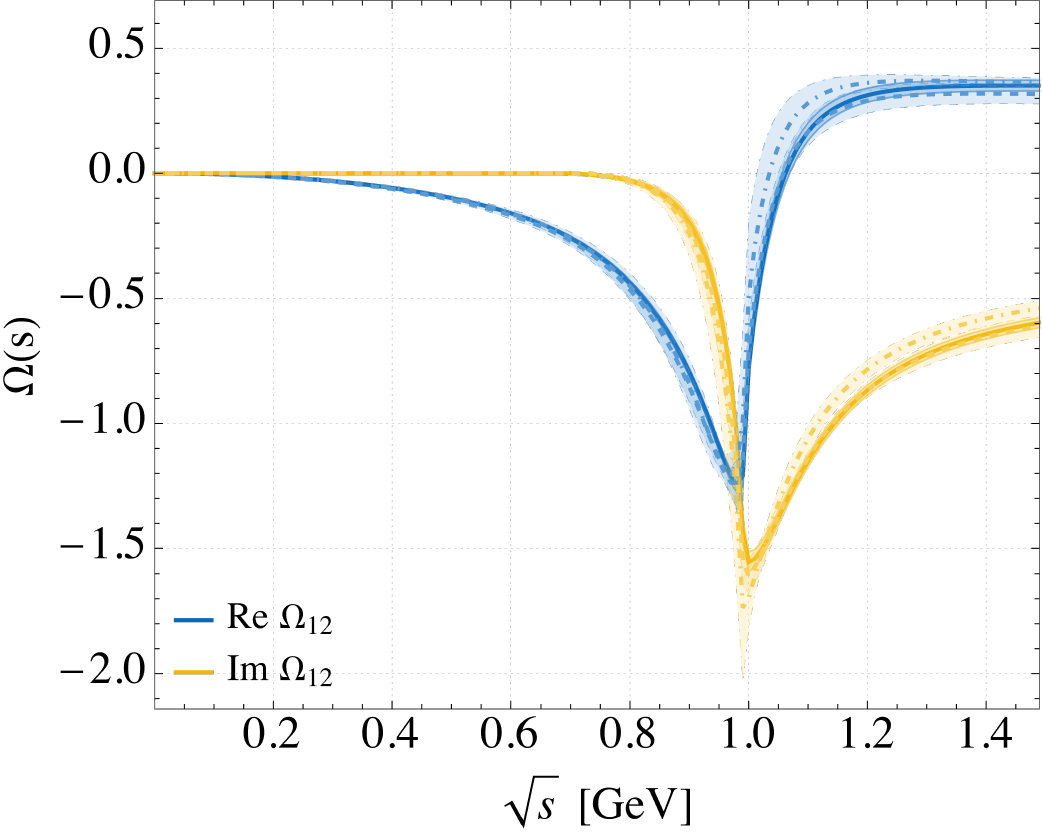}\\
	 \includegraphics[width =0.36\textwidth ]{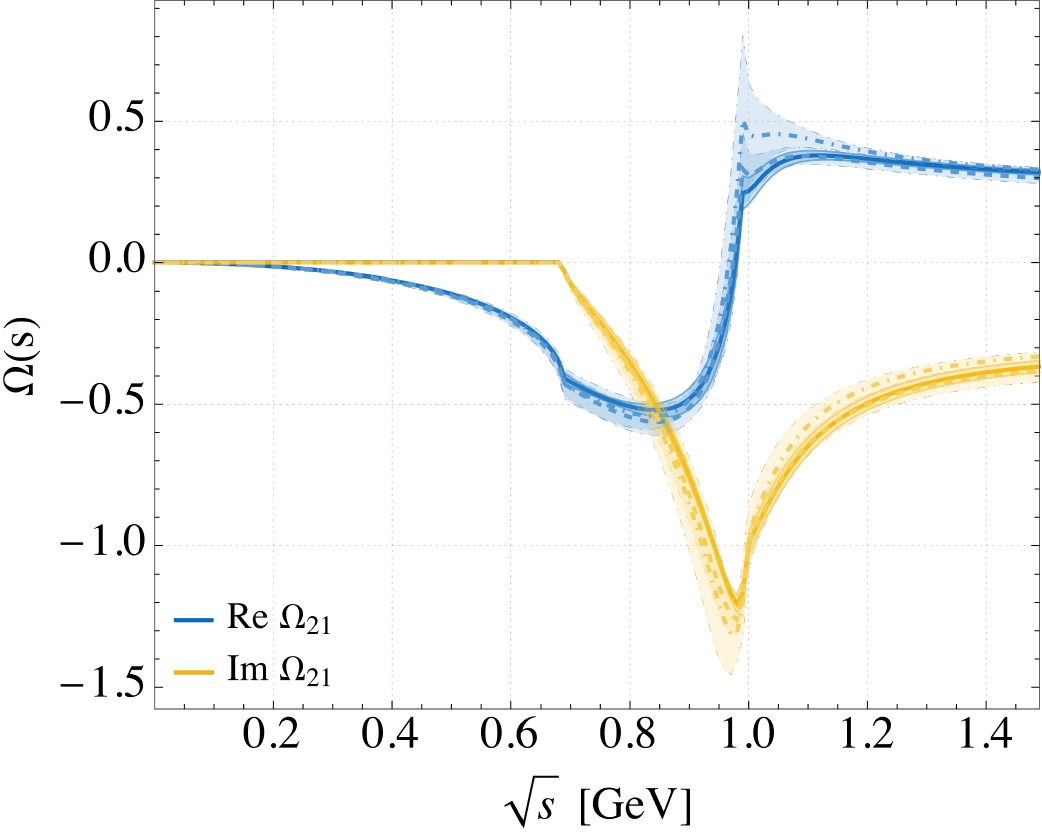}
	 \includegraphics[width =0.35\textwidth ]{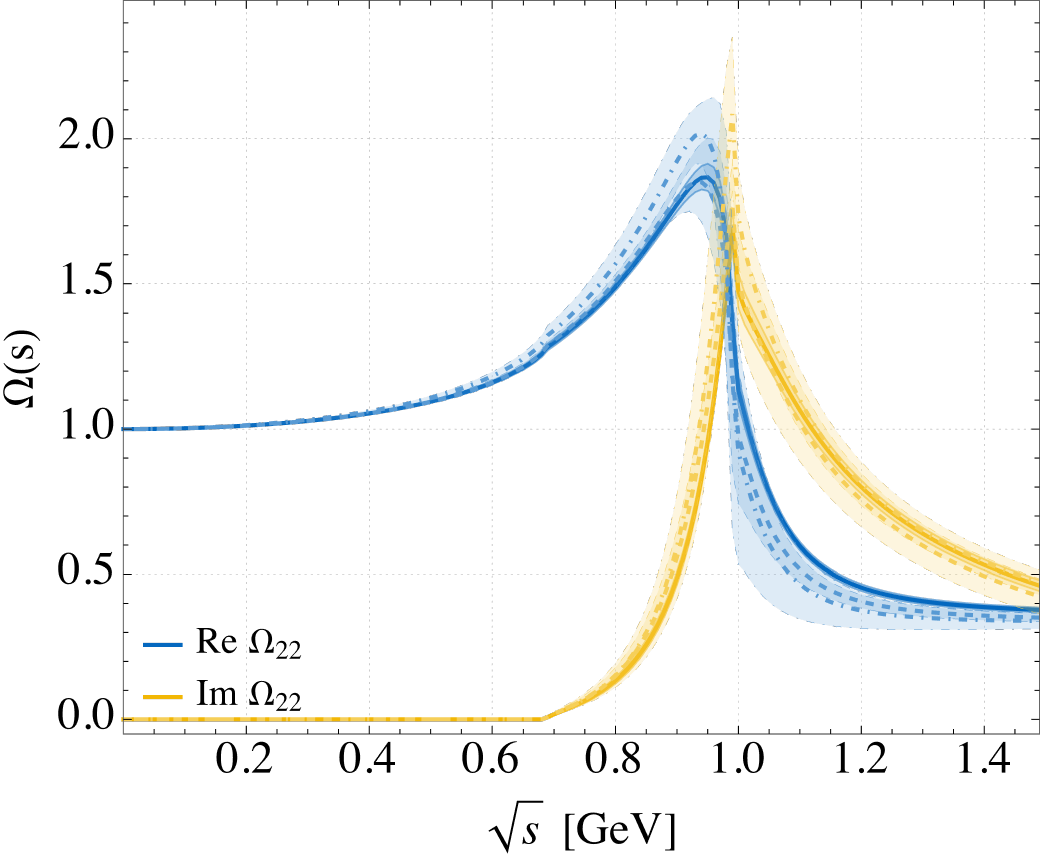}
	\caption{Comparison of the $S$-wave Omn\`es matrix obtained by fitting the \ggpieta and \ggksks   cross-sections to the experimental data from \cite{Belle:2009xpa} and \cite{Belle:2013eck}, respectively, using different treatments of the $\gamma\gamma$ left-hand cut in the dispersion relation. The solid line represents the results using DR in Eq.~(\ref{eq:drBorn}), dashed line - Eq.~(\ref{eq:drv}) and dot-dashed line Eq.~(\ref{eq:drv2}). All results coincide within error bands.}
	\label{fig:omnesdiffdr}
\end{figure*}

Our only experimental input comes from cross-section data for two-photon processes involving real photons, which are used to constrain the hadronic rescattering amplitudes.  It is therefore crucial to carefully examine the treatment of the left-hand cut in the $\gamma\gamma$ partial waves to avoid introducing model dependencies.

Firstly, similar to the treatment of the $\gamma\gamma \to \pi\pi/K\bar{K}_{I=0}$ process in \cite{Colangelo:2017qdm,*Colangelo:2017fiz, Danilkin:2021icn,Danilkin:2018qfn,*Danilkin:2019opj}, we focus solely on the left-hand cut contribution from the Born term. For describing the $S$-wave real-photon processes, it is convenient to work with the dispersion relation expressed in terms of the helicity partial-wave amplitude $h_{I,++}^{(0)}$. Accounting for the relation in Eq.~(\ref{Eq:constraints}), Eq.~(\ref{eq:DRCC}) can be rewritten in the real-photon case as:
\begin{align}\label{eq:drBorn}
&\left(\begin{array}{c}h^{(0)}_{1,++}(s)\\
k^{(0)}_{1,++}(s)\end{array}\right)=\left(\begin{array}{c}0\\
k^{(0), \text{ Born}}_{1,++}(s)\end{array}\right)+\Omega^{(0)}(s) \nonumber\\ &\hspace{4mm}\times\Bigg[-\frac{s}{\pi}\int\limits_{s_{th}}^{\infty}\frac{ds'}{s'}\,\frac{\text{Im}(\Omega^{(0)}(s'))^{-1}}{s'-s}\left(\begin{array}{c}0\\ k^{(0),\text{ Born}}_{1,++}(s')\end{array}\right)
\Bigg]\,,
\end{align}
where $\Omega^{(0)}$ is the hadronic Omnès matrix from Eq.(\ref{Omnes_Dfun}), and $k^{(0), \text{ Born}}_{1,++}$ represents the kaon Born contribution to the $S$-wave given in \cite{Danilkin:2017lyn,Lu:2020qeo}. 

Fitting the photon fusion experimental data reveals that the rescattering of kaon Born terms alone adequately describes the \azero resonance region (see Sec.~\ref{subsec:fits}). Similar to 
the dispersive description of $\gamma\gamma \to \pi\pi$ with only Born left-hand cuts \cite{Colangelo:2017qdm,*Colangelo:2017fiz, Danilkin:2021icn,Danilkin:2018qfn,*Danilkin:2019opj}, the obtained amplitude for $\gamma\gamma \to \pi\eta$ lacks the Adler zero which is predicted from chiral
dynamics. This constraint only slightly affects the behavior of the amplitude near the threshold. To assess the validity of this approximation and the importance of the Adler zero constraint, we additionally consider the contributions from vector-meson exchange left-hand cuts. These contributions are derived from the effective Lagrangian that couples the photon, vector meson ($V$), and pseudoscalar meson ($P$) fields,
\begin{equation}
\mathcal{L}_{VP\gamma} = e \,C_{VP\gamma}\,\epsilon^{\mu\nu\alpha\beta}\,F_{\mu\nu}\,\partial_\alpha P\,V_\beta\, , 
\end{equation}
where $F_{\mu\nu} = \partial_\mu A_\nu - \partial_\nu A_\mu$. The corresponding p.w. helicity amplitudes can be found in \cite{Garcia-Martin:2010kyn, Danilkin:2018qfn,*Danilkin:2019opj, Danilkin:2017lyn, Lu:2020qeo}. The radiative couplings $C_{VP\gamma}$ can be determined from the experimental values for the partial widths of light vector mesons using the relation
\begin{align}
\Gamma_{V\to P\gamma}=\alpha\,C_{VP\gamma}^2\,\frac{(M_V^2-m_P^2)^3}{6M_V^2}\,.
\end{align}
The modulus of the radiative couplings can be also estimated by the SU(3) relations as \footnote{Flavour symmetry also determines the relative sign, though this is not relevant for the subsequent calculations, as these couplings always appear squared.} 
\begin{align}
|g_{VP\gamma}|&\simeq |C_{\rho^{\pm,0}\pi^{\pm,0}\gamma}|\simeq \frac{|C_{\omega\pi^{0}\gamma}|}{3} \simeq \sqrt{3}\,|C_{\omega\eta\gamma}| \simeq \frac{|C_{\rho\eta\gamma}|}{\sqrt{3}}\nonumber\\
&\simeq |C_{K^{*\pm}K^{\pm}\gamma}|\simeq \frac{|C_{K^{*0}K^{0}\gamma}|}{2}\, .
\end{align}
In the following, we use the value of $|g_{VP\gamma}| =0.33(1)$ GeV$^{-1}$ from \cite{Danilkin:2018qfn,*Danilkin:2019opj}, which was the only parameter required for the dispersive description of the $\gamma\gamma\to\pi\pi$ cross-section data around the \ftwo resonance. This value also effectively accounts for contributions from other, heavier left-hand cuts. In Sec.~\ref{subsec:fits} we show, that this value does not require further adjustment in order to describe the $a_0(980)$ region. Moreover, it lies within the PDG average $|g^{\text{PDG}}_{VP\gamma}|=0.4(1)$ GeV$^{-1}$ \cite{ParticleDataGroup:2024cfk}, supporting the validity of approximating left-hand cuts with vector mesons.

The dispersive integral along the vector left-hand cuts remains formally convergent due to the asymptotically bounded behaviour of the $S$-wave Omn\`es function and the imaginary part of the amplitude $\text{Im } h_{1,++}^{(0),V}(-\infty)  \to\text{const}$:
\begin{align}\label{eq:dr}
&\left(\begin{array}{c}h^{(0)}_{1,++}(s)\\
k^{(0)}_{1,++}(s)\end{array}\right)=\left(\begin{array}{c}0\\
k^{(0), \text{ Born}}_{1,++}(s)\end{array}\right)+\Omega^{(0)}(s)\Bigg[ \mathcal{LHC}\nonumber\\ 
&\hspace{10mm}-\frac{s}{\pi}\int\limits_{s_{th}}^{\infty}\frac{ds'}{s'}\,\frac{\text{Im}(\Omega^{(0)}(s'))^{-1}}{s'-s}\left(\begin{array}{c}0\\ k^{(0),\text{ Born}}_{1,++}(s')\end{array}\right)
\Bigg]\,,
\end{align}
where the integral along the vector left-hand cuts are denoted as
\begin{align}\label{eq:lhcint}
\mathcal{LHC} \equiv \frac{s}{\pi}\int\limits_{-\infty}^{s_L^V}\frac{ds'}{s'}\,\frac{(\Omega^{(0)}(s'))^{-1}}{s'-s}\text{Im}\left(\begin{array}{c}h^{(0),V}_{1,++}(s')\\ k^{(0),V}_{1,++}(s')\end{array}\right)\, ,
\end{align}
and $s_L^V$ is the position of the closest left-hand cut branching point given by
\begin{equation}
s_L^V=-\frac{(M_V^2-m_{P,1}^2)(M_V^2-m_{P,2}^2)}{M_V^2}\,.
\end{equation}
However, the left-hand cut contributions acquire significant corrections from integration over large negative values of $s$. To address this, several approaches can be employed. A common method to improve the convergence of the left-hand cut integral in Eq.~(\ref{eq:lhcint}) involves introducing additional subtraction constants \cite{Lu:2020qeo}. The once-subtracted coupled-channel dispersion relation can be written as:
\begin{align}\label{eq:drv}
\left(\begin{array}{c}h^{(0)}_{1,++}(s)\\
k^{(0)}_{1,++}(s)\end{array}\right)&=\left(\begin{array}{c}0\\
k^{(0), \text{ Born}}_{1,++}(s)\end{array}\right)+\Omega^{(0)}(s)\Bigg[\left(\begin{array}{c}a\\
b\end{array}\right)s\nonumber\\ 
&\hspace{-10mm}+\frac{s^2}{\pi}\int\limits_{-\infty}^{s_L^V}\frac{ds'}{s'}\,\frac{(\Omega^{(0)}(s'))^{-1}}{s'(s'-s)}\text{Im}\left(\begin{array}{c}h^{(0),\text{ V}}_{1,++}(s')\\ k^{(0),\text{ V}}_{1,++}(s')\end{array}\right)\nonumber\\
&\hspace{-10mm}-\frac{s^2}{\pi}\int\limits_{s_{th}}^{\infty}\frac{ds'}{s'}\,\frac{\text{Im}(\Omega^{(0)}(s'))^{-1}}{s'(s'-s)}\left(\begin{array}{c}0\\ k^{(0),\text{ Born}}_{1,++}(s')\end{array}\right)
\Bigg]\,,
\end{align}
A formal drawback of this approach is the introduction of the subtraction constants $a$ and $b$, which impact the high-energy behavior of the partial-wave amplitude. Chiral dynamics requires the presence of an Adler zero in the $\gamma\gamma\to\pi^0\eta$ S-wave amplitude, which imposes the constraint on one of the subtraction constant:
\begin{equation}\label{eq:ggAdler}
h^{(0)}_{1,++}(m_\eta^2) = 0\,.
\end{equation}
However, the remaining subtraction constant must still be fixed using experimental data, given the limitations of the SU(3) chiral expansion for kaons.

Another approach involves implementing the conformal mapping technique, allowing us to approximate the contributions beyond the kaon-pole left-hand cut as follows \cite{Ermolina:2024daf}:
\begin{align}\label{eq:drv2}
&\left(\begin{array}{c}h^{(0)}_{1,++}(s)\\
k^{(0)}_{1,++}(s)\end{array}\right)=\left(\begin{array}{c}0\\
k^{(0), \text{ Born}}_{1,++}(s)\end{array}\right)+\Omega^{(0)}(s)\Bigg[ \left(\begin{array}{c}\phi_{\pi\eta}(s)\\
\phi_{K\bar{K}}(s)\end{array}\right)\nonumber\\ 
&\hspace{10mm}-\frac{s}{\pi}\int\limits_{s_{th}}^{\infty}\frac{ds'}{s'}\,\frac{\text{Im}(\Omega^{(0)}(s'))^{-1}}{s'-s}\left(\begin{array}{c}0\\ k^{(0),\text{ Born}}_{1,++}(s')\end{array}\right)
\Bigg]\,,
\end{align}
where 
\begin{equation}
\phi_a \equiv \sum_{n=1}^\infty \tilde{C}_{a,n}\,(\omega_{a}(s))^n\, , \quad a=\pi\eta, K\bar{K}\, ,
\end{equation}
and $\omega_a(s = 0) = 0$ ensures the soft-photon limit. Since all left-hand cuts for the real photon-fusion reaction lie on the real axis, we can utilize the following form $\omega(s)$ from Eq.~(\ref{xi-1})
\begin{equation}
\omega_a(s) = \frac{\sqrt{s-s^V_{L,a}} - \sqrt{-s^V_{L,a}}}{\sqrt{s-s^V_{L,a}} + \sqrt{-s^V_{L,a}}}\,.
\end{equation}
The parameters $\tilde{C}_{a,n}$ must be determined from experimental data on the $\gamma\gamma\to\pi^0\eta/K_S K_S$ cross sections. Similar to the previous method, one of the parameters can be fixed by imposing the Adler zero requirement of Eq.~(\ref{eq:ggAdler}). Another constraint can be derived from the vector meson exchange, leading to the following matching condition \cite{Ermolina:2024daf}:
\begin{align}\label{eq:matching_condition}
\left(\begin{array}{c}\phi_{\pi\eta}''(s)\\
\phi_{K\bar{K}}''(s)\end{array}\right)_{s=0} &= \mathcal{LHC}''\vert_{s=0}\nonumber\\
&\hspace{-5mm}=\frac{2}{\pi}\int\limits_{-\infty}^{s_L^V}\frac{ds'(\Omega^{(0)}(s'))^{-1}}{(s')^3}\text{Im}\left(\begin{array}{c}h^{(0),\text{ V}}_{1,++}(s')\\ k^{(0),\text{ V}}_{1,++}(s')\end{array}\right)\,.
\end{align}
where the right-hand side of the above equation is essentially a sum rule (multiplied by a factor of 2) for the over subtracted $\mathcal{LHC}$ at $s = 0$. The integrand in Eq.(\ref{eq:matching_condition}) is suppressed by an additional power of $s$ in the denominator compared to Eq.(\ref{eq:lhcint}). With these constraints, we find that only one parameter, $\tilde{C}_{K\bar{K},2}$ is needed to describe the data, making this approach equivalent to that of Eq.~(\ref{eq:drv}) in this regard.

In Fig.~\ref{fig:omnesdiffdr}, we show a comparison of the Omn\`es functions obtained from fitting the total cross sections to the $\gamma\gamma\to\pi^0\eta/K_S K_S$ data using Eqs.~(\ref{eq:drBorn}), (\ref{eq:drv}) and (\ref{eq:drv2}) up to 1.4 GeV. We discuss the treatment of the $D$-wave contribution in the next section. The resulting Omn\`es functions demonstrate excellent agreement with one another, thereby justifying the use of the unsubtracted version of the dispersion relation in Eq.~(\ref{eq:drBorn}) with Born left-hand cuts. This finding is significant because, in the absence of experimental data for the single-virtual photon case, there are no constraints on the additional parameters introduced by the vector-meson left-hand cuts. The form of Eqs.~(\ref{eq:drBorn}) for $\gamma\gamma \to \pi\eta/ K\bar{K}_{I=1}$ is analogous to that used for $\gamma\gamma\to\pi\pi/ K\bar{K}_{I=0}$ in \cite{Colangelo:2017qdm,*Colangelo:2017fiz, Danilkin:2021icn,Danilkin:2018qfn,*Danilkin:2019opj} allowing for a straightforward extension to the double virtual case via Eq.(\ref{eq:DRCC}) to calculate the $a_0(980)$ contribution to $(g-2)_\mu$.

\subsection{D-wave amplitudes for \ggpietakk}\label{subsec:dwave}
To describe the experimental data in the region up to 1.4 GeV it is necessary to incorporate the D-wave contributions. In contrast to the single-channel $\gamma\gamma \to \pi\pi$ scattering in the $J=2$ sector \cite{Danilkin:2018qfn,*Danilkin:2019opj}, where the \ftwo resonance was described within a dispersive framework including vector-meson exchange left-hand cuts, applying the same approach to the \atwo resonance is challenging due to the presence of multiple contributing channels. Instead, to account for both the \ftwo and \atwo resonances, we adopt a simple Breit-Wigner approximation, similar to its use for the \ftwo in \cite{Drechsel:1999rf, Hoferichter:2011wk} and for the \atwo in \cite{Danilkin:2017lyn,Lu:2020qeo, Deineka:2018nuh}.

We parametrize the coupling of a tensor resonance ($T$) to two pseudoscalar meson channels ($P$) and two photons using the following effective Lagrangians:
\begin{align}\label{eq:dwavelagrangian}
    \mathcal{L}_{TPP}=g_{T\to PP}T^{\mu\nu}\partial_\mu P\partial_\nu P\,, \nonumber\\
    \mathcal{L}_{T\gamma\gamma} = e^2 g_{T\to \gamma\gamma}T_{\mu\nu}F^{\mu\lambda}F_{\lambda}^\nu\,,
\end{align}
where $F^{\mu\lambda}$ is the electromagnetic tensor and $T_{\mu\nu}$ represents the massive spin-2 field. For this choice of Lagrangian, the physical amplitude with equal photon helicities vanishes\footnote{Alternatively,  a Lagrangian with two coupling constants that introduces a non-zero $\Lambda = 0$ contribution could be used, as in \cite{Lu:2020qeo}. However, we found this contribution to be negligible and chose not to include it, thus reducing the number of fitting parameters.}.

The expressions for the decay widths are then
\begin{align}\label{eq:dwavewidths}
\Gamma_{T\to a} &= \frac{g_{T\to a}^2}{60\pi}\frac{(p_{a}(m_T^2))^5}{m_T^2}\,,\nonumber\\
\Gamma_{T\to \gamma\gamma} &= \frac{e^4m_T^3 g_{T\to \gamma\gamma}^2}{80\pi}\, ,
\end{align}
where $m_T$ is the mass of the tensor resonance, and $a = \pi\eta, K\bar{K}$. Using the Particle Data Group (PDG) values \cite{ParticleDataGroup:2024cfk} for the branching fractions: $B^{\text{PDG}}_{a_2\to\pi\eta}=14.5(1.2)\times 10^{-2}$, 
$B^{\text{PDG}}_{a_2\to K\bar{K}}=4.9(8)\times 10^{-2}$, $B^{\text{PDG}}_{a_2\to\gamma\gamma}=9.4(7)\times 10^{-6}$, and total width $\Gamma^{\text{PDG}}_{\atwo}=0.107(5)$ GeV, the resulting coupling constants for the \atwo resonance are
\begin{align}\label{eq:a2coup}
|g^{\text{PDG}}_{a_2\to \pi\eta}| &= 10.8\pm 0.5 \, \text{GeV}^{-1}, \nonumber \\
|g^{\text{PDG}}_{a_2\to K\bar{K}}| &= 10.5\pm 0.6 \, \text{GeV}^{-1},\nonumber\\
|g^{\text{PDG}}_{a_2\to \gamma\gamma}| &= 0.115\pm 0.005 \, \text{GeV}^{-1}. 
\end{align}
Similarly, for the \ftwo resonance, using Eq.~(\ref{eq:dwavewidths}) and the PDG values for the branching fractions  $B^{\text{PDG}}_{f_2\to KK}=4.6(4)\times 10^{-2}$, $B^{\text{PDG}}_{f_2\to\gamma\gamma}=1.42(24)\times 10^{-5}$  and the total width $\Gamma^{\text{PDG}}_{\ftwo}=0.187(23)$ GeV, the coupling constants are found to be
\begin{align}\label{eq:f2coup}
|g^{\text{PDG}}_{f_2\to K\bar{K}}| &= 15.9\pm 0.9 \, \text{GeV}^{-1},\nonumber\\
|g^{\text{PDG}}_{f_2\to \gamma\gamma}| &= 0.19\pm0.02 \, \text{GeV}^{-1}\, .
\end{align}

In order to derive the Breit-Wigner approximation of the \atwo and \ftwo resonances, we modify the $J=2$ amplitudes obtained from Eq.~(\ref{eq:dwavelagrangian}) by incorporating the energy dependent width (following the approach of the Belle Collaboration \cite{Belle:2009xpa}) and the Blatt-Weisskopf factor
\begin{align}\label{eq:bwamplitude}
    h_{+-}^{(2),\text{BW}}(s)=&\frac{g_{T\to\gamma\gamma}g_{T\to a}}{10\sqrt{6}}\sqrt{\frac{D_2(p_{a}(s)r)}{D_2(p_{a}(m_T^2)r)}\frac{D_2(p_{\gamma\gamma}(s)r)}{D_2(p_{\gamma\gamma}(m_T^2)r)}}\nonumber\\
    &\times\frac{4\, s\, p_{a}^{2}(s)}{m_T^2-s-im_T\Gamma_T(s)}\,,
\end{align}
where  
\begin{align}
        D_2(x)&=\frac{1}{9+3x^2+x^4}\, ,
\end{align}
and $r$ is the effective interaction radius, which is treated as a fit parameter. The amplitude $k_{+-}^{(2),\text{BW}}(s)$ has the same form as in Eq.~(\ref{eq:bwamplitude}). Since the two-photon branching fractions of the \atwo and \ftwo resonances have been determined in the PDG, primarily from the photon-fusion data, we will fit the couplings $g_{a_2,f_2\to \gamma\gamma}$, using (\ref{eq:a2coup}) and (\ref{eq:f2coup}) only as guidance. On the other hand, we will fix the hadronic couplings $g_{a_2\to \pi\eta, KK}$, $g_{f_2\to KK}$ using the values provided by the PDG in (\ref{eq:a2coup}) and (\ref{eq:f2coup}).

The $\ggpieta$ process is purely an $I=1$ channel, with only the \atwo resonance contributing. In contrast, the $\ggkoko$ and $\ggkpkm$ channels receive contributions from both the \ftwo and \atwo resonances, corresponding to $I = 0$ and $I = 1$, respectively. Moreover, in the charged $\ggkpkm$ channel, there is an additional contribution from a $J = 2$ Born term. However, due to the absence of experimental data on the $K\bar{K}$ phase-shifts for $J=2,I=0$, a detailed discussion of the unitarization strategy lies beyond the scope of this study. Instead, we adopt a simplified parametrization for the total $\ggkk$ amplitude $k_{I,+-}^{(2)}(s)$ as follows:
\begin{equation}\label{eq:bornplusres}
    k_{I,+-}^{(2)}(s) = k_{I,+-}^{(2),\text{Born}}(s)+k_{I,+-}^{(2),\text{BW}}(s)\,,
\end{equation}
where the relative sign is fixed from the fit to the data.

\begin{figure}[t!]
\centering
\includegraphics[width =0.45\textwidth ]{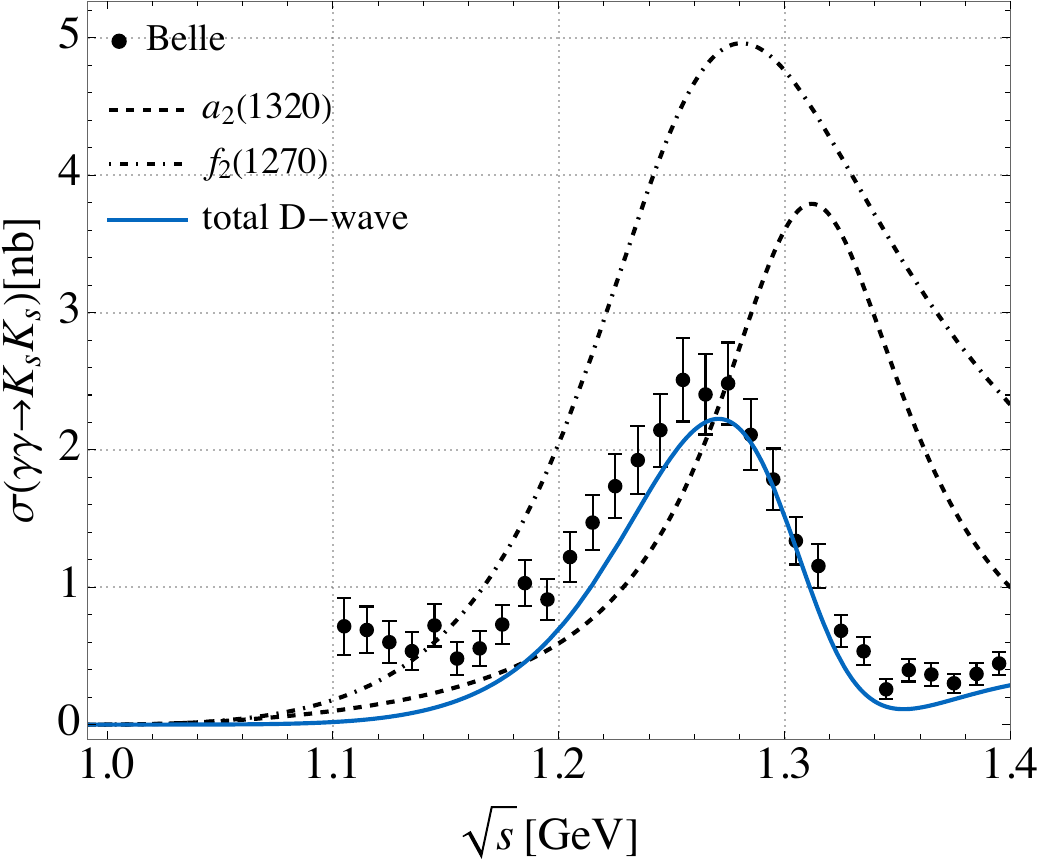}
\includegraphics[width =0.46\textwidth ]{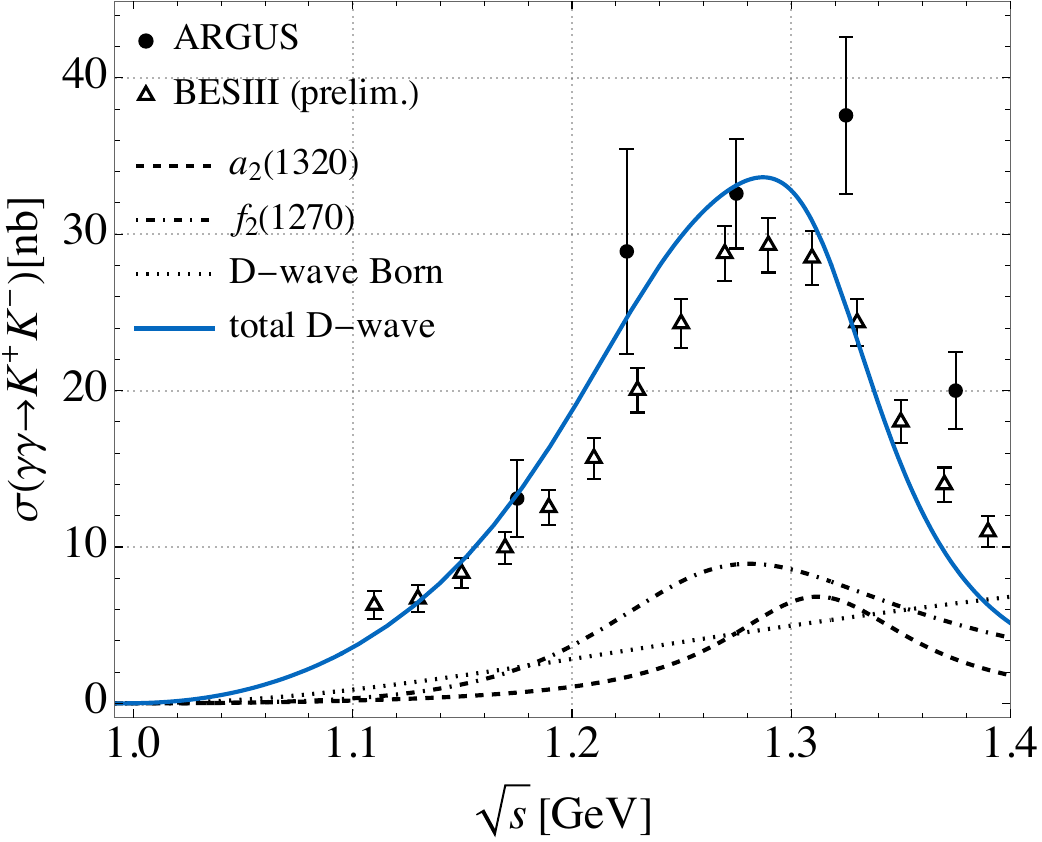}
\caption{D-wave cross sections for the \ggksks ($|\cos\theta|<0.8$) (upper panel) and \ggkpkm ($|\cos\theta|<0.6$) (lower panel) processes compared to the data from \cite{Belle:2013eck, ARGUS:1989ird, Kussner:2022dft,*Kussner:2024ryb}, where the separate \atwo and \ftwo resonances contributions are shown. The parameters are given in Table~\ref{tab:fitresults} using  Eq.~(\ref{eq:drBorn}) for the $S$-wave.}
\label{fig:resinterference}
\end{figure}

The isospin transformation for the kaon amplitudes is given by
\begin{align}\label{isospin}
k^{(J)}_{c,\lambda_1\lambda_2}(s)&=-\frac{1}{\sqrt{2}}\,\left(k^{(J)}_{0,\lambda_1\lambda_2}(s)+k^{(J)}_{1,\lambda_1\lambda_2}(s)\right)\, , \nonumber\\
k^{(J)}_{n,\lambda_1\lambda_2}(s) &= -\frac{1}{\sqrt{2}}\,\left(k^{(J)}_{0,\lambda_1\lambda_2}(s) - k^{(J)}_{1,\lambda_1\lambda_2}(s)\right)\,,
\end{align}
where $c$ denotes the charged channel and $n$ the neutral channel. Therefore, using Eq.~(\ref{eq:bornplusres}), we find that in the physical basis, the following relations hold:
\begin{align}\label{eq:a2f2interference}
    k_{c,+-}^{(2)}(s) &= k_{+-}^{(2),\text{Born}}(s)\nonumber\\
    &\hspace{0.35cm}-\frac{1}{\sqrt2}\left(k_{+-}^{(2),f_2}(s)+k_{+-}^{(2),a_2}(s)\right)\,,\nonumber\\
    k_{n,+-}^{(2)}(s) &= -\frac{1}{\sqrt2}\left(k_{+-}^{(2),f_2}(s)-k_{+-}^{(2),a_2}(s)\right)\,.
\end{align}
Since the couplings $g_{a_2\to K\bar{K}}$ and $g_{f_2\to K\bar{K}}$ have the same sign as expected from nonet symmetry arguments, Eq.~(\ref{eq:a2f2interference}) implies that in the charged channel, the \ftwo and \atwo interfere constructively, whereas in the neutral channel, they interfere destructively \cite{Oller:1997yg}, as shown in Fig. \ref{fig:resinterference}. We have chosen the coupling $g_{a_2 \to \pi\eta}$ to have a positive sign, and thus, by flavor symmetry, the coupling $g_{a_2 \to K\bar{K}}$ has a negative sign.

\section{Results and discussion}\label{sec:results}
\subsection{Experimental input}\label{subsec:experimental}

\begin{table}[t!]
\centering 
\renewcommand*{\arraystretch}{1.2}
\begin{tabular*}{\columnwidth}{@{\extracolsep{\fill}}c|ccc}
\hline\hline
        ~~~ & Eq.~(\ref{eq:drBorn}) & Eq.~(\ref{eq:drv}) & Eq.~(\ref{eq:drv2}) \\
        \midrule
        \multicolumn{4}{c}{$S$-wave parameters}\\
        \midrule
        $C_{0,11}$  & $-8.2(1.9)$  & $-11.8(2.3)$  & $-7.5(1.7)$  \\
        $C_{0,12}$  & $-23.0(2.2)$  & $-24.9(3.9)$  & $-23.0(3.5)$  \\
        $C_{1,12}$  & $38.0(5.6)$  & $40.8(7.6)$  & $35.8(2.6)$  \\
        $C_{0,22}$  & $-33.7(2.6)$ & $-36.4(3.4)$ & $-28.3(1.1)$ \\
        $C_{1,22}$  & $119.9(3.0)$ & $120.4(3.7)$ & $119.8(1.0)$ \\
        $\tilde{C}_{K\bar{K}, 2}$  & $-$ & $-$ & $0.08(1)$ \\
        $b$  & $-$ & $-0.045(2)$ & $-$ \\
        \midrule
        \multicolumn{4}{c}{$D$-wave parameters}\\
        \midrule
        $r_{f_2}$, GeV$^{-1}$ & $2.7(2)$ & $2.5(5)$ & $3.1(5)$\\
        $r_{a_2}$, GeV$^{-1}$  & 5.2(1.3) & $5.0(1.4)$ & $5.7(9)$ \\
        $g_{f_2\to\gamma\gamma}$, GeV$^{-1}$ & $0.188(2)$ & $0.192(7)$ & $0.188(7)$ \\
        $g_{a_2\to\gamma\gamma}$, GeV$^{-1}$  & $0.123(1)$ & $0.124(3)$ & $0.121(4)$ \\
        \midrule
        $\chi^2$/d.o.f & 1.06 & 0.99 & 1.03 \\
        \midrule
        \multicolumn{4}{c}{$\chi$PT constraints}\\
        \midrule
        $s_A$ & 0.279(33)& 0.281(35)& 0.268(34)\\
        $t^{(0)}_{\pi\eta\to\pi\eta}(s_{th})$ & 0.02(4)& 0.03(2)&0.03(3) \\
        $t^{(0)}_{\pi\eta\to KK}(s_{th})$ & -0.41(7)& -0.42(9)& -0.45(10)\\
       
\hline\hline
    \end{tabular*}
    \caption{The fit parameters entering $S$-wave conformal expansion in Eq.~(\ref{eq:confexpansion}) and $D$-wave parametrization in Eq.~(\ref{eq:bwamplitude}), along with the additional $S$-wave parameters needed for Eq.~(\ref{eq:drv}) and Eq.~(\ref{eq:drv2}), as discussed in Sec.~\ref{subsec:lhc}. We also show the resulting values for Eq.~(\ref{eq:chiral_const}), and the position of Adler zero, see Eq.~(\ref{eq:Adler}), (\ref{eq:AdlerNLO}).}
    \label{tab:fitresults}
\end{table}

As mentioned in Sec.~\ref{subsec:omnes}, in the absence of direct hadronic $\pi\eta/K\bar{K}_{I=1}$ data, the coefficients of the conformal expansion Eq.~(\ref{eq:confexpansion}) can be extracted from the data on two-photon collisions cross sections involving $\pi\eta$ and $K\bar{K}$ final states. 

Cross sections for the \ggpieta reaction were first determined by the Crystal Ball Collaboration \cite{CrystalBall:1985mzc}. Later, the Belle Collaboration provided measurements with statistics exceeding all previous experiments of the same reaction by more than two orders of magnitude \cite{Belle:2009xpa}. For the \ggkoko reaction, the early experimental results on cross sections were produced by TASSO \cite{TASSO:1985tme} and CELLO \cite{CELLO:1988xbx}. However, these datasets contained only a few points in the region of interest and feature significant uncertainties. The situation improved with the recent high-statistics data from the Belle Collaboration \cite{Belle:2013eck} in the $K_S K_S$ channel. 

Unfortunately, the data available for the charged $\gamma\gamma\to K^+K^-$, channel, which is needed to better constrain the coupled-channel dynamics, is limited. Results from ARGUS \cite{ARGUS:1989ird} have very low statistics. Much more precise data from the Belle Collaboration \cite{Belle:2003xlt} is available only in the higher energy region above 1.4 GeV and therefore does not provide substantial constraints near the $K\bar{K}$ threshold. In contrast, preliminary results from the BESIII Collaboration \cite{Kussner:2022dft,*Kussner:2024ryb} are promising, providing important input around 1.1 GeV for the \ggkpkm channel, along with additional data for the \ggpieta channel. 

In our fitting procedure, we primarily rely on the high-precision data from the Belle Collaboration on the differential cross sections for the $\gamma\gamma\to\pi^0\eta/K_S K_S$ processes \cite{Belle:2009xpa, Belle:2013eck}. We limit our analysis to the energy region up to 1.4 GeV. Specifically, for the $\ggpieta$ process, we use 28 differential cross sections ranging from 0.85 to 1.39 GeV, and for the $\ggksks$ process, we rely on 30 cross sections covering the energy range from 1.105 to 1.395 GeV. While input from the charged $K^+K^-$ channel is crucial for constraining the coupled-channel hadronic Omn\`es matrix, the preliminary data from the BESIII Collaboration \cite{Kussner:2022dft,*Kussner:2024ryb} is concentrated in the region where interference between the \atwo and \ftwo resonances is expected to dominate. As mentioned in Sec.~\ref{subsec:dwave}, a proper unitarization procedure must be applied in the charged channel to account for the non-resonant background. In our simplified parametrization, shown in Eq.~(\ref{eq:bornplusres}), we do not include the $\gamma\gamma\to K^+K^-$ data in the fits. Instead, we show the prediction of our cross section fit against the BESIII $\ggkpkm$ data.

\subsection{Fit results}\label{subsec:fits}

\begin{figure*}[t!]
\centering
\includegraphics[width =0.45\textwidth ]{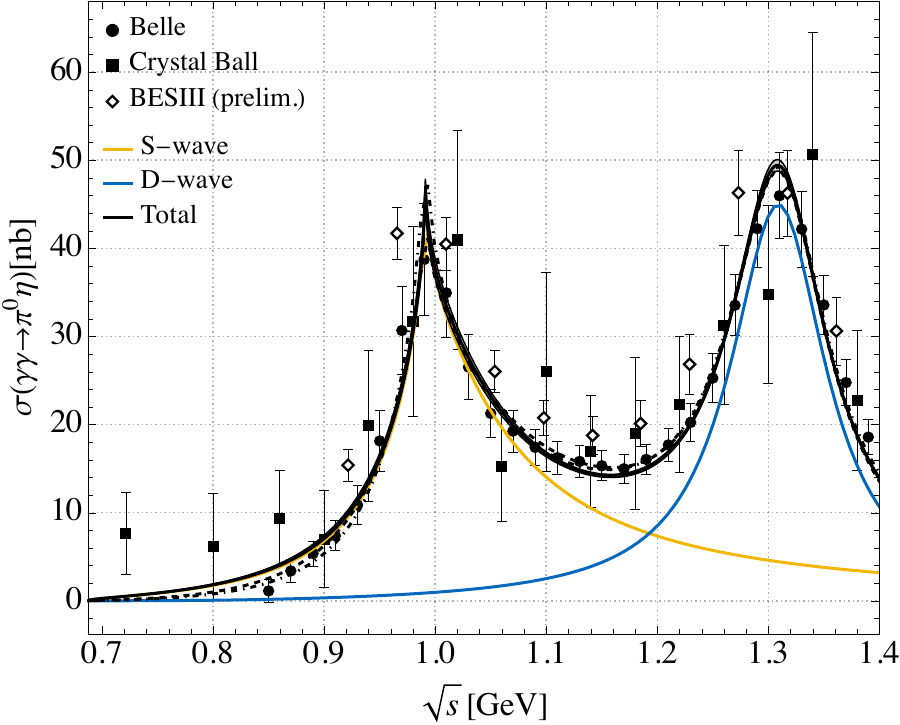}
\includegraphics[width =0.45\textwidth ]{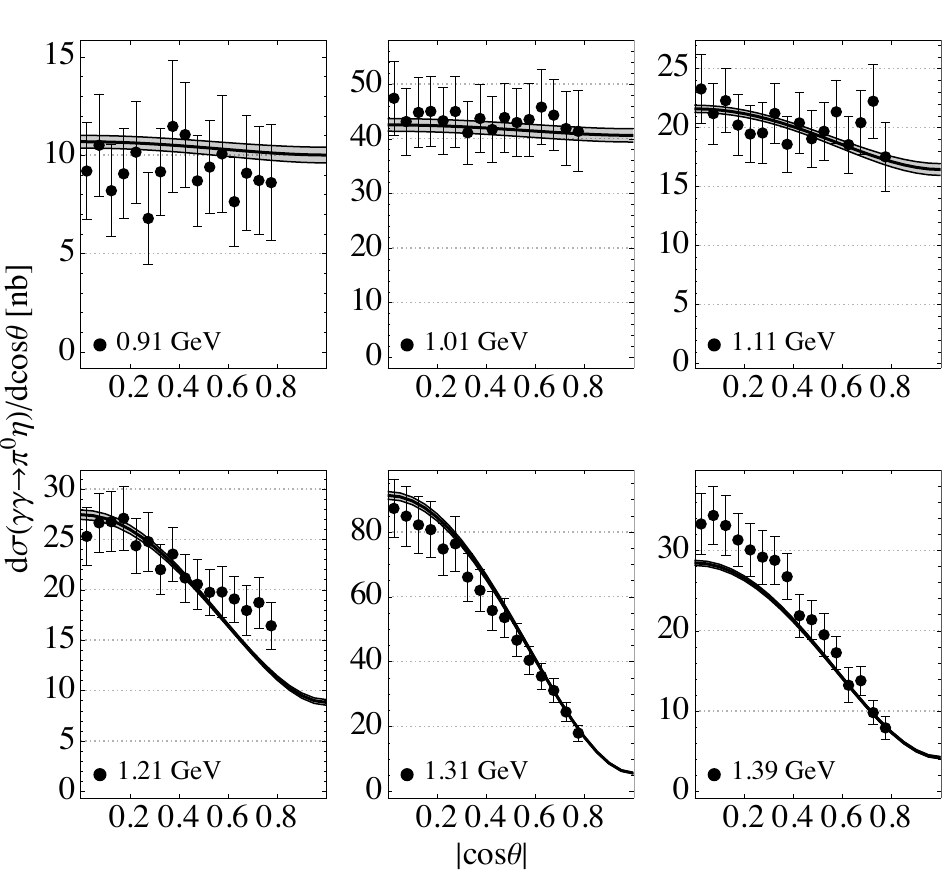}
\caption{Total cross section ($|\cos\theta|<0.8$) (left) and differential cross section (right) for the \ggpieta process compared to the fit results. The data are taken from \cite{Belle:2009xpa, Kussner:2022dft,*Kussner:2024ryb}. Solid line shows the result using Eq.~(\ref{eq:drBorn}), dashed line - Eq.~(\ref{eq:drv}), and dot-dashed line - Eq.~(\ref{eq:drv2})}
\label{fig:ggpieta}
\end{figure*}

\begin{figure*}[t!]
\centering
\includegraphics[width =0.45\textwidth ]{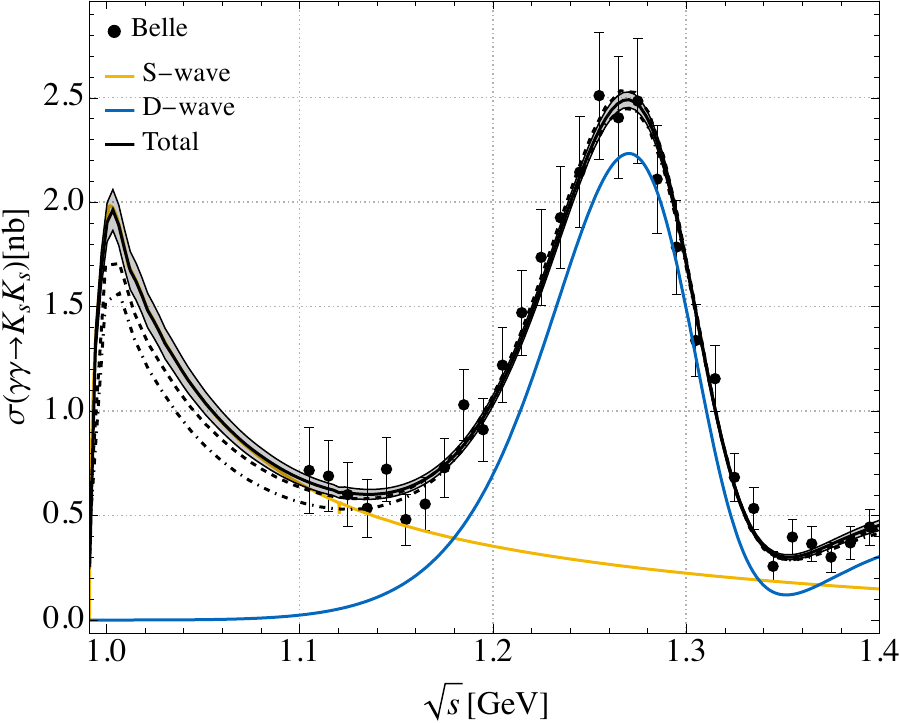}
\includegraphics[width =0.45\textwidth ]{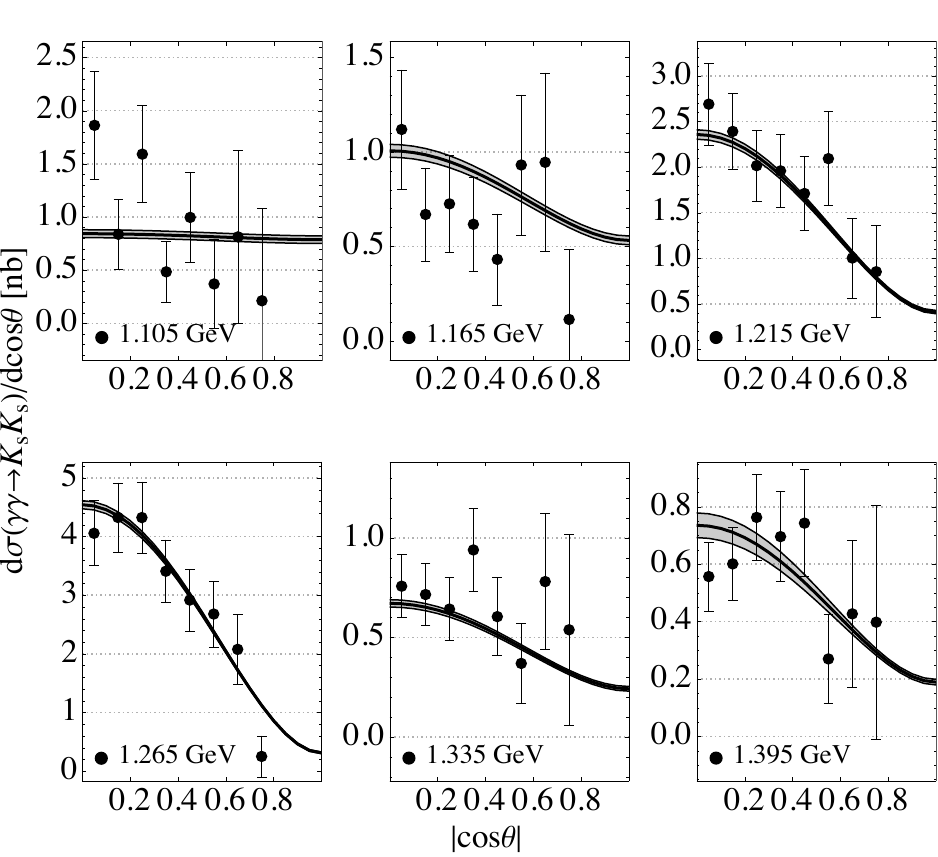}
\caption{Total cross section ($|\cos\theta|<0.8$) (left) and differential cross section (right) for the \ggksks process compared to the fit results. The data are taken from \cite{Belle:2013eck}. Solid line shows the result using Eq.~(\ref{eq:drBorn}), dashed line - Eq.~(\ref{eq:drv}), and dot-dashed line - Eq.~(\ref{eq:drv2})}
\label{fig:ggksks}
\end{figure*}

In general, for the double virtual process $\gamma^*\gamma^*\to\pi\eta/K \bar{K}_{I=1}$ the cross sections can involve two transverse (TT) photon polarizations, two longitudinal (LL) photon polarizations, or a combination of one transverse and one longitudinal (TL) photon polarization \cite{Pascalutsa:2012pr}. In this study, we fit the differential cross section for the real process, which is given by:
\begin{equation}\label{Eq:Cross_section}
    \frac{d\sigma}{d\cos\theta}=\frac{\beta_{a}(s)}{64\,\pi\,s}\,\left(|H_{++}|^2+\,|H_{+-}|^2\right)\,,\quad a = \pi\eta, K\bar{K}
\end{equation}
where $\beta_a = \sqrt{\lambda_a}/s$ and we omit the $TT$ index for clarity. 
To reconstruct the physical $\gamma\gamma\to K_S K_S$\footnote{Ignoring the tiny effect of CP violation, we have $|K_S\rangle=\frac{1}{\sqrt{2}}(|K^0\rangle+|\bar{K}^0\rangle)$ and therefore $\sigma_{\gamma\gamma\to K_S K_S}=\frac{1}{2} \sigma_{\gamma\gamma\to K_0 \bar{K}_0}$} and $\gamma\gamma\to K^+K^-$ cross sections, we use the $I=0$, $S$-wave amplitude $k^{(0)}_{0,++}(s)$ obtained from the coupled-channel $\pi\pi/K\bar{K}_{I=0}$ analysis \cite{Danilkin:2018qfn,*Danilkin:2019opj,Danilkin:2020pak}. For the $I=1$ component, when accounting only for the rescattering of the kaon Born terms following Eq.~(\ref{eq:drBorn}), the only fit parameters in the $S$-wave are the coefficients in the conformal expansion Eq.~(\ref{eq:confexpansion}). The inclusion of vector meson left-hand cuts, as introduced in Eqs.~(\ref{eq:drv}) and (\ref{eq:drv2}) adds an additional parameter that must be constrained by the data, as discussed in Sec.~\ref{subsec:lhc}.
Moreover, additional parameters are required for the $D$-wave description. These include the $\gamma\gamma$ couplings $g^{a_2}_{\gamma\gamma}\,, g^{f_2}_{\gamma\gamma}$ for the \atwo and \ftwo resonances, respectively, as well as the effective interaction radius $r$ for each resonances. Notably, the fitted values of the two-photon couplings for the \atwo and \ftwo resonances show only slight deviations from those provided in Eq.(\ref{eq:a2coup}) and Eq.(\ref{eq:f2coup}).

\begin{figure}[t!]
\centering
\includegraphics[width =0.45\textwidth ]{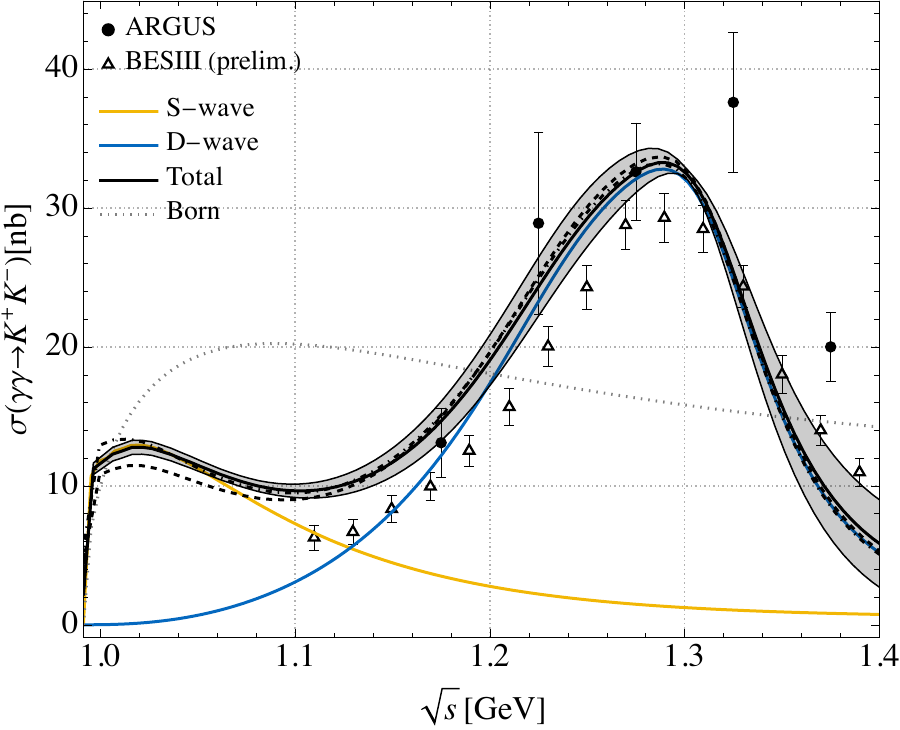}
\caption{Total cross section ($|\cos\theta|<0.6$) for the \ggkpkm processes compared to the predictions of out fits. The data are taken from \cite{ARGUS:1989ird, Kussner:2022dft,*Kussner:2024ryb} and was not used in our fit. Solid line shows the result using Eq.~(\ref{eq:drBorn}), dashed line - Eq.~(\ref{eq:drv}), and dot-dashed line - Eq.~(\ref{eq:drv2})}
\label{fig:ggkpkm}
\end{figure}

The resulting fit parameters are summarized in Table \ref{tab:fitresults}, with statistical uncertainties determined using a bootstrap approach. It is important to note that the treatment of the two-photon left-hand cuts does not significantly influence the parameters of the hadronic conformal expansion given in Eq.~(\ref{eq:confexpansion}), as discussed in Sec.~\ref{subsec:lhc}. The inclusion of the heavier left-hand cuts, along with the Adler zero for the $\gamma\gamma \to \pi^0\eta$ process, only slightly improves the description at low energies (as seen by comparing the dashed/dot-dashed lines to the solid line in Fig.~\ref{fig:ggpieta}). A similar effect was observed in the $\gamma\gamma \to \pi^0\pi^0$ process \cite{Ermolina:2024daf}. We achieve an excellent agreement with the experimental data on differential cross sections using as few as $\{1,2,2\}$ parameters in the $\{11,12,22\}$ channels respectively, as shown in Figs.~\ref{fig:ggpieta},~\ref{fig:ggksks}. The prediction for the total cross section in the $\gamma\gamma\to K^+K^-$ channel, which was not included in our fit shown in Fig.~\ref{fig:ggkpkm}. Similar to the findings in \cite{Oller:1998hw, Achasov:2009ee,*Achasov:2012sc}, we observe a significant reduction in the Born contribution at low energy, attributed to the impact of final state interactions.

At this stage, we would like to comment on other studies that have employed similar modified MO representations. The present work improves upon our earlier analysis presented in \cite{Danilkin:2017lyn,Deineka:2018nuh} in several ways. Firstly, in the present work we included the soft-photon constraint or the Adler zero condition for the $\gamma\gamma \to \pi^0\eta$ process. While the Adler zero serves more as a fine-tuning adjustment, the soft-photon constraint is essential for extending the analysis to the double-virtual case. Secondly, the Omn\`es matrix in \cite{Danilkin:2017lyn,Deineka:2018nuh} was entirely determined by the $\chi$PT input. In the present paper, we implemented all the necessary kinematic constraints and performed a more data-driven analysis. We use $\chi$PT results only as constraints at the threshold and Adler zero values (see Eqs.(\ref{eq:chiral_const}) and (\ref{eq:AdlerNLO}). The work presented in \cite{Lu:2020qeo} employs a chiral K-matrix type representation \cite{Albaladejo:2015aca,Albaladejo:2016mad,*Albaladejo:2017hhj} for $t^{(0)}(s)$ that heavily relies on the choice of $O(p^4)$ chiral parameters. Additionally, six phenomenological polynomial parameters were fitted to the two-photon data, with the Omn\`es matrix obtained from the solution of the homogeneous set of coupled-channel equations \cite{Donoghue:1990xh, Moussallam:1999aq}
\begin{align}\label{eq:Omnes_direct}
    \Omega^{(0)}(s)&=\int_{s_{th}}^{\infty}\frac{ds'}{\pi}\frac{1}{s'-s}\,t^{(0)}(s')\,\rho(s')\,\Omega^{(0)*}(s')\,.\\
    \Omega^{(0)}(0)&=\text{diag}\{1,1\}\nonumber
\end{align}
The unsubtracted form of (\ref{eq:Omnes_direct}) is motivated by the perturbative behaviour of the scalar form factors ($\sim O(1/s)$). However, this constraint imposes particular asymptotic conditions for the phase shifts\footnote{In the n-channel case, it would be $\sum_{n} \delta_{nn}(\infty) \to n\pi$ \cite{FritzNoether,Muskhelishvili}.} $\delta_{11}+\delta_{22}\to 2\pi$ and requires the introduction of an excited $a_0'$ resonance to satisfy it. Moreover, the behaviour $\Omega(s)\sim O(1/s)$ necessitate the use of a subtracted dispersion relation for the $\gamma\gamma \to \pi\eta/K\bar{K}$ process \cite{Lu:2020qeo}. The present study solves a once-subtracted version of (\ref{eq:Omnes_direct}) (see Eq.(\ref{eq:Omnes_direct_sub})) using the $N/D$ method \cite{Chew:1960iv}, which produces an asymptotically bounded Omn\`es matrix, and does not require the inclusion of an excited $a_0'$. Our MO representation accommodates both subtracted and unsubtracted versions, yielding remarkably similar results for the hadronic Omn\`es matrix while still enforcing the necessary soft-photon and soft-pion constraints.

\subsection{\azero resonance parameters}\label{subsec:a0}

In order to identify the parameters of the $a_0(980)$ resonance, we determine the poles in the complex $s$-plane. In the two-channel $\pi\eta/ K\bar{K}$ case there are four Riemann sheets, corresponding to different signs of the imaginary parts of the center of mass momenta $p_a$ \cite{Badalian:1981xj}. The \azero exhibits a cusp-like appearance to its proximity to the $K\bar{K}$ threshold, indicating that multiple Riemann sheets are relevant for determining the pole position.  
In the vicinity of the pole, the elements of the hadronic $t$-matrix can be written as
\begin{equation}\label{t-pole}
t_{ab}^{(0),\text{sheet}}(s)\simeq \frac{g^\text{sheet}_a\,g^\text{sheet}_b}{s^{\text{sheet}}_{a_0}-s}\,,
\end{equation}
where $s^{\text{sheet}}_{a_0}$ denotes the pole position in the corresponding sheet. The couplings $g_{a}$ and $g_{b}$ indicate the strength of the coupling of the resonance to each channel and can be extracted as the residues of the $t$-matrix poles.

\begin{table*}
    \begin{tabular*}{\textwidth}[t]{@{\extracolsep{\fill}}c|ccc}
\hline\hline
        ~~~ & Eq.~(\ref{eq:drBorn}) & Eq.~(\ref{eq:drv}) & Eq.~(\ref{eq:drv2}) \\
        \midrule
        $\sqrt{s_{a_0}^{II}}$,\text{ GeV}  & $1.060(4)-i\,0.072(8)$  & $1.043(11)-i\,0.082(19)$  & $1.040(25)-i\,0.061(17)$  \\
        $|g^{II}_{a_0\to\pi\eta}|$,\text{ GeV}  & $3.81(19)$  & $3.89(46)$  & $3.78(49)$  \\
        $|g^{II}_{a_0\to K\bar{K}}|$,\text{ GeV}  & $5.01(17)$ & $5.18(48)$ & $5.31(60)$ \\
        $|g^{II}_{a_0\to\gamma\gamma}|,\text{ MeV}$  & $8.19(24)$ & $7.22(37)$ & $7.50(1.11)$ \\
        \midrule
        $\sqrt{s_{a_0}^{III}}$,\text{ GeV}  & $0.929(6)-i\,0.085(5)$  & $0.930(21)-i\,0.077(13)$  & $0.931(34)-i\,0.070(11)$  \\
        $|g^{III}_{a_0\to\pi\eta}|$,\text{ GeV}  & $2.99(6)$  & $2.88(17)$  & $2.82(31)$  \\
        $|g^{III}_{a_0\to K\bar{K}}|$,\text{ GeV}  & $2.03(3)$ & $1.91(21)$ & $1.99(18)$ \\
        $|g^{III}_{a_0\to\gamma\gamma}|,\text{ MeV}$  & $9.16(7)$ & $8.70(33)$ & $8.81(63)$ \\
\hline\hline
    \end{tabular*}
    \caption{List of pole positions and couplings. The error is statistical and originates from the fit. The spread between different approaches provides an indication of systematic effects. The averaged value, calculated using the procedure outlined in \cite{Rodas:2021tyb}, is shown in the main text.}
    \label{tab:fitresults_poles}
\end{table*}

In our formalism, we find the poles on the II and III Riemann sheets. The full list of our results using Eqs.~(\ref{eq:drBorn}), (\ref{eq:drv}) and (\ref{eq:drv2}) is shown in Table \ref{tab:fitresults_poles}. For the pole on the II Riemann sheet, $(\text{Sign}[\text{Im}p_{\pi\eta}],\text{Sign}[\text{Im}p_{K\bar{K}}]) = (-+)$, we obtain the following averaged values\footnote{To obtain averaged pole parameters from the three methods, we employed the averaging approach described in \cite{Rodas:2021tyb}. In this approach, central values and errors are calculated as the mean and variance of the combined pseudodatasets resulting from a simultaneous bootstrap analysis of all models.}:
\begin{align}
\sqrt{s_{a_0}^{II}} &= 1.047(18)-i\,0.072(17)\text{ GeV}\, , \nonumber\\
|g^{II}_{a_0\to\pi\eta}| &= 3.82(33)\text{ GeV}\, , \nonumber\\
|g^{II}_{a_0\to K\bar{K}}| &= 5.16(43)\text{ GeV}\, ,\nonumber\\
|g^{II}_{a_0\to\gamma\gamma}|&=7.27(46)\text{ MeV}\,,
\end{align}
whereas for the III Riemann sheet, $(--)$, it holds:
\begin{align}
\sqrt{s_{a_0}^{III}} &= 0.930(25)-i\,0.080(10)\text{ GeV}\, , \nonumber\\
|g^{III}_{a_0\to\pi\eta}| &= 2.91(13)\text{ GeV}\, , \nonumber\\
|g^{III}_{a_0\to K\bar{K}}| &= 1.99(12)\text{ GeV}\, ,\nonumber\\
|g^{III}_{a_0\to\gamma\gamma}|&=8.94(28)\text{ MeV}\, .
\end{align}

It can be seen from the values of $|g_{a_0\to K\bar{K}}|$ and $|g_{a_0\to K\bar{K}}|$, that \azero exhibits strong coupling to both to $\pi\eta$ and $K\bar{K}$ channels, as expected. In our result, both poles on the second and third Riemann sheets are near the physical region: the pole on the RSII sheet is slightly above the $K\bar{K}$ threshold, while the pole on the RSIII sheet is slightly below it. This differs from the recent analyses in \cite{Lu:2020qeo}, where the pole on the RSII sheet was found below the threshold. While both analyses use the same cross-section data as input, there are significant methodological distinctions, as discussed in the previous paragraph. To resolve the differences in the pole position, $\gamma\gamma \to K^+K^-, K_S K_S$ cross-section data near the $K\bar{K}$ thresholds is needed, as its behavior is highly sensitive to the position of the $a_0(980)$ resonance. This can be seen by comparing, for example, the $\gamma\gamma \to K_S K_S$ cross-section, where our result is several times larger than that of \cite{Lu:2020qeo}, but at the same time, it is two times smaller than the result of \cite{Dai:2014zta}.
We also note, that the pole on the II sheet lies somewhat above the range estimated by PDG \cite{ParticleDataGroup:2024cfk}:
\begin{equation}
    \sqrt{s_{a_0}^{\text{PDG}}} = (970 - 1020) - i(30-70)\text{ MeV}\, .
\end{equation}
The broad range of values reflects the current lack of consensus on the precise nature of the \azero resonance, particularly regarding its width \cite{ParticleDataGroup:2024cfk}. Numerous theoretical studies have sought to clarify the properties of the \azero \cite{Oller:1998hw, Oller:1998zr, GomezNicola:2001as, Guo:2011pa,Baru:2004xg, Albaladejo:2015aca, Lu:2020qeo}, with the pole position often found on the II or III Riemann sheets or both. Additionally, lattice QCD studies of $\pi\eta$ scattering \cite{Dudek:2016cru,Wilson:2016rid} have identified a pole on the IV Riemann sheet, which remains stable across variations in K-matrix parametrizations. However, these results have so far been obtained for unphysical pion mass values of $m_\pi = 391$ MeV.

\subsection{\azero contribution to HLbL}\label{subsec:a0HLbL}

Using the hadronic Omn\`es matrix within the unsubtracted dispersion relation framework in Eq.~(\ref{eq:drBorn}), we can now proceed to calculate the amplitudes for $\gamma^*\gamma^*\to \pi\eta/ K\bar{K}$ using Eq.~(\ref{eq:DRCC}) and by considering only the kaon-pole left-hand cut. The generalization of the kaon-pole left-hand cut contribution $k_i^{(0),\text{Born}}$ to the case involving off-shell photons is achieved by multiplying the scalar QED result by the electromagnetic kaon form factor \cite{Fearing:1996gs, Colangelo:2015ama}. The latter is parametrized using the vector meson dominance (VMD) model,
\begin{equation}
    F_K^V(Q^2)=\frac{1}{2}\frac{M_\rho^2}{M_\rho^2+Q^2}+\frac{1}{6}\frac{M_\omega^2}{M_\omega^2+Q^2}+\frac{1}{3}\frac{M_\phi^2}{M_\phi^2+Q^2}
\end{equation}
We have verified that within the $Q^2\lesssim 1$ GeV$^2$ range, which is crucial for the $a_\mu$ calculation, VMD is consistent with a simple monopole fit to the existing data \cite{Deineka:2019bey} and with the dispersive estimate from \cite{Stamen:2022uqh}. In the $a_\mu$ calculation, we account for the spread between these models as an additional source of uncertainty due to the mismatch in form factors for heavier-than-kaon-pole left-hand cuts.

In principle, there also the contribution from the rescattering inside the neutral kaon box. However, this effect is significantly suppressed.
The neutral kaon box itself is already at the level of $10^{-15}$ \cite{Aoyama:2020ynm,*Aoyama:2012wk,*Aoyama:2019ryr,*Czarnecki:2002nt,*Gnendiger:2013pva,*Davier:2017zfy,*Keshavarzi:2018mgv,*Colangelo:2018mtw,*Hoferichter:2019mqg,*Davier:2019can,*Keshavarzi:2019abf,*Kurz:2014wya,*Melnikov:2003xd,*Masjuan:2017tvw,*Hoferichter:2018kwz,*Gerardin:2019vio,*Bijnens:2019ghy,*Colangelo:2019uex,*Blum:2019ugy,*Colangelo:2014qya}. Rescattering corrections to this process would naturally be even smaller and therefore negligible for the purposes of $(g-2)_\mu$. As such, this effect is not included in the present work.

\begin{figure*}[t!]
\centering
\includegraphics[width =0.45\textwidth ]{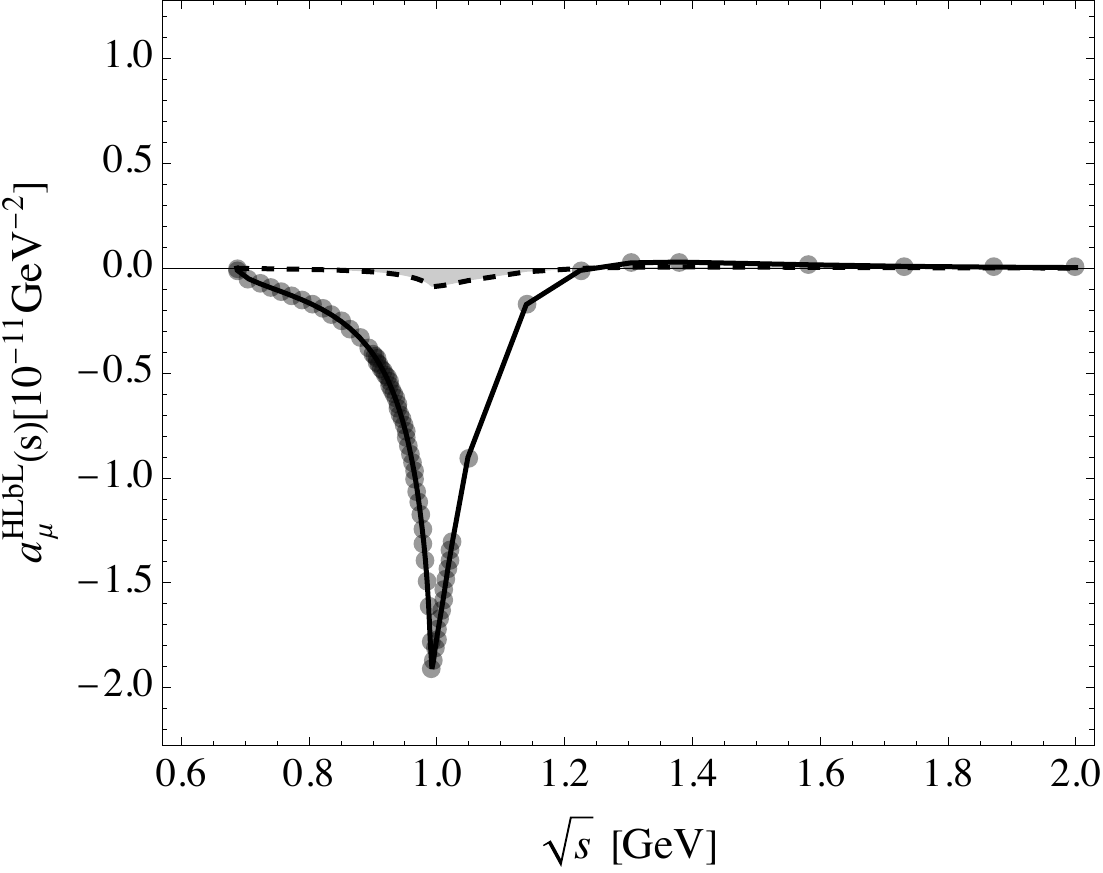}\quad\quad
\includegraphics[width =0.45\textwidth ]{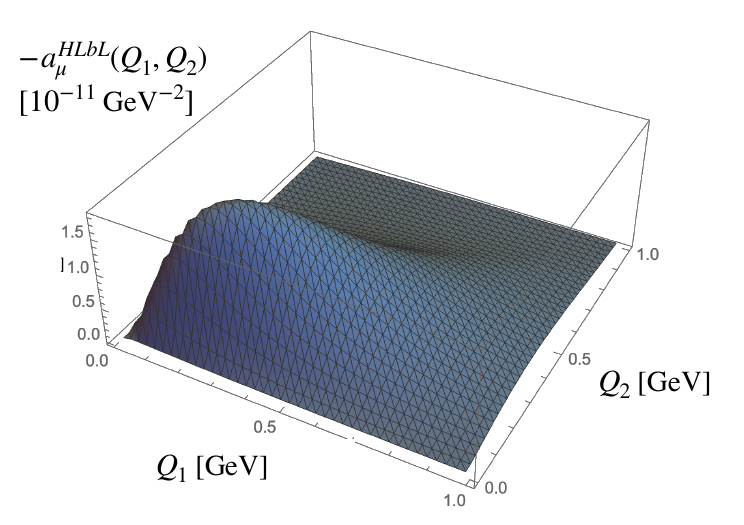}
\caption{Left: the integrand in Eq.~(\ref{eq:amuint_2}), where gray shaded area shows the sum rule violation, reflecting the choice of the HLbL basis \cite{Colangelo:2017qdm,*Colangelo:2017fiz}. Right: the integrand in Eq.~(\ref{eq:amuint_1}) showing the range of relevant photon virtualities for $a_\mu^{\text{HLbL}}[a_0(980)]_{\text{resc.}}$.}
\label{fig:amu}
\end{figure*}

Using the input from $\gamma^*\gamma^*\to \pi\eta/ K\bar{K}$ partial wave helicity amplitudes, the $a_0(980)$ contribution to the HLbL in $(g-2)_\mu$ can be calculated using Eqs.~(\ref{eq:master_formula}), (\ref{eq:Pi3Pi9}) and (\ref{Eq:HLbL_Unitarity}).
It is instructive to first display the integrands before proceeding to the full contribution:
\begin{align}\label{eq:amuint_1}
a_\mu^\text{HLbL}&=\int\limits_0^\infty dQ_1\int\limits_0^\infty dQ_2\, \frac{d\,a_\mu^\text{HLbL}}{dQ_1\,dQ_2} \\
\label{eq:amuint_2}
&=\int\limits_{s_{th}}^\infty ds'\, \frac{d\,a_\mu^\text{HLbL}}{d s'}
\end{align}
As shown in Fig.~\ref{fig:amu}, the largest contribution to $a_\mu^\text{HLbL}$ comes from the region of low photon virtualities. In this context, the forthcoming data from the BESIII Collaboration, which will cover the range $Q^2 = 0.1 - 3.0$ GeV$^2$, is especially valuable \cite{Redmer:2018gah}. Additionally, in Fig.~\ref{fig:amu} one can observe the line-shape of the $a_0(980)$ contribution along with sum rule violation, reflecting the choice of the HLBL basis \cite{Colangelo:2017qdm,*Colangelo:2017fiz}. 
After performing the integration, the contribution of \azero to the muon anomalous magnetic moment is given by
\begin{equation}\label{Eq:amu_final}
    a_\mu^{\text{HLbL}}[a_0(980)]_{\text{resc.}}=-0.44(3)(3)(2)\times 10^{-11}\,.
\end{equation}
The central value arises from the arithmetic average of $-0.43(1)$, $-0.44(2)$, and $-0.46(3)$, which correspond to the Omn\`es matrix input obtained using Eqs.(\ref{eq:drBorn}), (\ref{eq:drv}), and (\ref{eq:drv2}), respectively (see Fig.\ref{fig:omnesdiffdr} and Table~\ref{tab:fitresults} for more details). The first uncertainty is taken as the largest individual uncertainty among these three results. The second uncertainty arises from the spread among different models for the kaon form factor. The third uncertainty stems from sum rule violation. This result in Eq.(\ref{Eq:amu_final}) is one order of magnitude more precise than the outcome from the narrow-width approximation $a_\mu^{\text{HLbL}}[a_0(980)]_{\text{NWA}}=-\left([0.3,0.6]^{+0.2}_{-0.1}\right)\times 10^{-11}$ \cite{Danilkin:2021icn},
where the range reflects variations in the scale of the transition form factor parametrization, taken from the quark model \cite{Schuler:1997yw}, with the two-photon width input from \cite{Lu:2020qeo}.

\section{Conclusion and outlook}\label{sec:conc}

In this work, we presented a dispersive analysis of the $\gamma\gamma\to \pi^0\eta/K\bar{K}$ reactions from threshold up to 1.4 GeV. The hadronic Omn\`es matrix was directly constrained using data from two-photon differential cross sections with $\pi^0\eta$ and $K_S K_S$ final states. We systematically compared subtracted and unsubtracted dispersive formalisms to minimize model dependence in the extracted parameters, allowing us to investigate the properties of the $a_0(980)$ resonance. Notably, we identified two corresponding poles on the second and third Riemann sheets.

The use of dispersion relations allowed to estimate the \azero HLbL contribution with an order of magnitude better precision compared to the narrow resonance approximation. The final result is of the same order as the charged kaon box contribution in HLbL scattering \cite{Stamen:2022uqh}, resolving another small but important piece of the overall puzzle. Additionally, the produced amplitude serves as a Monte Carlo generator for analyzing photon-fusion processes with a single tagged photon by the BESIII Collaboration \cite{Redmer:2018gah}.

Future work will involve incorporating new experimental data into the current analysis. The hadronic $\pi\eta/ K\bar{K}_{I=1}$ rescattering will be further constrained by including existing data from the $\phi\to\gamma\pi\eta$ \cite{KLOE:2009ehb} and $\eta' \to \pi\pi\eta$ \cite{BESIII:2017djm} decays.
Furthermore, we look forward to the final results on the 
$\gamma\gamma\to K^+K^-$ process from BESIII, which will include both total and differential cross sections \cite{Kussner:2022dft,*Kussner:2024ryb}.

\section*{Acknowledgements}
This work was supported by the Deutsche Forschungsgemeinschaft (DFG, German Research Foundation) within the Research Unit [Photon-photon interactions in the Standard Model and beyond, Projektnummer 458854507 - FOR 5327].
	
\bibliographystyle{apsrevM}
\bibliography{bibliography.bib}

\ifx\mcitethebibliography\mciteundefinedmacro
\PackageError{apsrevM.bst}{mciteplus.sty has not been loaded}
{This bibstyle requires the use of the mciteplus package.}\fi
\begin{mcitethebibliography}{110}
\expandafter\ifx\csname natexlab\endcsname\relax\def\natexlab#1{#1}\fi
\expandafter\ifx\csname bibnamefont\endcsname\relax
  \def\bibnamefont#1{#1}\fi
\expandafter\ifx\csname bibfnamefont\endcsname\relax
  \def\bibfnamefont#1{#1}\fi
\expandafter\ifx\csname citenamefont\endcsname\relax
  \def\citenamefont#1{#1}\fi
\expandafter\ifx\csname url\endcsname\relax
  \def\url#1{\texttt{#1}}\fi
\expandafter\ifx\csname urlprefix\endcsname\relax\def\urlprefix{URL }\fi
\providecommand{\bibinfo}[2]{#2}
\providecommand{\eprint}[2][]{\url{#2}}

\bibitem[{\citenamefont{Aguillard et~al.}(2023)}]{Muong-2:2023cdq}
\bibinfo{author}{\bibfnamefont{D.~P.} \bibnamefont{Aguillard}} \bibnamefont{et~al.} (\bibinfo{collaboration}{Muon g-2}), \bibinfo{journal}{Phys. Rev. Lett.} \textbf{\bibinfo{volume}{131}}, \bibinfo{pages}{161802} (\bibinfo{year}{2023}), \eprint{2308.06230}\relax
\mciteBstWouldAddEndPuncttrue
\mciteSetBstMidEndSepPunct{\mcitedefaultmidpunct}
{\mcitedefaultendpunct}{\mcitedefaultseppunct}\relax
\EndOfBibitem
\bibitem[{\citenamefont{Aguillard et~al.}(2024)}]{Muong-2:2024hpx}
\bibinfo{author}{\bibfnamefont{D.~P.} \bibnamefont{Aguillard}} \bibnamefont{et~al.} (\bibinfo{collaboration}{Muon g-2}), \bibinfo{journal}{Phys. Rev. D} \textbf{\bibinfo{volume}{110}}, \bibinfo{pages}{032009} (\bibinfo{year}{2024}), \eprint{2402.15410}\relax
\mciteBstWouldAddEndPuncttrue
\mciteSetBstMidEndSepPunct{\mcitedefaultmidpunct}
{\mcitedefaultendpunct}{\mcitedefaultseppunct}\relax
\EndOfBibitem
\bibitem[{\citenamefont{Colangelo et~al.}(2022)}]{Colangelo:2022jxc}
\bibinfo{author}{\bibfnamefont{G.}~\bibnamefont{Colangelo}} \bibnamefont{et~al.}, \bibinfo{journal}{Contribution to Snowmass 2021}  (\bibinfo{year}{2022}), \eprint{2203.15810}\relax
\mciteBstWouldAddEndPuncttrue
\mciteSetBstMidEndSepPunct{\mcitedefaultmidpunct}
{\mcitedefaultendpunct}{\mcitedefaultseppunct}\relax
\EndOfBibitem
\bibitem[{\citenamefont{Aoyama et~al.}(2020)}]{Aoyama:2020ynm}
\bibinfo{author}{\bibfnamefont{T.}~\bibnamefont{Aoyama}} \bibnamefont{et~al.}, \bibinfo{journal}{Phys. Rept.} \textbf{\bibinfo{volume}{887}}, \bibinfo{pages}{1} (\bibinfo{year}{2020}), \eprint{2006.04822}\relax
\mciteBstWouldAddEndPuncttrue
\mciteSetBstMidEndSepPunct{\mcitedefaultmidpunct}
{\mcitedefaultendpunct}{\mcitedefaultseppunct}\relax
\EndOfBibitem
\bibitem[{\citenamefont{Aoyama et~al.}(2012)\citenamefont{Aoyama, Hayakawa, Kinoshita, and Nio}}]{Aoyama:2012wk}
\bibinfo{author}{\bibfnamefont{T.}~\bibnamefont{Aoyama}}, \bibinfo{author}{\bibfnamefont{M.}~\bibnamefont{Hayakawa}}, \bibinfo{author}{\bibfnamefont{T.}~\bibnamefont{Kinoshita}}, \bibnamefont{and} \bibinfo{author}{\bibfnamefont{M.}~\bibnamefont{Nio}}, \bibinfo{journal}{Phys. Rev. Lett.} \textbf{\bibinfo{volume}{109}}, \bibinfo{pages}{111808} (\bibinfo{year}{2012}), \eprint{1205.5370}\relax
\mciteBstWouldAddEndPuncttrue
\mciteSetBstMidEndSepPunct{\mcitedefaultmidpunct}
{\mcitedefaultendpunct}{\mcitedefaultseppunct}\relax
\EndOfBibitem
\bibitem[{\citenamefont{Aoyama et~al.}(2019)\citenamefont{Aoyama, Kinoshita, and Nio}}]{Aoyama:2019ryr}
\bibinfo{author}{\bibfnamefont{T.}~\bibnamefont{Aoyama}}, \bibinfo{author}{\bibfnamefont{T.}~\bibnamefont{Kinoshita}}, \bibnamefont{and} \bibinfo{author}{\bibfnamefont{M.}~\bibnamefont{Nio}}, \bibinfo{journal}{Atoms} \textbf{\bibinfo{volume}{7}}, \bibinfo{pages}{28} (\bibinfo{year}{2019})\relax
\mciteBstWouldAddEndPuncttrue
\mciteSetBstMidEndSepPunct{\mcitedefaultmidpunct}
{\mcitedefaultendpunct}{\mcitedefaultseppunct}\relax
\EndOfBibitem
\bibitem[{\citenamefont{Czarnecki et~al.}(2003)\citenamefont{Czarnecki, Marciano, and Vainshtein}}]{Czarnecki:2002nt}
\bibinfo{author}{\bibfnamefont{A.}~\bibnamefont{Czarnecki}}, \bibinfo{author}{\bibfnamefont{W.~J.} \bibnamefont{Marciano}}, \bibnamefont{and} \bibinfo{author}{\bibfnamefont{A.}~\bibnamefont{Vainshtein}}, \bibinfo{journal}{Phys. Rev.} \textbf{\bibinfo{volume}{D67}}, \bibinfo{pages}{073006} (\bibinfo{year}{2003}), \bibinfo{note}{[Erratum: Phys. Rev. {\bf D73}, 119901 (2006)]}, \eprint{hep-ph/0212229}\relax
\mciteBstWouldAddEndPuncttrue
\mciteSetBstMidEndSepPunct{\mcitedefaultmidpunct}
{\mcitedefaultendpunct}{\mcitedefaultseppunct}\relax
\EndOfBibitem
\bibitem[{\citenamefont{Gnendiger et~al.}(2013)\citenamefont{Gnendiger, St{\"o}ckinger, and St{\"o}ckinger-Kim}}]{Gnendiger:2013pva}
\bibinfo{author}{\bibfnamefont{C.}~\bibnamefont{Gnendiger}}, \bibinfo{author}{\bibfnamefont{D.}~\bibnamefont{St{\"o}ckinger}}, \bibnamefont{and} \bibinfo{author}{\bibfnamefont{H.}~\bibnamefont{St{\"o}ckinger-Kim}}, \bibinfo{journal}{Phys. Rev.} \textbf{\bibinfo{volume}{D88}}, \bibinfo{pages}{053005} (\bibinfo{year}{2013}), \eprint{1306.5546}\relax
\mciteBstWouldAddEndPuncttrue
\mciteSetBstMidEndSepPunct{\mcitedefaultmidpunct}
{\mcitedefaultendpunct}{\mcitedefaultseppunct}\relax
\EndOfBibitem
\bibitem[{\citenamefont{Davier et~al.}(2017)\citenamefont{Davier, Hoecker, Malaescu, and Zhang}}]{Davier:2017zfy}
\bibinfo{author}{\bibfnamefont{M.}~\bibnamefont{Davier}}, \bibinfo{author}{\bibfnamefont{A.}~\bibnamefont{Hoecker}}, \bibinfo{author}{\bibfnamefont{B.}~\bibnamefont{Malaescu}}, \bibnamefont{and} \bibinfo{author}{\bibfnamefont{Z.}~\bibnamefont{Zhang}}, \bibinfo{journal}{Eur. Phys. J.} \textbf{\bibinfo{volume}{C77}}, \bibinfo{pages}{827} (\bibinfo{year}{2017}), \eprint{1706.09436}\relax
\mciteBstWouldAddEndPuncttrue
\mciteSetBstMidEndSepPunct{\mcitedefaultmidpunct}
{\mcitedefaultendpunct}{\mcitedefaultseppunct}\relax
\EndOfBibitem
\bibitem[{\citenamefont{Keshavarzi et~al.}(2018)\citenamefont{Keshavarzi, Nomura, and Teubner}}]{Keshavarzi:2018mgv}
\bibinfo{author}{\bibfnamefont{A.}~\bibnamefont{Keshavarzi}}, \bibinfo{author}{\bibfnamefont{D.}~\bibnamefont{Nomura}}, \bibnamefont{and} \bibinfo{author}{\bibfnamefont{T.}~\bibnamefont{Teubner}}, \bibinfo{journal}{Phys. Rev.} \textbf{\bibinfo{volume}{D97}}, \bibinfo{pages}{114025} (\bibinfo{year}{2018}), \eprint{1802.02995}\relax
\mciteBstWouldAddEndPuncttrue
\mciteSetBstMidEndSepPunct{\mcitedefaultmidpunct}
{\mcitedefaultendpunct}{\mcitedefaultseppunct}\relax
\EndOfBibitem
\bibitem[{\citenamefont{Colangelo et~al.}(2019)\citenamefont{Colangelo, Hoferichter, and Stoffer}}]{Colangelo:2018mtw}
\bibinfo{author}{\bibfnamefont{G.}~\bibnamefont{Colangelo}}, \bibinfo{author}{\bibfnamefont{M.}~\bibnamefont{Hoferichter}}, \bibnamefont{and} \bibinfo{author}{\bibfnamefont{P.}~\bibnamefont{Stoffer}}, \bibinfo{journal}{JHEP} \textbf{\bibinfo{volume}{02}}, \bibinfo{pages}{006} (\bibinfo{year}{2019}), \eprint{1810.00007}\relax
\mciteBstWouldAddEndPuncttrue
\mciteSetBstMidEndSepPunct{\mcitedefaultmidpunct}
{\mcitedefaultendpunct}{\mcitedefaultseppunct}\relax
\EndOfBibitem
\bibitem[{\citenamefont{Hoferichter et~al.}(2019)\citenamefont{Hoferichter, Hoid, and Kubis}}]{Hoferichter:2019mqg}
\bibinfo{author}{\bibfnamefont{M.}~\bibnamefont{Hoferichter}}, \bibinfo{author}{\bibfnamefont{B.-L.} \bibnamefont{Hoid}}, \bibnamefont{and} \bibinfo{author}{\bibfnamefont{B.}~\bibnamefont{Kubis}}, \bibinfo{journal}{JHEP} \textbf{\bibinfo{volume}{08}}, \bibinfo{pages}{137} (\bibinfo{year}{2019}), \eprint{1907.01556}\relax
\mciteBstWouldAddEndPuncttrue
\mciteSetBstMidEndSepPunct{\mcitedefaultmidpunct}
{\mcitedefaultendpunct}{\mcitedefaultseppunct}\relax
\EndOfBibitem
\bibitem[{\citenamefont{Davier et~al.}(2020)\citenamefont{Davier, Hoecker, Malaescu, and Zhang}}]{Davier:2019can}
\bibinfo{author}{\bibfnamefont{M.}~\bibnamefont{Davier}}, \bibinfo{author}{\bibfnamefont{A.}~\bibnamefont{Hoecker}}, \bibinfo{author}{\bibfnamefont{B.}~\bibnamefont{Malaescu}}, \bibnamefont{and} \bibinfo{author}{\bibfnamefont{Z.}~\bibnamefont{Zhang}}, \bibinfo{journal}{Eur. Phys. J.} \textbf{\bibinfo{volume}{C80}}, \bibinfo{pages}{241} (\bibinfo{year}{2020}), \bibinfo{note}{[Erratum: Eur. Phys. J. {\bf C80}, 410 (2020)]}, \eprint{1908.00921}\relax
\mciteBstWouldAddEndPuncttrue
\mciteSetBstMidEndSepPunct{\mcitedefaultmidpunct}
{\mcitedefaultendpunct}{\mcitedefaultseppunct}\relax
\EndOfBibitem
\bibitem[{\citenamefont{Keshavarzi et~al.}(2020)\citenamefont{Keshavarzi, Nomura, and Teubner}}]{Keshavarzi:2019abf}
\bibinfo{author}{\bibfnamefont{A.}~\bibnamefont{Keshavarzi}}, \bibinfo{author}{\bibfnamefont{D.}~\bibnamefont{Nomura}}, \bibnamefont{and} \bibinfo{author}{\bibfnamefont{T.}~\bibnamefont{Teubner}}, \bibinfo{journal}{Phys. Rev.} \textbf{\bibinfo{volume}{D101}}, \bibinfo{pages}{014029} (\bibinfo{year}{2020}), \eprint{1911.00367}\relax
\mciteBstWouldAddEndPuncttrue
\mciteSetBstMidEndSepPunct{\mcitedefaultmidpunct}
{\mcitedefaultendpunct}{\mcitedefaultseppunct}\relax
\EndOfBibitem
\bibitem[{\citenamefont{Kurz et~al.}(2014)\citenamefont{Kurz, Liu, Marquard, and Steinhauser}}]{Kurz:2014wya}
\bibinfo{author}{\bibfnamefont{A.}~\bibnamefont{Kurz}}, \bibinfo{author}{\bibfnamefont{T.}~\bibnamefont{Liu}}, \bibinfo{author}{\bibfnamefont{P.}~\bibnamefont{Marquard}}, \bibnamefont{and} \bibinfo{author}{\bibfnamefont{M.}~\bibnamefont{Steinhauser}}, \bibinfo{journal}{Phys. Lett.} \textbf{\bibinfo{volume}{B734}}, \bibinfo{pages}{144} (\bibinfo{year}{2014}), \eprint{1403.6400}\relax
\mciteBstWouldAddEndPuncttrue
\mciteSetBstMidEndSepPunct{\mcitedefaultmidpunct}
{\mcitedefaultendpunct}{\mcitedefaultseppunct}\relax
\EndOfBibitem
\bibitem[{\citenamefont{Melnikov and Vainshtein}(2004)}]{Melnikov:2003xd}
\bibinfo{author}{\bibfnamefont{K.}~\bibnamefont{Melnikov}} \bibnamefont{and} \bibinfo{author}{\bibfnamefont{A.}~\bibnamefont{Vainshtein}}, \bibinfo{journal}{Phys. Rev.} \textbf{\bibinfo{volume}{D70}}, \bibinfo{pages}{113006} (\bibinfo{year}{2004}), \eprint{hep-ph/0312226}\relax
\mciteBstWouldAddEndPuncttrue
\mciteSetBstMidEndSepPunct{\mcitedefaultmidpunct}
{\mcitedefaultendpunct}{\mcitedefaultseppunct}\relax
\EndOfBibitem
\bibitem[{\citenamefont{Masjuan and S{\'a}nchez-Puertas}(2017)}]{Masjuan:2017tvw}
\bibinfo{author}{\bibfnamefont{P.}~\bibnamefont{Masjuan}} \bibnamefont{and} \bibinfo{author}{\bibfnamefont{P.}~\bibnamefont{S{\'a}nchez-Puertas}}, \bibinfo{journal}{Phys. Rev.} \textbf{\bibinfo{volume}{D95}}, \bibinfo{pages}{054026} (\bibinfo{year}{2017}), \eprint{1701.05829}\relax
\mciteBstWouldAddEndPuncttrue
\mciteSetBstMidEndSepPunct{\mcitedefaultmidpunct}
{\mcitedefaultendpunct}{\mcitedefaultseppunct}\relax
\EndOfBibitem
\bibitem[{\citenamefont{Hoferichter et~al.}(2018)\citenamefont{Hoferichter, Hoid, Kubis, Leupold, and Schneider}}]{Hoferichter:2018kwz}
\bibinfo{author}{\bibfnamefont{M.}~\bibnamefont{Hoferichter}}, \bibinfo{author}{\bibfnamefont{B.-L.} \bibnamefont{Hoid}}, \bibinfo{author}{\bibfnamefont{B.}~\bibnamefont{Kubis}}, \bibinfo{author}{\bibfnamefont{S.}~\bibnamefont{Leupold}}, \bibnamefont{and} \bibinfo{author}{\bibfnamefont{S.~P.} \bibnamefont{Schneider}}, \bibinfo{journal}{JHEP} \textbf{\bibinfo{volume}{10}}, \bibinfo{pages}{141} (\bibinfo{year}{2018}), \eprint{1808.04823}\relax
\mciteBstWouldAddEndPuncttrue
\mciteSetBstMidEndSepPunct{\mcitedefaultmidpunct}
{\mcitedefaultendpunct}{\mcitedefaultseppunct}\relax
\EndOfBibitem
\bibitem[{\citenamefont{G{\'e}rardin et~al.}(2019)\citenamefont{G{\'e}rardin, Meyer, and Nyffeler}}]{Gerardin:2019vio}
\bibinfo{author}{\bibfnamefont{A.}~\bibnamefont{G{\'e}rardin}}, \bibinfo{author}{\bibfnamefont{H.~B.} \bibnamefont{Meyer}}, \bibnamefont{and} \bibinfo{author}{\bibfnamefont{A.}~\bibnamefont{Nyffeler}}, \bibinfo{journal}{Phys. Rev.} \textbf{\bibinfo{volume}{D100}}, \bibinfo{pages}{034520} (\bibinfo{year}{2019}), \eprint{1903.09471}\relax
\mciteBstWouldAddEndPuncttrue
\mciteSetBstMidEndSepPunct{\mcitedefaultmidpunct}
{\mcitedefaultendpunct}{\mcitedefaultseppunct}\relax
\EndOfBibitem
\bibitem[{\citenamefont{Bijnens et~al.}(2019)\citenamefont{Bijnens, Hermansson-Truedsson, and Rodr{\'i}guez-S{\'a}nchez}}]{Bijnens:2019ghy}
\bibinfo{author}{\bibfnamefont{J.}~\bibnamefont{Bijnens}}, \bibinfo{author}{\bibfnamefont{N.}~\bibnamefont{Hermansson-Truedsson}}, \bibnamefont{and} \bibinfo{author}{\bibfnamefont{A.}~\bibnamefont{Rodr{\'i}guez-S{\'a}nchez}}, \bibinfo{journal}{Phys. Lett.} \textbf{\bibinfo{volume}{B798}}, \bibinfo{pages}{134994} (\bibinfo{year}{2019}), \eprint{1908.03331}\relax
\mciteBstWouldAddEndPuncttrue
\mciteSetBstMidEndSepPunct{\mcitedefaultmidpunct}
{\mcitedefaultendpunct}{\mcitedefaultseppunct}\relax
\EndOfBibitem
\bibitem[{\citenamefont{Colangelo et~al.}(2020)\citenamefont{Colangelo, Hagelstein, Hoferichter, Laub, and Stoffer}}]{Colangelo:2019uex}
\bibinfo{author}{\bibfnamefont{G.}~\bibnamefont{Colangelo}}, \bibinfo{author}{\bibfnamefont{F.}~\bibnamefont{Hagelstein}}, \bibinfo{author}{\bibfnamefont{M.}~\bibnamefont{Hoferichter}}, \bibinfo{author}{\bibfnamefont{L.}~\bibnamefont{Laub}}, \bibnamefont{and} \bibinfo{author}{\bibfnamefont{P.}~\bibnamefont{Stoffer}}, \bibinfo{journal}{JHEP} \textbf{\bibinfo{volume}{03}}, \bibinfo{pages}{101} (\bibinfo{year}{2020}), \eprint{1910.13432}\relax
\mciteBstWouldAddEndPuncttrue
\mciteSetBstMidEndSepPunct{\mcitedefaultmidpunct}
{\mcitedefaultendpunct}{\mcitedefaultseppunct}\relax
\EndOfBibitem
\bibitem[{\citenamefont{Blum et~al.}(2020)\citenamefont{Blum, Christ, Hayakawa, Izubuchi, Jin, Jung, and Lehner}}]{Blum:2019ugy}
\bibinfo{author}{\bibfnamefont{T.}~\bibnamefont{Blum}}, \bibinfo{author}{\bibfnamefont{N.}~\bibnamefont{Christ}}, \bibinfo{author}{\bibfnamefont{M.}~\bibnamefont{Hayakawa}}, \bibinfo{author}{\bibfnamefont{T.}~\bibnamefont{Izubuchi}}, \bibinfo{author}{\bibfnamefont{L.}~\bibnamefont{Jin}}, \bibinfo{author}{\bibfnamefont{C.}~\bibnamefont{Jung}}, \bibnamefont{and} \bibinfo{author}{\bibfnamefont{C.}~\bibnamefont{Lehner}}, \bibinfo{journal}{Phys. Rev. Lett.} \textbf{\bibinfo{volume}{124}}, \bibinfo{pages}{132002} (\bibinfo{year}{2020}), \eprint{1911.08123}\relax
\mciteBstWouldAddEndPuncttrue
\mciteSetBstMidEndSepPunct{\mcitedefaultmidpunct}
{\mcitedefaultendpunct}{\mcitedefaultseppunct}\relax
\EndOfBibitem
\bibitem[{\citenamefont{Colangelo et~al.}(2014)\citenamefont{Colangelo, Hoferichter, Nyffeler, Passera, and Stoffer}}]{Colangelo:2014qya}
\bibinfo{author}{\bibfnamefont{G.}~\bibnamefont{Colangelo}}, \bibinfo{author}{\bibfnamefont{M.}~\bibnamefont{Hoferichter}}, \bibinfo{author}{\bibfnamefont{A.}~\bibnamefont{Nyffeler}}, \bibinfo{author}{\bibfnamefont{M.}~\bibnamefont{Passera}}, \bibnamefont{and} \bibinfo{author}{\bibfnamefont{P.}~\bibnamefont{Stoffer}}, \bibinfo{journal}{Phys. Lett.} \textbf{\bibinfo{volume}{B735}}, \bibinfo{pages}{90} (\bibinfo{year}{2014}), \eprint{1403.7512}\relax
\mciteBstWouldAddEndPuncttrue
\mciteSetBstMidEndSepPunct{\mcitedefaultmidpunct}
{\mcitedefaultendpunct}{\mcitedefaultseppunct}\relax
\EndOfBibitem
\bibitem[{\citenamefont{Colangelo et~al.}(2017{\natexlab{a}})\citenamefont{Colangelo, Hoferichter, Procura, and Stoffer}}]{Colangelo:2017qdm}
\bibinfo{author}{\bibfnamefont{G.}~\bibnamefont{Colangelo}}, \bibinfo{author}{\bibfnamefont{M.}~\bibnamefont{Hoferichter}}, \bibinfo{author}{\bibfnamefont{M.}~\bibnamefont{Procura}}, \bibnamefont{and} \bibinfo{author}{\bibfnamefont{P.}~\bibnamefont{Stoffer}}, \bibinfo{journal}{Phys. Rev. Lett.} \textbf{\bibinfo{volume}{118}}, \bibinfo{pages}{232001} (\bibinfo{year}{2017}{\natexlab{a}}), \eprint{1701.06554}\relax
\mciteBstWouldAddEndPuncttrue
\mciteSetBstMidEndSepPunct{\mcitedefaultmidpunct}
{\mcitedefaultendpunct}{\mcitedefaultseppunct}\relax
\EndOfBibitem
\bibitem[{\citenamefont{Colangelo et~al.}(2017{\natexlab{b}})\citenamefont{Colangelo, Hoferichter, Procura, and Stoffer}}]{Colangelo:2017fiz}
\bibinfo{author}{\bibfnamefont{G.}~\bibnamefont{Colangelo}}, \bibinfo{author}{\bibfnamefont{M.}~\bibnamefont{Hoferichter}}, \bibinfo{author}{\bibfnamefont{M.}~\bibnamefont{Procura}}, \bibnamefont{and} \bibinfo{author}{\bibfnamefont{P.}~\bibnamefont{Stoffer}}, \bibinfo{journal}{JHEP} \textbf{\bibinfo{volume}{04}}, \bibinfo{pages}{161} (\bibinfo{year}{2017}{\natexlab{b}}), \eprint{1702.07347}\relax
\mciteBstWouldAddEndPuncttrue
\mciteSetBstMidEndSepPunct{\mcitedefaultmidpunct}
{\mcitedefaultendpunct}{\mcitedefaultseppunct}\relax
\EndOfBibitem
\bibitem[{\citenamefont{Ignatov et~al.}(2024{\natexlab{a}})}]{CMD-3:2023rfe}
\bibinfo{author}{\bibfnamefont{F.~V.} \bibnamefont{Ignatov}} \bibnamefont{et~al.} (\bibinfo{collaboration}{CMD-3}), \bibinfo{journal}{Phys. Rev. Lett.} \textbf{\bibinfo{volume}{132}}, \bibinfo{pages}{231903} (\bibinfo{year}{2024}{\natexlab{a}}), \eprint{2309.12910}\relax
\mciteBstWouldAddEndPuncttrue
\mciteSetBstMidEndSepPunct{\mcitedefaultmidpunct}
{\mcitedefaultendpunct}{\mcitedefaultseppunct}\relax
\EndOfBibitem
\bibitem[{\citenamefont{Ignatov et~al.}(2024{\natexlab{b}})}]{CMD-3:2023alj}
\bibinfo{author}{\bibfnamefont{F.~V.} \bibnamefont{Ignatov}} \bibnamefont{et~al.} (\bibinfo{collaboration}{CMD-3}), \bibinfo{journal}{Phys. Rev. D} \textbf{\bibinfo{volume}{109}}, \bibinfo{pages}{112002} (\bibinfo{year}{2024}{\natexlab{b}}), \eprint{2302.08834}\relax
\mciteBstWouldAddEndPuncttrue
\mciteSetBstMidEndSepPunct{\mcitedefaultmidpunct}
{\mcitedefaultendpunct}{\mcitedefaultseppunct}\relax
\EndOfBibitem
\bibitem[{\citenamefont{Boccaletti et~al.}(2024)}]{Boccaletti:2024guq}
\bibinfo{author}{\bibfnamefont{A.}~\bibnamefont{Boccaletti}} \bibnamefont{et~al.} (\bibinfo{year}{2024}), \eprint{2407.10913}\relax
\mciteBstWouldAddEndPuncttrue
\mciteSetBstMidEndSepPunct{\mcitedefaultmidpunct}
{\mcitedefaultendpunct}{\mcitedefaultseppunct}\relax
\EndOfBibitem
\bibitem[{\citenamefont{Borsanyi et~al.}(2021)}]{Borsanyi:2020mff}
\bibinfo{author}{\bibfnamefont{S.}~\bibnamefont{Borsanyi}} \bibnamefont{et~al.}, \bibinfo{journal}{Nature} \textbf{\bibinfo{volume}{593}}, \bibinfo{pages}{51} (\bibinfo{year}{2021}), \eprint{2002.12347}\relax
\mciteBstWouldAddEndPuncttrue
\mciteSetBstMidEndSepPunct{\mcitedefaultmidpunct}
{\mcitedefaultendpunct}{\mcitedefaultseppunct}\relax
\EndOfBibitem
\bibitem[{\citenamefont{Blum et~al.}(2024)}]{RBC:2024fic}
\bibinfo{author}{\bibfnamefont{T.}~\bibnamefont{Blum}} \bibnamefont{et~al.} (\bibinfo{collaboration}{RBC, UKQCD}) (\bibinfo{year}{2024}), \eprint{2410.20590}\relax
\mciteBstWouldAddEndPuncttrue
\mciteSetBstMidEndSepPunct{\mcitedefaultmidpunct}
{\mcitedefaultendpunct}{\mcitedefaultseppunct}\relax
\EndOfBibitem
\bibitem[{\citenamefont{Djukanovic et~al.}(2024)\citenamefont{Djukanovic, von Hippel, Kuberski, Meyer, Miller, Ottnad, Parrino, Risch, and Wittig}}]{Djukanovic:2024cmq}
\bibinfo{author}{\bibfnamefont{D.}~\bibnamefont{Djukanovic}}, \bibinfo{author}{\bibfnamefont{G.}~\bibnamefont{von Hippel}}, \bibinfo{author}{\bibfnamefont{S.}~\bibnamefont{Kuberski}}, \bibinfo{author}{\bibfnamefont{H.~B.} \bibnamefont{Meyer}}, \bibinfo{author}{\bibfnamefont{N.}~\bibnamefont{Miller}}, \bibinfo{author}{\bibfnamefont{K.}~\bibnamefont{Ottnad}}, \bibinfo{author}{\bibfnamefont{J.}~\bibnamefont{Parrino}}, \bibinfo{author}{\bibfnamefont{A.}~\bibnamefont{Risch}}, \bibnamefont{and} \bibinfo{author}{\bibfnamefont{H.}~\bibnamefont{Wittig}} (\bibinfo{year}{2024}), \eprint{2411.07969}\relax
\mciteBstWouldAddEndPuncttrue
\mciteSetBstMidEndSepPunct{\mcitedefaultmidpunct}
{\mcitedefaultendpunct}{\mcitedefaultseppunct}\relax
\EndOfBibitem
\bibitem[{\citenamefont{Bazavov et~al.}(2024)}]{Bazavov:2024eou}
\bibinfo{author}{\bibfnamefont{A.}~\bibnamefont{Bazavov}} \bibnamefont{et~al.} (\bibinfo{year}{2024}), \eprint{2412.18491}\relax
\mciteBstWouldAddEndPuncttrue
\mciteSetBstMidEndSepPunct{\mcitedefaultmidpunct}
{\mcitedefaultendpunct}{\mcitedefaultseppunct}\relax
\EndOfBibitem
\bibitem[{\citenamefont{Grange et~al.}(2015)}]{Muong-2:2015xgu}
\bibinfo{author}{\bibfnamefont{J.}~\bibnamefont{Grange}} \bibnamefont{et~al.} (\bibinfo{collaboration}{Muon g-2}) (\bibinfo{year}{2015}), \eprint{1501.06858}\relax
\mciteBstWouldAddEndPuncttrue
\mciteSetBstMidEndSepPunct{\mcitedefaultmidpunct}
{\mcitedefaultendpunct}{\mcitedefaultseppunct}\relax
\EndOfBibitem
\bibitem[{\citenamefont{Danilkin et~al.}(2021{\natexlab{a}})\citenamefont{Danilkin, Hoferichter, and Stoffer}}]{Danilkin:2021icn}
\bibinfo{author}{\bibfnamefont{I.}~\bibnamefont{Danilkin}}, \bibinfo{author}{\bibfnamefont{M.}~\bibnamefont{Hoferichter}}, \bibnamefont{and} \bibinfo{author}{\bibfnamefont{P.}~\bibnamefont{Stoffer}}, \bibinfo{journal}{Phys. Lett. B} \textbf{\bibinfo{volume}{820}}, \bibinfo{pages}{136502} (\bibinfo{year}{2021}{\natexlab{a}}), \eprint{2105.01666}\relax
\mciteBstWouldAddEndPuncttrue
\mciteSetBstMidEndSepPunct{\mcitedefaultmidpunct}
{\mcitedefaultendpunct}{\mcitedefaultseppunct}\relax
\EndOfBibitem
\bibitem[{\citenamefont{Uehara et~al.}(2009)}]{Belle:2009xpa}
\bibinfo{author}{\bibfnamefont{S.}~\bibnamefont{Uehara}} \bibnamefont{et~al.} (\bibinfo{collaboration}{Belle}), \bibinfo{journal}{Phys. Rev. D} \textbf{\bibinfo{volume}{80}}, \bibinfo{pages}{032001} (\bibinfo{year}{2009}), \eprint{0906.1464}\relax
\mciteBstWouldAddEndPuncttrue
\mciteSetBstMidEndSepPunct{\mcitedefaultmidpunct}
{\mcitedefaultendpunct}{\mcitedefaultseppunct}\relax
\EndOfBibitem
\bibitem[{\citenamefont{Uehara et~al.}(2013)}]{Belle:2013eck}
\bibinfo{author}{\bibfnamefont{S.}~\bibnamefont{Uehara}} \bibnamefont{et~al.} (\bibinfo{collaboration}{Belle}), \bibinfo{journal}{PTEP} \textbf{\bibinfo{volume}{2013}}, \bibinfo{pages}{123C01} (\bibinfo{year}{2013}), \eprint{1307.7457}\relax
\mciteBstWouldAddEndPuncttrue
\mciteSetBstMidEndSepPunct{\mcitedefaultmidpunct}
{\mcitedefaultendpunct}{\mcitedefaultseppunct}\relax
\EndOfBibitem
\bibitem[{\citenamefont{Redmer}(2018)}]{Redmer:2018gah}
\bibinfo{author}{\bibfnamefont{C.~F.} \bibnamefont{Redmer}} (\bibinfo{collaboration}{BESIII}), \bibinfo{journal}{EPJ Web Conf.} \textbf{\bibinfo{volume}{166}}, \bibinfo{pages}{00017} (\bibinfo{year}{2018})\relax
\mciteBstWouldAddEndPuncttrue
\mciteSetBstMidEndSepPunct{\mcitedefaultmidpunct}
{\mcitedefaultendpunct}{\mcitedefaultseppunct}\relax
\EndOfBibitem
\bibitem[{\citenamefont{Garcia-Martin and Moussallam}(2010)}]{Garcia-Martin:2010kyn}
\bibinfo{author}{\bibfnamefont{R.}~\bibnamefont{Garcia-Martin}} \bibnamefont{and} \bibinfo{author}{\bibfnamefont{B.}~\bibnamefont{Moussallam}}, \bibinfo{journal}{Eur. Phys. J. C} \textbf{\bibinfo{volume}{70}}, \bibinfo{pages}{155} (\bibinfo{year}{2010}), \eprint{1006.5373}\relax
\mciteBstWouldAddEndPuncttrue
\mciteSetBstMidEndSepPunct{\mcitedefaultmidpunct}
{\mcitedefaultendpunct}{\mcitedefaultseppunct}\relax
\EndOfBibitem
\bibitem[{\citenamefont{Moussallam}(2011)}]{Moussallam:2011zg}
\bibinfo{author}{\bibfnamefont{B.}~\bibnamefont{Moussallam}}, \bibinfo{journal}{Eur. Phys. J. C} \textbf{\bibinfo{volume}{71}}, \bibinfo{pages}{1814} (\bibinfo{year}{2011}), \eprint{1110.6074}\relax
\mciteBstWouldAddEndPuncttrue
\mciteSetBstMidEndSepPunct{\mcitedefaultmidpunct}
{\mcitedefaultendpunct}{\mcitedefaultseppunct}\relax
\EndOfBibitem
\bibitem[{\citenamefont{Dai and Pennington}(2014{\natexlab{a}})}]{Dai:2014zta}
\bibinfo{author}{\bibfnamefont{L.-Y.} \bibnamefont{Dai}} \bibnamefont{and} \bibinfo{author}{\bibfnamefont{M.~R.} \bibnamefont{Pennington}}, \bibinfo{journal}{Phys. Rev. D} \textbf{\bibinfo{volume}{90}}, \bibinfo{pages}{036004} (\bibinfo{year}{2014}{\natexlab{a}}), \eprint{1404.7524}\relax
\mciteBstWouldAddEndPuncttrue
\mciteSetBstMidEndSepPunct{\mcitedefaultmidpunct}
{\mcitedefaultendpunct}{\mcitedefaultseppunct}\relax
\EndOfBibitem
\bibitem[{\citenamefont{Dai and Pennington}(2014{\natexlab{b}})}]{Dai:2014lza}
\bibinfo{author}{\bibfnamefont{L.-Y.} \bibnamefont{Dai}} \bibnamefont{and} \bibinfo{author}{\bibfnamefont{M.~R.} \bibnamefont{Pennington}}, \bibinfo{journal}{Phys. Lett. B} \textbf{\bibinfo{volume}{736}}, \bibinfo{pages}{11} (\bibinfo{year}{2014}{\natexlab{b}}), \eprint{1403.7514}\relax
\mciteBstWouldAddEndPuncttrue
\mciteSetBstMidEndSepPunct{\mcitedefaultmidpunct}
{\mcitedefaultendpunct}{\mcitedefaultseppunct}\relax
\EndOfBibitem
\bibitem[{\citenamefont{Danilkin and Vanderhaeghen}(2019)}]{Danilkin:2018qfn}
\bibinfo{author}{\bibfnamefont{I.}~\bibnamefont{Danilkin}} \bibnamefont{and} \bibinfo{author}{\bibfnamefont{M.}~\bibnamefont{Vanderhaeghen}}, \bibinfo{journal}{Phys. Lett. B} \textbf{\bibinfo{volume}{789}}, \bibinfo{pages}{366} (\bibinfo{year}{2019}), \eprint{1810.03669}\relax
\mciteBstWouldAddEndPuncttrue
\mciteSetBstMidEndSepPunct{\mcitedefaultmidpunct}
{\mcitedefaultendpunct}{\mcitedefaultseppunct}\relax
\EndOfBibitem
\bibitem[{\citenamefont{Danilkin et~al.}(2020)\citenamefont{Danilkin, Deineka, and Vanderhaeghen}}]{Danilkin:2019opj}
\bibinfo{author}{\bibfnamefont{I.}~\bibnamefont{Danilkin}}, \bibinfo{author}{\bibfnamefont{O.}~\bibnamefont{Deineka}}, \bibnamefont{and} \bibinfo{author}{\bibfnamefont{M.}~\bibnamefont{Vanderhaeghen}}, \bibinfo{journal}{Phys. Rev. D} \textbf{\bibinfo{volume}{101}}, \bibinfo{pages}{054008} (\bibinfo{year}{2020}), \eprint{1909.04158}\relax
\mciteBstWouldAddEndPuncttrue
\mciteSetBstMidEndSepPunct{\mcitedefaultmidpunct}
{\mcitedefaultendpunct}{\mcitedefaultseppunct}\relax
\EndOfBibitem
\bibitem[{\citenamefont{Stamen et~al.}(2024)\citenamefont{Stamen, Dammann, Korte, and Kubis}}]{Stamen:2024ocm}
\bibinfo{author}{\bibfnamefont{D.}~\bibnamefont{Stamen}}, \bibinfo{author}{\bibfnamefont{J.~L.} \bibnamefont{Dammann}}, \bibinfo{author}{\bibfnamefont{Y.}~\bibnamefont{Korte}}, \bibnamefont{and} \bibinfo{author}{\bibfnamefont{B.}~\bibnamefont{Kubis}} (\bibinfo{year}{2024}), \eprint{2409.05955}\relax
\mciteBstWouldAddEndPuncttrue
\mciteSetBstMidEndSepPunct{\mcitedefaultmidpunct}
{\mcitedefaultendpunct}{\mcitedefaultseppunct}\relax
\EndOfBibitem
\bibitem[{\citenamefont{Danilkin et~al.}(2017)\citenamefont{Danilkin, Deineka, and Vanderhaeghen}}]{Danilkin:2017lyn}
\bibinfo{author}{\bibfnamefont{I.}~\bibnamefont{Danilkin}}, \bibinfo{author}{\bibfnamefont{O.}~\bibnamefont{Deineka}}, \bibnamefont{and} \bibinfo{author}{\bibfnamefont{M.}~\bibnamefont{Vanderhaeghen}}, \bibinfo{journal}{Phys. Rev. D} \textbf{\bibinfo{volume}{96}}, \bibinfo{pages}{114018} (\bibinfo{year}{2017}), \eprint{1709.08595}\relax
\mciteBstWouldAddEndPuncttrue
\mciteSetBstMidEndSepPunct{\mcitedefaultmidpunct}
{\mcitedefaultendpunct}{\mcitedefaultseppunct}\relax
\EndOfBibitem
\bibitem[{\citenamefont{Lu and Moussallam}(2020)}]{Lu:2020qeo}
\bibinfo{author}{\bibfnamefont{J.}~\bibnamefont{Lu}} \bibnamefont{and} \bibinfo{author}{\bibfnamefont{B.}~\bibnamefont{Moussallam}}, \bibinfo{journal}{Eur. Phys. J. C} \textbf{\bibinfo{volume}{80}}, \bibinfo{pages}{436} (\bibinfo{year}{2020}), \eprint{2002.04441}\relax
\mciteBstWouldAddEndPuncttrue
\mciteSetBstMidEndSepPunct{\mcitedefaultmidpunct}
{\mcitedefaultendpunct}{\mcitedefaultseppunct}\relax
\EndOfBibitem
\bibitem[{\citenamefont{Sch\"afer et~al.}(2023)\citenamefont{Sch\"afer, Zanke, Korte, and Kubis}}]{Schafer:2023qtl}
\bibinfo{author}{\bibfnamefont{H.}~\bibnamefont{Sch\"afer}}, \bibinfo{author}{\bibfnamefont{M.}~\bibnamefont{Zanke}}, \bibinfo{author}{\bibfnamefont{Y.}~\bibnamefont{Korte}}, \bibnamefont{and} \bibinfo{author}{\bibfnamefont{B.}~\bibnamefont{Kubis}}, \bibinfo{journal}{Phys. Rev. D} \textbf{\bibinfo{volume}{108}}, \bibinfo{pages}{074025} (\bibinfo{year}{2023}), \eprint{2307.10357}\relax
\mciteBstWouldAddEndPuncttrue
\mciteSetBstMidEndSepPunct{\mcitedefaultmidpunct}
{\mcitedefaultendpunct}{\mcitedefaultseppunct}\relax
\EndOfBibitem
\bibitem[{\citenamefont{Jacob and Wick}(1959)}]{Jacob:1959at}
\bibinfo{author}{\bibfnamefont{M.}~\bibnamefont{Jacob}} \bibnamefont{and} \bibinfo{author}{\bibfnamefont{G.~C.} \bibnamefont{Wick}}, \bibinfo{journal}{Annals Phys.} \textbf{\bibinfo{volume}{7}}, \bibinfo{pages}{404} (\bibinfo{year}{1959})\relax
\mciteBstWouldAddEndPuncttrue
\mciteSetBstMidEndSepPunct{\mcitedefaultmidpunct}
{\mcitedefaultendpunct}{\mcitedefaultseppunct}\relax
\EndOfBibitem
\bibitem[{\citenamefont{Low}(1958)}]{Low:1958sn}
\bibinfo{author}{\bibfnamefont{F.~E.} \bibnamefont{Low}}, \bibinfo{journal}{Phys. Rev.} \textbf{\bibinfo{volume}{110}}, \bibinfo{pages}{974} (\bibinfo{year}{1958})\relax
\mciteBstWouldAddEndPuncttrue
\mciteSetBstMidEndSepPunct{\mcitedefaultmidpunct}
{\mcitedefaultendpunct}{\mcitedefaultseppunct}\relax
\EndOfBibitem
\bibitem[{\citenamefont{Hoferichter and Stoffer}(2019)}]{Hoferichter:2019nlq}
\bibinfo{author}{\bibfnamefont{M.}~\bibnamefont{Hoferichter}} \bibnamefont{and} \bibinfo{author}{\bibfnamefont{P.}~\bibnamefont{Stoffer}}, \bibinfo{journal}{JHEP} \textbf{\bibinfo{volume}{07}}, \bibinfo{pages}{073} (\bibinfo{year}{2019})\relax
\mciteBstWouldAddEndPuncttrue
\mciteSetBstMidEndSepPunct{\mcitedefaultmidpunct}
{\mcitedefaultendpunct}{\mcitedefaultseppunct}\relax
\EndOfBibitem
\bibitem[{\citenamefont{Omnes}(1958)}]{Omnes:1958hv}
\bibinfo{author}{\bibfnamefont{R.}~\bibnamefont{Omnes}}, \bibinfo{journal}{Nuovo Cim.} \textbf{\bibinfo{volume}{8}}, \bibinfo{pages}{316} (\bibinfo{year}{1958})\relax
\mciteBstWouldAddEndPuncttrue
\mciteSetBstMidEndSepPunct{\mcitedefaultmidpunct}
{\mcitedefaultendpunct}{\mcitedefaultseppunct}\relax
\EndOfBibitem
\bibitem[{\citenamefont{Pennington}(2006)}]{Pennington:2006dg}
\bibinfo{author}{\bibfnamefont{M.~R.} \bibnamefont{Pennington}}, \bibinfo{journal}{Phys. Rev. Lett.} \textbf{\bibinfo{volume}{97}}, \bibinfo{pages}{011601} (\bibinfo{year}{2006})\relax
\mciteBstWouldAddEndPuncttrue
\mciteSetBstMidEndSepPunct{\mcitedefaultmidpunct}
{\mcitedefaultendpunct}{\mcitedefaultseppunct}\relax
\EndOfBibitem
\bibitem[{\citenamefont{Moussallam}(2021)}]{Moussallam:2021dpk}
\bibinfo{author}{\bibfnamefont{B.}~\bibnamefont{Moussallam}}, \bibinfo{journal}{Eur. Phys. J. C} \textbf{\bibinfo{volume}{81}}, \bibinfo{pages}{993} (\bibinfo{year}{2021}), \eprint{2107.14147}\relax
\mciteBstWouldAddEndPuncttrue
\mciteSetBstMidEndSepPunct{\mcitedefaultmidpunct}
{\mcitedefaultendpunct}{\mcitedefaultseppunct}\relax
\EndOfBibitem
\bibitem[{\citenamefont{Chew and Mandelstam}(1960)}]{Chew:1960iv}
\bibinfo{author}{\bibfnamefont{G.~F.} \bibnamefont{Chew}} \bibnamefont{and} \bibinfo{author}{\bibfnamefont{S.}~\bibnamefont{Mandelstam}}, \bibinfo{journal}{Phys. Rev.} \textbf{\bibinfo{volume}{119}}, \bibinfo{pages}{467} (\bibinfo{year}{1960})\relax
\mciteBstWouldAddEndPuncttrue
\mciteSetBstMidEndSepPunct{\mcitedefaultmidpunct}
{\mcitedefaultendpunct}{\mcitedefaultseppunct}\relax
\EndOfBibitem
\bibitem[{\citenamefont{Castillejo et~al.}(1956)\citenamefont{Castillejo, Dalitz, and Dyson}}]{Castillejo:1955ed}
\bibinfo{author}{\bibfnamefont{L.}~\bibnamefont{Castillejo}}, \bibinfo{author}{\bibfnamefont{R.~H.} \bibnamefont{Dalitz}}, \bibnamefont{and} \bibinfo{author}{\bibfnamefont{F.~J.} \bibnamefont{Dyson}}, \bibinfo{journal}{Phys. Rev.} \textbf{\bibinfo{volume}{101}}, \bibinfo{pages}{453} (\bibinfo{year}{1956})\relax
\mciteBstWouldAddEndPuncttrue
\mciteSetBstMidEndSepPunct{\mcitedefaultmidpunct}
{\mcitedefaultendpunct}{\mcitedefaultseppunct}\relax
\EndOfBibitem
\bibitem[{\citenamefont{Danilkin et~al.}(2021{\natexlab{b}})\citenamefont{Danilkin, Deineka, and Vanderhaeghen}}]{Danilkin:2020pak}
\bibinfo{author}{\bibfnamefont{I.}~\bibnamefont{Danilkin}}, \bibinfo{author}{\bibfnamefont{O.}~\bibnamefont{Deineka}}, \bibnamefont{and} \bibinfo{author}{\bibfnamefont{M.}~\bibnamefont{Vanderhaeghen}}, \bibinfo{journal}{Phys. Rev. D} \textbf{\bibinfo{volume}{103}}, \bibinfo{pages}{114023} (\bibinfo{year}{2021}{\natexlab{b}}), \eprint{2012.11636}\relax
\mciteBstWouldAddEndPuncttrue
\mciteSetBstMidEndSepPunct{\mcitedefaultmidpunct}
{\mcitedefaultendpunct}{\mcitedefaultseppunct}\relax
\EndOfBibitem
\bibitem[{\citenamefont{Luming}(1964)}]{Luming:1964}
\bibinfo{author}{\bibfnamefont{M.}~\bibnamefont{Luming}}, \bibinfo{journal}{Phys. Rev.} \textbf{\bibinfo{volume}{136}}, \bibinfo{pages}{B1120} (\bibinfo{year}{1964})\relax
\mciteBstWouldAddEndPuncttrue
\mciteSetBstMidEndSepPunct{\mcitedefaultmidpunct}
{\mcitedefaultendpunct}{\mcitedefaultseppunct}\relax
\EndOfBibitem
\bibitem[{\citenamefont{Gasparyan and Lutz}(2010)}]{Gasparyan:2010xz}
\bibinfo{author}{\bibfnamefont{A.}~\bibnamefont{Gasparyan}} \bibnamefont{and} \bibinfo{author}{\bibfnamefont{M.~F.~M.} \bibnamefont{Lutz}}, \bibinfo{journal}{Nucl. Phys. A} \textbf{\bibinfo{volume}{848}}, \bibinfo{pages}{126} (\bibinfo{year}{2010}), \eprint{1003.3426}\relax
\mciteBstWouldAddEndPuncttrue
\mciteSetBstMidEndSepPunct{\mcitedefaultmidpunct}
{\mcitedefaultendpunct}{\mcitedefaultseppunct}\relax
\EndOfBibitem
\bibitem[{\citenamefont{Danilkin et~al.}(2011{\natexlab{a}})\citenamefont{Danilkin, Gasparyan, and Lutz}}]{Danilkin:2010xd}
\bibinfo{author}{\bibfnamefont{I.~V.} \bibnamefont{Danilkin}}, \bibinfo{author}{\bibfnamefont{A.~M.} \bibnamefont{Gasparyan}}, \bibnamefont{and} \bibinfo{author}{\bibfnamefont{M.~F.~M.} \bibnamefont{Lutz}}, \bibinfo{journal}{Phys. Lett. B} \textbf{\bibinfo{volume}{697}}, \bibinfo{pages}{147} (\bibinfo{year}{2011}{\natexlab{a}}), \eprint{1009.5928}\relax
\mciteBstWouldAddEndPuncttrue
\mciteSetBstMidEndSepPunct{\mcitedefaultmidpunct}
{\mcitedefaultendpunct}{\mcitedefaultseppunct}\relax
\EndOfBibitem
\bibitem[{\citenamefont{Gasparyan et~al.}(2011)\citenamefont{Gasparyan, Lutz, and Pasquini}}]{Gasparyan:2011yw}
\bibinfo{author}{\bibfnamefont{A.~M.} \bibnamefont{Gasparyan}}, \bibinfo{author}{\bibfnamefont{M.~F.~M.} \bibnamefont{Lutz}}, \bibnamefont{and} \bibinfo{author}{\bibfnamefont{B.}~\bibnamefont{Pasquini}}, \bibinfo{journal}{Nucl. Phys. A} \textbf{\bibinfo{volume}{866}}, \bibinfo{pages}{79} (\bibinfo{year}{2011}), \eprint{1102.3375}\relax
\mciteBstWouldAddEndPuncttrue
\mciteSetBstMidEndSepPunct{\mcitedefaultmidpunct}
{\mcitedefaultendpunct}{\mcitedefaultseppunct}\relax
\EndOfBibitem
\bibitem[{\citenamefont{Gasparyan et~al.}(2013)\citenamefont{Gasparyan, Lutz, and Epelbaum}}]{Gasparyan:2012km}
\bibinfo{author}{\bibfnamefont{A.~M.} \bibnamefont{Gasparyan}}, \bibinfo{author}{\bibfnamefont{M.~F.~M.} \bibnamefont{Lutz}}, \bibnamefont{and} \bibinfo{author}{\bibfnamefont{E.}~\bibnamefont{Epelbaum}}, \bibinfo{journal}{Eur. Phys. J. A} \textbf{\bibinfo{volume}{49}}, \bibinfo{pages}{115} (\bibinfo{year}{2013}), \eprint{1212.3057}\relax
\mciteBstWouldAddEndPuncttrue
\mciteSetBstMidEndSepPunct{\mcitedefaultmidpunct}
{\mcitedefaultendpunct}{\mcitedefaultseppunct}\relax
\EndOfBibitem
\bibitem[{\citenamefont{Frazer}(1961)}]{Frazer:1961zz}
\bibinfo{author}{\bibfnamefont{W.~R.} \bibnamefont{Frazer}}, \bibinfo{journal}{Phys. Rev.} \textbf{\bibinfo{volume}{123}}, \bibinfo{pages}{2180} (\bibinfo{year}{1961})\relax
\mciteBstWouldAddEndPuncttrue
\mciteSetBstMidEndSepPunct{\mcitedefaultmidpunct}
{\mcitedefaultendpunct}{\mcitedefaultseppunct}\relax
\EndOfBibitem
\bibitem[{\citenamefont{Albaladejo and Moussallam}(2015)}]{Albaladejo:2015aca}
\bibinfo{author}{\bibfnamefont{M.}~\bibnamefont{Albaladejo}} \bibnamefont{and} \bibinfo{author}{\bibfnamefont{B.}~\bibnamefont{Moussallam}}, \bibinfo{journal}{Eur. Phys. J. C} \textbf{\bibinfo{volume}{75}}, \bibinfo{pages}{488} (\bibinfo{year}{2015}), \eprint{1507.04526}\relax
\mciteBstWouldAddEndPuncttrue
\mciteSetBstMidEndSepPunct{\mcitedefaultmidpunct}
{\mcitedefaultendpunct}{\mcitedefaultseppunct}\relax
\EndOfBibitem
\bibitem[{\citenamefont{Kennedy and Spearman}(1962)}]{PhysRev.126.1596}
\bibinfo{author}{\bibfnamefont{J.}~\bibnamefont{Kennedy}} \bibnamefont{and} \bibinfo{author}{\bibfnamefont{T.~D.} \bibnamefont{Spearman}}, \bibinfo{journal}{Phys. Rev.} \textbf{\bibinfo{volume}{126}}, \bibinfo{pages}{1596} (\bibinfo{year}{1962}), \urlprefix\url{https://link.aps.org/doi/10.1103/PhysRev.126.1596}\relax
\mciteBstWouldAddEndPuncttrue
\mciteSetBstMidEndSepPunct{\mcitedefaultmidpunct}
{\mcitedefaultendpunct}{\mcitedefaultseppunct}\relax
\EndOfBibitem
\bibitem[{\citenamefont{Garcia-Martin et~al.}(2011{\natexlab{a}})\citenamefont{Garcia-Martin, Kaminski, Pelaez, Ruiz~de Elvira, and Yndurain}}]{Garcia-Martin:2011iqs}
\bibinfo{author}{\bibfnamefont{R.}~\bibnamefont{Garcia-Martin}}, \bibinfo{author}{\bibfnamefont{R.}~\bibnamefont{Kaminski}}, \bibinfo{author}{\bibfnamefont{J.~R.} \bibnamefont{Pelaez}}, \bibinfo{author}{\bibfnamefont{J.}~\bibnamefont{Ruiz~de Elvira}}, \bibnamefont{and} \bibinfo{author}{\bibfnamefont{F.~J.} \bibnamefont{Yndurain}}, \bibinfo{journal}{Phys. Rev. D} \textbf{\bibinfo{volume}{83}}, \bibinfo{pages}{074004} (\bibinfo{year}{2011}{\natexlab{a}}), \eprint{1102.2183}\relax
\mciteBstWouldAddEndPuncttrue
\mciteSetBstMidEndSepPunct{\mcitedefaultmidpunct}
{\mcitedefaultendpunct}{\mcitedefaultseppunct}\relax
\EndOfBibitem
\bibitem[{\citenamefont{Pel\'aez and Rodas}(2022)}]{Pelaez:2020gnd}
\bibinfo{author}{\bibfnamefont{J.~R.} \bibnamefont{Pel\'aez}} \bibnamefont{and} \bibinfo{author}{\bibfnamefont{A.}~\bibnamefont{Rodas}}, \bibinfo{journal}{Phys. Rept.} \textbf{\bibinfo{volume}{969}}, \bibinfo{pages}{1} (\bibinfo{year}{2022}), \eprint{2010.11222}\relax
\mciteBstWouldAddEndPuncttrue
\mciteSetBstMidEndSepPunct{\mcitedefaultmidpunct}
{\mcitedefaultendpunct}{\mcitedefaultseppunct}\relax
\EndOfBibitem
\bibitem[{\citenamefont{Caprini et~al.}(2006)\citenamefont{Caprini, Colangelo, and Leutwyler}}]{Caprini:2005zr}
\bibinfo{author}{\bibfnamefont{I.}~\bibnamefont{Caprini}}, \bibinfo{author}{\bibfnamefont{G.}~\bibnamefont{Colangelo}}, \bibnamefont{and} \bibinfo{author}{\bibfnamefont{H.}~\bibnamefont{Leutwyler}}, \bibinfo{journal}{Phys. Rev. Lett.} \textbf{\bibinfo{volume}{96}}, \bibinfo{pages}{132001} (\bibinfo{year}{2006}), \eprint{hep-ph/0512364}\relax
\mciteBstWouldAddEndPuncttrue
\mciteSetBstMidEndSepPunct{\mcitedefaultmidpunct}
{\mcitedefaultendpunct}{\mcitedefaultseppunct}\relax
\EndOfBibitem
\bibitem[{\citenamefont{Garcia-Martin et~al.}(2011{\natexlab{b}})\citenamefont{Garcia-Martin, Kaminski, Pelaez, and Ruiz~de Elvira}}]{Garcia-Martin:2011nna}
\bibinfo{author}{\bibfnamefont{R.}~\bibnamefont{Garcia-Martin}}, \bibinfo{author}{\bibfnamefont{R.}~\bibnamefont{Kaminski}}, \bibinfo{author}{\bibfnamefont{J.~R.} \bibnamefont{Pelaez}}, \bibnamefont{and} \bibinfo{author}{\bibfnamefont{J.}~\bibnamefont{Ruiz~de Elvira}}, \bibinfo{journal}{Phys. Rev. Lett.} \textbf{\bibinfo{volume}{107}}, \bibinfo{pages}{072001} (\bibinfo{year}{2011}{\natexlab{b}}), \eprint{1107.1635}\relax
\mciteBstWouldAddEndPuncttrue
\mciteSetBstMidEndSepPunct{\mcitedefaultmidpunct}
{\mcitedefaultendpunct}{\mcitedefaultseppunct}\relax
\EndOfBibitem
\bibitem[{\citenamefont{Danilkin et~al.}(2011{\natexlab{b}})\citenamefont{Danilkin, Gil, and Lutz}}]{Danilkin:2011fz}
\bibinfo{author}{\bibfnamefont{I.~V.} \bibnamefont{Danilkin}}, \bibinfo{author}{\bibfnamefont{L.~I.~R.} \bibnamefont{Gil}}, \bibnamefont{and} \bibinfo{author}{\bibfnamefont{M.~F.~M.} \bibnamefont{Lutz}}, \bibinfo{journal}{Phys. Lett. B} \textbf{\bibinfo{volume}{703}}, \bibinfo{pages}{504} (\bibinfo{year}{2011}{\natexlab{b}}), \eprint{1106.2230}\relax
\mciteBstWouldAddEndPuncttrue
\mciteSetBstMidEndSepPunct{\mcitedefaultmidpunct}
{\mcitedefaultendpunct}{\mcitedefaultseppunct}\relax
\EndOfBibitem
\bibitem[{\citenamefont{Danilkin et~al.}(2013)\citenamefont{Danilkin, Lutz, Leupold, and Terschlusen}}]{Danilkin:2012ua}
\bibinfo{author}{\bibfnamefont{I.~V.} \bibnamefont{Danilkin}}, \bibinfo{author}{\bibfnamefont{M.~F.~M.} \bibnamefont{Lutz}}, \bibinfo{author}{\bibfnamefont{S.}~\bibnamefont{Leupold}}, \bibnamefont{and} \bibinfo{author}{\bibfnamefont{C.}~\bibnamefont{Terschlusen}}, \bibinfo{journal}{Eur. Phys. J. C} \textbf{\bibinfo{volume}{73}}, \bibinfo{pages}{2358} (\bibinfo{year}{2013}), \eprint{1211.1503}\relax
\mciteBstWouldAddEndPuncttrue
\mciteSetBstMidEndSepPunct{\mcitedefaultmidpunct}
{\mcitedefaultendpunct}{\mcitedefaultseppunct}\relax
\EndOfBibitem
\bibitem[{\citenamefont{Gomez~Nicola and Pelaez}(2002)}]{GomezNicola:2001as}
\bibinfo{author}{\bibfnamefont{A.}~\bibnamefont{Gomez~Nicola}} \bibnamefont{and} \bibinfo{author}{\bibfnamefont{J.~R.} \bibnamefont{Pelaez}}, \bibinfo{journal}{Phys. Rev. D} \textbf{\bibinfo{volume}{65}}, \bibinfo{pages}{054009} (\bibinfo{year}{2002}), \eprint{hep-ph/0109056}\relax
\mciteBstWouldAddEndPuncttrue
\mciteSetBstMidEndSepPunct{\mcitedefaultmidpunct}
{\mcitedefaultendpunct}{\mcitedefaultseppunct}\relax
\EndOfBibitem
\bibitem[{\citenamefont{Bijnens and Ecker}(2014)}]{Bijnens:2014lea}
\bibinfo{author}{\bibfnamefont{J.}~\bibnamefont{Bijnens}} \bibnamefont{and} \bibinfo{author}{\bibfnamefont{G.}~\bibnamefont{Ecker}}, \bibinfo{journal}{Ann. Rev. Nucl. Part. Sci.} \textbf{\bibinfo{volume}{64}}, \bibinfo{pages}{149} (\bibinfo{year}{2014}), \eprint{1405.6488}\relax
\mciteBstWouldAddEndPuncttrue
\mciteSetBstMidEndSepPunct{\mcitedefaultmidpunct}
{\mcitedefaultendpunct}{\mcitedefaultseppunct}\relax
\EndOfBibitem
\bibitem[{\citenamefont{Gasser and Leutwyler}(1985)}]{Gasser:1984gg}
\bibinfo{author}{\bibfnamefont{J.}~\bibnamefont{Gasser}} \bibnamefont{and} \bibinfo{author}{\bibfnamefont{H.}~\bibnamefont{Leutwyler}}, \bibinfo{journal}{Nucl. Phys. B} \textbf{\bibinfo{volume}{250}}, \bibinfo{pages}{465} (\bibinfo{year}{1985})\relax
\mciteBstWouldAddEndPuncttrue
\mciteSetBstMidEndSepPunct{\mcitedefaultmidpunct}
{\mcitedefaultendpunct}{\mcitedefaultseppunct}\relax
\EndOfBibitem
\bibitem[{\citenamefont{Navas et~al.}(2024)}]{ParticleDataGroup:2024cfk}
\bibinfo{author}{\bibfnamefont{S.}~\bibnamefont{Navas}} \bibnamefont{et~al.} (\bibinfo{collaboration}{Particle Data Group}), \bibinfo{journal}{Phys. Rev. D} \textbf{\bibinfo{volume}{110}}, \bibinfo{pages}{030001} (\bibinfo{year}{2024})\relax
\mciteBstWouldAddEndPuncttrue
\mciteSetBstMidEndSepPunct{\mcitedefaultmidpunct}
{\mcitedefaultendpunct}{\mcitedefaultseppunct}\relax
\EndOfBibitem
\bibitem[{\citenamefont{Ermolina et~al.}(2024)\citenamefont{Ermolina, Danilkin, and Vanderhaeghen}}]{Ermolina:2024daf}
\bibinfo{author}{\bibfnamefont{V.}~\bibnamefont{Ermolina}}, \bibinfo{author}{\bibfnamefont{I.}~\bibnamefont{Danilkin}}, \bibnamefont{and} \bibinfo{author}{\bibfnamefont{M.}~\bibnamefont{Vanderhaeghen}}, \bibinfo{journal}{EPJ Web Conf.} \textbf{\bibinfo{volume}{303}}, \bibinfo{pages}{01007} (\bibinfo{year}{2024}), \eprint{2407.21471}\relax
\mciteBstWouldAddEndPuncttrue
\mciteSetBstMidEndSepPunct{\mcitedefaultmidpunct}
{\mcitedefaultendpunct}{\mcitedefaultseppunct}\relax
\EndOfBibitem
\bibitem[{\citenamefont{Drechsel et~al.}(1999)\citenamefont{Drechsel, Gorchtein, Pasquini, and Vanderhaeghen}}]{Drechsel:1999rf}
\bibinfo{author}{\bibfnamefont{D.}~\bibnamefont{Drechsel}}, \bibinfo{author}{\bibfnamefont{M.}~\bibnamefont{Gorchtein}}, \bibinfo{author}{\bibfnamefont{B.}~\bibnamefont{Pasquini}}, \bibnamefont{and} \bibinfo{author}{\bibfnamefont{M.}~\bibnamefont{Vanderhaeghen}}, \bibinfo{journal}{Phys. Rev. C} \textbf{\bibinfo{volume}{61}}, \bibinfo{pages}{015204} (\bibinfo{year}{1999}), \eprint{hep-ph/9904290}\relax
\mciteBstWouldAddEndPuncttrue
\mciteSetBstMidEndSepPunct{\mcitedefaultmidpunct}
{\mcitedefaultendpunct}{\mcitedefaultseppunct}\relax
\EndOfBibitem
\bibitem[{\citenamefont{Hoferichter et~al.}(2011)\citenamefont{Hoferichter, Phillips, and Schat}}]{Hoferichter:2011wk}
\bibinfo{author}{\bibfnamefont{M.}~\bibnamefont{Hoferichter}}, \bibinfo{author}{\bibfnamefont{D.~R.} \bibnamefont{Phillips}}, \bibnamefont{and} \bibinfo{author}{\bibfnamefont{C.}~\bibnamefont{Schat}}, \bibinfo{journal}{Eur. Phys. J. C} \textbf{\bibinfo{volume}{71}}, \bibinfo{pages}{1743} (\bibinfo{year}{2011}), \eprint{1106.4147}\relax
\mciteBstWouldAddEndPuncttrue
\mciteSetBstMidEndSepPunct{\mcitedefaultmidpunct}
{\mcitedefaultendpunct}{\mcitedefaultseppunct}\relax
\EndOfBibitem
\bibitem[{\citenamefont{Deineka et~al.}(2019{\natexlab{a}})\citenamefont{Deineka, Danilkin, and Vanderhaeghen}}]{Deineka:2018nuh}
\bibinfo{author}{\bibfnamefont{O.}~\bibnamefont{Deineka}}, \bibinfo{author}{\bibfnamefont{I.}~\bibnamefont{Danilkin}}, \bibnamefont{and} \bibinfo{author}{\bibfnamefont{M.}~\bibnamefont{Vanderhaeghen}}, \bibinfo{journal}{EPJ Web Conf.} \textbf{\bibinfo{volume}{199}}, \bibinfo{pages}{02005} (\bibinfo{year}{2019}{\natexlab{a}}), \eprint{1808.04117}\relax
\mciteBstWouldAddEndPuncttrue
\mciteSetBstMidEndSepPunct{\mcitedefaultmidpunct}
{\mcitedefaultendpunct}{\mcitedefaultseppunct}\relax
\EndOfBibitem
\bibitem[{\citenamefont{Albrecht et~al.}(1990)}]{ARGUS:1989ird}
\bibinfo{author}{\bibfnamefont{H.}~\bibnamefont{Albrecht}} \bibnamefont{et~al.} (\bibinfo{collaboration}{ARGUS}), \bibinfo{journal}{Z. Phys. C} \textbf{\bibinfo{volume}{48}}, \bibinfo{pages}{183} (\bibinfo{year}{1990})\relax
\mciteBstWouldAddEndPuncttrue
\mciteSetBstMidEndSepPunct{\mcitedefaultmidpunct}
{\mcitedefaultendpunct}{\mcitedefaultseppunct}\relax
\EndOfBibitem
\bibitem[{\citenamefont{K\"u\ss{}ner}(2022)}]{Kussner:2022dft}
\bibinfo{author}{\bibfnamefont{M.}~\bibnamefont{K\"u\ss{}ner}}, Ph.D. thesis, \bibinfo{school}{Ruhr U., Bochum} (\bibinfo{year}{2022})\relax
\mciteBstWouldAddEndPuncttrue
\mciteSetBstMidEndSepPunct{\mcitedefaultmidpunct}
{\mcitedefaultendpunct}{\mcitedefaultseppunct}\relax
\EndOfBibitem
\bibitem[{\citenamefont{K\"u\ss{}ner}(2024)}]{Kussner:2024ryb}
\bibinfo{author}{\bibfnamefont{M.}~\bibnamefont{K\"u\ss{}ner}} (\bibinfo{collaboration}{BESIII}), \bibinfo{journal}{EPJ Web Conf.} \textbf{\bibinfo{volume}{291}}, \bibinfo{pages}{01002} (\bibinfo{year}{2024})\relax
\mciteBstWouldAddEndPuncttrue
\mciteSetBstMidEndSepPunct{\mcitedefaultmidpunct}
{\mcitedefaultendpunct}{\mcitedefaultseppunct}\relax
\EndOfBibitem
\bibitem[{\citenamefont{Oller and Oset}(1998)}]{Oller:1997yg}
\bibinfo{author}{\bibfnamefont{J.~A.} \bibnamefont{Oller}} \bibnamefont{and} \bibinfo{author}{\bibfnamefont{E.}~\bibnamefont{Oset}}, \bibinfo{journal}{Nucl. Phys. A} \textbf{\bibinfo{volume}{629}}, \bibinfo{pages}{739} (\bibinfo{year}{1998}), \eprint{hep-ph/9706487}\relax
\mciteBstWouldAddEndPuncttrue
\mciteSetBstMidEndSepPunct{\mcitedefaultmidpunct}
{\mcitedefaultendpunct}{\mcitedefaultseppunct}\relax
\EndOfBibitem
\bibitem[{\citenamefont{Antreasyan et~al.}(1986)}]{CrystalBall:1985mzc}
\bibinfo{author}{\bibfnamefont{D.}~\bibnamefont{Antreasyan}} \bibnamefont{et~al.} (\bibinfo{collaboration}{Crystal Ball}), \bibinfo{journal}{Phys. Rev. D} \textbf{\bibinfo{volume}{33}}, \bibinfo{pages}{1847} (\bibinfo{year}{1986})\relax
\mciteBstWouldAddEndPuncttrue
\mciteSetBstMidEndSepPunct{\mcitedefaultmidpunct}
{\mcitedefaultendpunct}{\mcitedefaultseppunct}\relax
\EndOfBibitem
\bibitem[{\citenamefont{Althoff et~al.}(1985)}]{TASSO:1985tme}
\bibinfo{author}{\bibfnamefont{M.}~\bibnamefont{Althoff}} \bibnamefont{et~al.} (\bibinfo{collaboration}{TASSO}), \bibinfo{journal}{Z. Phys. C} \textbf{\bibinfo{volume}{29}}, \bibinfo{pages}{189} (\bibinfo{year}{1985})\relax
\mciteBstWouldAddEndPuncttrue
\mciteSetBstMidEndSepPunct{\mcitedefaultmidpunct}
{\mcitedefaultendpunct}{\mcitedefaultseppunct}\relax
\EndOfBibitem
\bibitem[{\citenamefont{Behrend et~al.}(1989)}]{CELLO:1988xbx}
\bibinfo{author}{\bibfnamefont{H.~J.} \bibnamefont{Behrend}} \bibnamefont{et~al.} (\bibinfo{collaboration}{CELLO}), \bibinfo{journal}{Z. Phys. C} \textbf{\bibinfo{volume}{43}}, \bibinfo{pages}{91} (\bibinfo{year}{1989})\relax
\mciteBstWouldAddEndPuncttrue
\mciteSetBstMidEndSepPunct{\mcitedefaultmidpunct}
{\mcitedefaultendpunct}{\mcitedefaultseppunct}\relax
\EndOfBibitem
\bibitem[{\citenamefont{Abe et~al.}(2003)}]{Belle:2003xlt}
\bibinfo{author}{\bibfnamefont{K.}~\bibnamefont{Abe}} \bibnamefont{et~al.} (\bibinfo{collaboration}{Belle}), \bibinfo{journal}{Eur. Phys. J. C} \textbf{\bibinfo{volume}{32}}, \bibinfo{pages}{323} (\bibinfo{year}{2003}), \eprint{hep-ex/0309077}\relax
\mciteBstWouldAddEndPuncttrue
\mciteSetBstMidEndSepPunct{\mcitedefaultmidpunct}
{\mcitedefaultendpunct}{\mcitedefaultseppunct}\relax
\EndOfBibitem
\bibitem[{\citenamefont{Pascalutsa et~al.}(2012)\citenamefont{Pascalutsa, Pauk, and Vanderhaeghen}}]{Pascalutsa:2012pr}
\bibinfo{author}{\bibfnamefont{V.}~\bibnamefont{Pascalutsa}}, \bibinfo{author}{\bibfnamefont{V.}~\bibnamefont{Pauk}}, \bibnamefont{and} \bibinfo{author}{\bibfnamefont{M.}~\bibnamefont{Vanderhaeghen}}, \bibinfo{journal}{Phys. Rev.} \textbf{\bibinfo{volume}{D85}}, \bibinfo{pages}{116001} (\bibinfo{year}{2012})\relax
\mciteBstWouldAddEndPuncttrue
\mciteSetBstMidEndSepPunct{\mcitedefaultmidpunct}
{\mcitedefaultendpunct}{\mcitedefaultseppunct}\relax
\EndOfBibitem
\bibitem[{\citenamefont{Oller et~al.}(1999)\citenamefont{Oller, Oset, and Pelaez}}]{Oller:1998hw}
\bibinfo{author}{\bibfnamefont{J.~A.} \bibnamefont{Oller}}, \bibinfo{author}{\bibfnamefont{E.}~\bibnamefont{Oset}}, \bibnamefont{and} \bibinfo{author}{\bibfnamefont{J.~R.} \bibnamefont{Pelaez}}, \bibinfo{journal}{Phys. Rev. D} \textbf{\bibinfo{volume}{59}}, \bibinfo{pages}{074001} (\bibinfo{year}{1999}), \bibinfo{note}{[Erratum: Phys.Rev.D 60, 099906 (1999), Erratum: Phys.Rev.D 75, 099903 (2007)]}, \eprint{hep-ph/9804209}\relax
\mciteBstWouldAddEndPuncttrue
\mciteSetBstMidEndSepPunct{\mcitedefaultmidpunct}
{\mcitedefaultendpunct}{\mcitedefaultseppunct}\relax
\EndOfBibitem
\bibitem[{\citenamefont{Achasov and Shestakov}(2011)}]{Achasov:2009ee}
\bibinfo{author}{\bibfnamefont{N.~N.} \bibnamefont{Achasov}} \bibnamefont{and} \bibinfo{author}{\bibfnamefont{G.~N.} \bibnamefont{Shestakov}}, \bibinfo{journal}{Phys. Usp.} \textbf{\bibinfo{volume}{54}}, \bibinfo{pages}{799} (\bibinfo{year}{2011}), \eprint{0905.2017}\relax
\mciteBstWouldAddEndPuncttrue
\mciteSetBstMidEndSepPunct{\mcitedefaultmidpunct}
{\mcitedefaultendpunct}{\mcitedefaultseppunct}\relax
\EndOfBibitem
\bibitem[{\citenamefont{Achasov and Shestakov}(2012)}]{Achasov:2012sc}
\bibinfo{author}{\bibfnamefont{N.~N.} \bibnamefont{Achasov}} \bibnamefont{and} \bibinfo{author}{\bibfnamefont{G.~N.} \bibnamefont{Shestakov}}, \bibinfo{journal}{JETP Lett.} \textbf{\bibinfo{volume}{96}}, \bibinfo{pages}{493} (\bibinfo{year}{2012}), \eprint{1210.0739}\relax
\mciteBstWouldAddEndPuncttrue
\mciteSetBstMidEndSepPunct{\mcitedefaultmidpunct}
{\mcitedefaultendpunct}{\mcitedefaultseppunct}\relax
\EndOfBibitem
\bibitem[{\citenamefont{Albaladejo et~al.}(2017)\citenamefont{Albaladejo, Daub, Hanhart, Kubis, and Moussallam}}]{Albaladejo:2016mad}
\bibinfo{author}{\bibfnamefont{M.}~\bibnamefont{Albaladejo}}, \bibinfo{author}{\bibfnamefont{J.~T.} \bibnamefont{Daub}}, \bibinfo{author}{\bibfnamefont{C.}~\bibnamefont{Hanhart}}, \bibinfo{author}{\bibfnamefont{B.}~\bibnamefont{Kubis}}, \bibnamefont{and} \bibinfo{author}{\bibfnamefont{B.}~\bibnamefont{Moussallam}}, \bibinfo{journal}{JHEP} \textbf{\bibinfo{volume}{04}}, \bibinfo{pages}{010} (\bibinfo{year}{2017}), \eprint{1611.03502}\relax
\mciteBstWouldAddEndPuncttrue
\mciteSetBstMidEndSepPunct{\mcitedefaultmidpunct}
{\mcitedefaultendpunct}{\mcitedefaultseppunct}\relax
\EndOfBibitem
\bibitem[{\citenamefont{Albaladejo and Moussallam}(2017)}]{Albaladejo:2017hhj}
\bibinfo{author}{\bibfnamefont{M.}~\bibnamefont{Albaladejo}} \bibnamefont{and} \bibinfo{author}{\bibfnamefont{B.}~\bibnamefont{Moussallam}}, \bibinfo{journal}{Eur. Phys. J. C} \textbf{\bibinfo{volume}{77}}, \bibinfo{pages}{508} (\bibinfo{year}{2017}), \eprint{1702.04931}\relax
\mciteBstWouldAddEndPuncttrue
\mciteSetBstMidEndSepPunct{\mcitedefaultmidpunct}
{\mcitedefaultendpunct}{\mcitedefaultseppunct}\relax
\EndOfBibitem
\bibitem[{\citenamefont{Donoghue et~al.}(1990)\citenamefont{Donoghue, Gasser, and Leutwyler}}]{Donoghue:1990xh}
\bibinfo{author}{\bibfnamefont{J.~F.} \bibnamefont{Donoghue}}, \bibinfo{author}{\bibfnamefont{J.}~\bibnamefont{Gasser}}, \bibnamefont{and} \bibinfo{author}{\bibfnamefont{H.}~\bibnamefont{Leutwyler}}, \bibinfo{journal}{Nucl. Phys. B} \textbf{\bibinfo{volume}{343}}, \bibinfo{pages}{341} (\bibinfo{year}{1990})\relax
\mciteBstWouldAddEndPuncttrue
\mciteSetBstMidEndSepPunct{\mcitedefaultmidpunct}
{\mcitedefaultendpunct}{\mcitedefaultseppunct}\relax
\EndOfBibitem
\bibitem[{\citenamefont{Moussallam}(2000)}]{Moussallam:1999aq}
\bibinfo{author}{\bibfnamefont{B.}~\bibnamefont{Moussallam}}, \bibinfo{journal}{Eur. Phys. J. C} \textbf{\bibinfo{volume}{14}}, \bibinfo{pages}{111} (\bibinfo{year}{2000}), \eprint{hep-ph/9909292}\relax
\mciteBstWouldAddEndPuncttrue
\mciteSetBstMidEndSepPunct{\mcitedefaultmidpunct}
{\mcitedefaultendpunct}{\mcitedefaultseppunct}\relax
\EndOfBibitem
\bibitem[{\citenamefont{Noether}(1921)}]{FritzNoether}
\bibinfo{author}{\bibfnamefont{F.}~\bibnamefont{Noether}}, \bibinfo{journal}{Math. Ann.} \textbf{\bibinfo{volume}{82}}, \bibinfo{pages}{42} (\bibinfo{year}{1921})\relax
\mciteBstWouldAddEndPuncttrue
\mciteSetBstMidEndSepPunct{\mcitedefaultmidpunct}
{\mcitedefaultendpunct}{\mcitedefaultseppunct}\relax
\EndOfBibitem
\bibitem[{\citenamefont{Muskhelishvili}(1953)}]{Muskhelishvili}
\bibinfo{author}{\bibfnamefont{N.~L.} \bibnamefont{Muskhelishvili}}, \bibinfo{journal}{Singular Integral Equations (P. Noordhoff, Groningen)}  (\bibinfo{year}{1953})\relax
\mciteBstWouldAddEndPuncttrue
\mciteSetBstMidEndSepPunct{\mcitedefaultmidpunct}
{\mcitedefaultendpunct}{\mcitedefaultseppunct}\relax
\EndOfBibitem
\bibitem[{\citenamefont{Badalian et~al.}(1982)\citenamefont{Badalian, Kok, Polikarpov, and Simonov}}]{Badalian:1981xj}
\bibinfo{author}{\bibfnamefont{A.~M.} \bibnamefont{Badalian}}, \bibinfo{author}{\bibfnamefont{L.~P.} \bibnamefont{Kok}}, \bibinfo{author}{\bibfnamefont{M.~I.} \bibnamefont{Polikarpov}}, \bibnamefont{and} \bibinfo{author}{\bibfnamefont{Y.~A.} \bibnamefont{Simonov}}, \bibinfo{journal}{Phys. Rept.} \textbf{\bibinfo{volume}{82}}, \bibinfo{pages}{31} (\bibinfo{year}{1982})\relax
\mciteBstWouldAddEndPuncttrue
\mciteSetBstMidEndSepPunct{\mcitedefaultmidpunct}
{\mcitedefaultendpunct}{\mcitedefaultseppunct}\relax
\EndOfBibitem
\bibitem[{\citenamefont{Rodas et~al.}(2022)\citenamefont{Rodas, Pilloni, Albaladejo, Fernandez-Ramirez, Mathieu, and Szczepaniak}}]{Rodas:2021tyb}
\bibinfo{author}{\bibfnamefont{A.}~\bibnamefont{Rodas}}, \bibinfo{author}{\bibfnamefont{A.}~\bibnamefont{Pilloni}}, \bibinfo{author}{\bibfnamefont{M.}~\bibnamefont{Albaladejo}}, \bibinfo{author}{\bibfnamefont{C.}~\bibnamefont{Fernandez-Ramirez}}, \bibinfo{author}{\bibfnamefont{V.}~\bibnamefont{Mathieu}}, \bibnamefont{and} \bibinfo{author}{\bibfnamefont{A.~P.} \bibnamefont{Szczepaniak}} (\bibinfo{collaboration}{Joint Physics Analysis Center}), \bibinfo{journal}{Eur. Phys. J. C} \textbf{\bibinfo{volume}{82}}, \bibinfo{pages}{80} (\bibinfo{year}{2022}), \eprint{2110.00027}\relax
\mciteBstWouldAddEndPuncttrue
\mciteSetBstMidEndSepPunct{\mcitedefaultmidpunct}
{\mcitedefaultendpunct}{\mcitedefaultseppunct}\relax
\EndOfBibitem
\bibitem[{\citenamefont{Oller and Oset}(1999)}]{Oller:1998zr}
\bibinfo{author}{\bibfnamefont{J.~A.} \bibnamefont{Oller}} \bibnamefont{and} \bibinfo{author}{\bibfnamefont{E.}~\bibnamefont{Oset}}, \bibinfo{journal}{Phys. Rev. D} \textbf{\bibinfo{volume}{60}}, \bibinfo{pages}{074023} (\bibinfo{year}{1999}), \eprint{hep-ph/9809337}\relax
\mciteBstWouldAddEndPuncttrue
\mciteSetBstMidEndSepPunct{\mcitedefaultmidpunct}
{\mcitedefaultendpunct}{\mcitedefaultseppunct}\relax
\EndOfBibitem
\bibitem[{\citenamefont{Guo and Oller}(2011)}]{Guo:2011pa}
\bibinfo{author}{\bibfnamefont{Z.-H.} \bibnamefont{Guo}} \bibnamefont{and} \bibinfo{author}{\bibfnamefont{J.~A.} \bibnamefont{Oller}}, \bibinfo{journal}{Phys. Rev. D} \textbf{\bibinfo{volume}{84}}, \bibinfo{pages}{034005} (\bibinfo{year}{2011}), \eprint{1104.2849}\relax
\mciteBstWouldAddEndPuncttrue
\mciteSetBstMidEndSepPunct{\mcitedefaultmidpunct}
{\mcitedefaultendpunct}{\mcitedefaultseppunct}\relax
\EndOfBibitem
\bibitem[{\citenamefont{Baru et~al.}(2005)\citenamefont{Baru, Haidenbauer, Hanhart, Kudryavtsev, and Meissner}}]{Baru:2004xg}
\bibinfo{author}{\bibfnamefont{V.}~\bibnamefont{Baru}}, \bibinfo{author}{\bibfnamefont{J.}~\bibnamefont{Haidenbauer}}, \bibinfo{author}{\bibfnamefont{C.}~\bibnamefont{Hanhart}}, \bibinfo{author}{\bibfnamefont{A.~E.} \bibnamefont{Kudryavtsev}}, \bibnamefont{and} \bibinfo{author}{\bibfnamefont{U.-G.} \bibnamefont{Meissner}}, \bibinfo{journal}{Eur. Phys. J. A} \textbf{\bibinfo{volume}{23}}, \bibinfo{pages}{523} (\bibinfo{year}{2005}), \eprint{nucl-th/0410099}\relax
\mciteBstWouldAddEndPuncttrue
\mciteSetBstMidEndSepPunct{\mcitedefaultmidpunct}
{\mcitedefaultendpunct}{\mcitedefaultseppunct}\relax
\EndOfBibitem
\bibitem[{\citenamefont{Dudek et~al.}(2016)\citenamefont{Dudek, Edwards, and Wilson}}]{Dudek:2016cru}
\bibinfo{author}{\bibfnamefont{J.~J.} \bibnamefont{Dudek}}, \bibinfo{author}{\bibfnamefont{R.~G.} \bibnamefont{Edwards}}, \bibnamefont{and} \bibinfo{author}{\bibfnamefont{D.~J.} \bibnamefont{Wilson}} (\bibinfo{collaboration}{Hadron Spectrum}), \bibinfo{journal}{Phys. Rev. D} \textbf{\bibinfo{volume}{93}}, \bibinfo{pages}{094506} (\bibinfo{year}{2016}), \eprint{1602.05122}\relax
\mciteBstWouldAddEndPuncttrue
\mciteSetBstMidEndSepPunct{\mcitedefaultmidpunct}
{\mcitedefaultendpunct}{\mcitedefaultseppunct}\relax
\EndOfBibitem
\bibitem[{\citenamefont{Wilson}(2016)}]{Wilson:2016rid}
\bibinfo{author}{\bibfnamefont{D.~J.} \bibnamefont{Wilson}}, \bibinfo{journal}{PoS} \textbf{\bibinfo{volume}{LATTICE2016}}, \bibinfo{pages}{016} (\bibinfo{year}{2016}), \eprint{1611.07281}\relax
\mciteBstWouldAddEndPuncttrue
\mciteSetBstMidEndSepPunct{\mcitedefaultmidpunct}
{\mcitedefaultendpunct}{\mcitedefaultseppunct}\relax
\EndOfBibitem
\bibitem[{\citenamefont{Fearing and Scherer}(1998)}]{Fearing:1996gs}
\bibinfo{author}{\bibfnamefont{H.~W.} \bibnamefont{Fearing}} \bibnamefont{and} \bibinfo{author}{\bibfnamefont{S.}~\bibnamefont{Scherer}}, \bibinfo{journal}{Few Body Syst.} \textbf{\bibinfo{volume}{23}}, \bibinfo{pages}{111} (\bibinfo{year}{1998}), \eprint{nucl-th/9607056}\relax
\mciteBstWouldAddEndPuncttrue
\mciteSetBstMidEndSepPunct{\mcitedefaultmidpunct}
{\mcitedefaultendpunct}{\mcitedefaultseppunct}\relax
\EndOfBibitem
\bibitem[{\citenamefont{Colangelo et~al.}(2015)\citenamefont{Colangelo, Hoferichter, Procura, and Stoffer}}]{Colangelo:2015ama}
\bibinfo{author}{\bibfnamefont{G.}~\bibnamefont{Colangelo}}, \bibinfo{author}{\bibfnamefont{M.}~\bibnamefont{Hoferichter}}, \bibinfo{author}{\bibfnamefont{M.}~\bibnamefont{Procura}}, \bibnamefont{and} \bibinfo{author}{\bibfnamefont{P.}~\bibnamefont{Stoffer}}, \bibinfo{journal}{JHEP} \textbf{\bibinfo{volume}{09}}, \bibinfo{pages}{074} (\bibinfo{year}{2015}), \eprint{1506.01386}\relax
\mciteBstWouldAddEndPuncttrue
\mciteSetBstMidEndSepPunct{\mcitedefaultmidpunct}
{\mcitedefaultendpunct}{\mcitedefaultseppunct}\relax
\EndOfBibitem
\bibitem[{\citenamefont{Deineka et~al.}(2019{\natexlab{b}})\citenamefont{Deineka, Danilkin, and Vanderhaeghen}}]{Deineka:2019bey}
\bibinfo{author}{\bibfnamefont{O.}~\bibnamefont{Deineka}}, \bibinfo{author}{\bibfnamefont{I.}~\bibnamefont{Danilkin}}, \bibnamefont{and} \bibinfo{author}{\bibfnamefont{M.}~\bibnamefont{Vanderhaeghen}}, \bibinfo{journal}{Acta Phys. Polon. B} \textbf{\bibinfo{volume}{50}}, \bibinfo{pages}{1901} (\bibinfo{year}{2019}{\natexlab{b}})\relax
\mciteBstWouldAddEndPuncttrue
\mciteSetBstMidEndSepPunct{\mcitedefaultmidpunct}
{\mcitedefaultendpunct}{\mcitedefaultseppunct}\relax
\EndOfBibitem
\bibitem[{\citenamefont{Stamen et~al.}(2022)\citenamefont{Stamen, Hariharan, Hoferichter, Kubis, and Stoffer}}]{Stamen:2022uqh}
\bibinfo{author}{\bibfnamefont{D.}~\bibnamefont{Stamen}}, \bibinfo{author}{\bibfnamefont{D.}~\bibnamefont{Hariharan}}, \bibinfo{author}{\bibfnamefont{M.}~\bibnamefont{Hoferichter}}, \bibinfo{author}{\bibfnamefont{B.}~\bibnamefont{Kubis}}, \bibnamefont{and} \bibinfo{author}{\bibfnamefont{P.}~\bibnamefont{Stoffer}}, \bibinfo{journal}{Eur. Phys. J. C} \textbf{\bibinfo{volume}{82}}, \bibinfo{pages}{432} (\bibinfo{year}{2022}), \eprint{2202.11106}\relax
\mciteBstWouldAddEndPuncttrue
\mciteSetBstMidEndSepPunct{\mcitedefaultmidpunct}
{\mcitedefaultendpunct}{\mcitedefaultseppunct}\relax
\EndOfBibitem
\bibitem[{\citenamefont{Schuler et~al.}(1998)\citenamefont{Schuler, Berends, and van Gulik}}]{Schuler:1997yw}
\bibinfo{author}{\bibfnamefont{G.~A.} \bibnamefont{Schuler}}, \bibinfo{author}{\bibfnamefont{F.~A.} \bibnamefont{Berends}}, \bibnamefont{and} \bibinfo{author}{\bibfnamefont{R.}~\bibnamefont{van Gulik}}, \bibinfo{journal}{Nucl. Phys. B} \textbf{\bibinfo{volume}{523}}, \bibinfo{pages}{423} (\bibinfo{year}{1998}), \eprint{hep-ph/9710462}\relax
\mciteBstWouldAddEndPuncttrue
\mciteSetBstMidEndSepPunct{\mcitedefaultmidpunct}
{\mcitedefaultendpunct}{\mcitedefaultseppunct}\relax
\EndOfBibitem
\bibitem[{\citenamefont{Ambrosino et~al.}(2009)}]{KLOE:2009ehb}
\bibinfo{author}{\bibfnamefont{F.}~\bibnamefont{Ambrosino}} \bibnamefont{et~al.} (\bibinfo{collaboration}{KLOE}), \bibinfo{journal}{Phys. Lett. B} \textbf{\bibinfo{volume}{681}}, \bibinfo{pages}{5} (\bibinfo{year}{2009}), \eprint{0904.2539}\relax
\mciteBstWouldAddEndPuncttrue
\mciteSetBstMidEndSepPunct{\mcitedefaultmidpunct}
{\mcitedefaultendpunct}{\mcitedefaultseppunct}\relax
\EndOfBibitem
\bibitem[{\citenamefont{Ablikim et~al.}(2018)}]{BESIII:2017djm}
\bibinfo{author}{\bibfnamefont{M.}~\bibnamefont{Ablikim}} \bibnamefont{et~al.} (\bibinfo{collaboration}{BESIII}), \bibinfo{journal}{Phys. Rev. D} \textbf{\bibinfo{volume}{97}}, \bibinfo{pages}{012003} (\bibinfo{year}{2018}), \eprint{1709.04627}\relax
\mciteBstWouldAddEndPuncttrue
\mciteSetBstMidEndSepPunct{\mcitedefaultmidpunct}
{\mcitedefaultendpunct}{\mcitedefaultseppunct}\relax
\EndOfBibitem
\end{mcitethebibliography}
	
\end{document}